\newif\ifrevtex
\IfFileExists{revtex4-2.cls}{%
  \revtextrue
  \documentclass[aps,prb,twocolumn,superscriptaddress,nofootinbib,floatfix]{revtex4-2}
}{%
  \revtexfalse
  \documentclass[10pt,twocolumn]{article}
}
\usepackage{amsmath,amssymb,bm}
\usepackage[hidelinks]{hyperref}
\usepackage{graphicx}
\usepackage{xcolor}
\usepackage{booktabs}
\usepackage{tikz}
\usetikzlibrary{arrows.meta,calc,positioning,shapes.geometric}
\ifrevtex\else
  \usepackage[letterpaper,margin=0.67in,columnsep=0.25in]{geometry}
  \usepackage{authblk}
  \providecommand{\onlinecite}[1]{\cite{#1}}
\fi

\newcommand{\Astr}{\mathcal A_{\mathrm{str}}}
\newcommand{\Lamy}{\Lambda_y}
\newcommand{\Lstat}{\Lambda_y^{\mathrm{stat}}}
\newcommand{\Ldyn}{\Lambda_y^{\mathrm{dyn}}}
\newcommand{\Mbar}{\overline{\mathcal M}_{3}}
\newcommand{\Ffab}{\mathcal F_y}
\newcommand{\Me}{\mathsf M_y}
\newcommand{\ii}{\mathrm i}
\newcommand{\nn}{\nonumber}
\newcommand{\pdv}[2]{\frac{\partial #1}{\partial #2}}
\newcommand{\ket}[1]{\left|#1\right\rangle}
\newcommand{\ev}[1]{\left\langle #1\right\rangle}
\newcommand{\mel}[3]{\left\langle #1\middle|#2\middle|#3\right\rangle}
\newcommand{\innerproduct}[2]{\left\langle #1\middle|#2\right\rangle}
\newcommand{\1}{\mathbb{1}}
\DeclareMathOperator{\Tr}{Tr}

\begin{document}

\title{Mirror-Constrained Optical Hall Rectification in Asymmetric Gold Gratings:\\
Edelstein Conversion and a Phase-Seeded Kerr Four-Wave Mixing Instability}

\ifrevtex
  \author{Michael N. Leuenberger}
  \affiliation{NanoScience Technology Center, Department of Physics, and College of Optics and Photonics (CREOL), University of Central Florida, Orlando, Florida 32826, USA}
  \author{Richard M. Osgood III}
  \affiliation{U.S. Army Combat Capabilities Development Command Soldier Center, Natick, Massachusetts 01760, USA}
\else
  \author[1]{Michael N. Leuenberger}
  \author[2]{Richard M. Osgood III}
  \affil[1]{NanoScience Technology Center, Department of Physics, and College of Optics and Photonics (CREOL), University of Central Florida, Orlando, Florida 32826, USA}
  \affil[2]{U.S. Army Combat Capabilities Development Command Soldier Center, Natick, Massachusetts 01760, USA}
  \date{}
\fi

\newcommand{\manuscriptabstract}{%
Transverse optical rectification has been reported in a one-dimensional asymmetric Au grating even though an ideal profile $h(x)$ retains the $xz$ mirror $\Me:y\to-y$.  Because $J_y$ is odd under $\Me$, whereas the $p$-polarized bilinears $|E_x|^2$, $|E_z|^2$, and $\mathrm{Re}(E_xE_z^*)$ are even, the transverse dc current requires a distinct mirror-odd factor $\Lamy$.  We formulate a symmetry-complete effective model combining the origin-independent cell invariant $\Astr=\mathrm{Im}(h_1^2h_2^*)$, Drude--Edelstein spin polarization, and spin--orbit conversion.  We then propose a dynamical realization of $\Lamy$ for a finite-width device.  A driven pump-like plasmonic mode at $k_y\simeq0$ generates a mirror pair of transverse surface-plasmon-polariton quasinormal modes $(K_x,\pm q_y)$ through the same third-order susceptibility that produces self- and cross-Kerr shifts.  Four-wave mixing creates the pair and locks its phase sum but, by itself, cannot change the pair-population difference.  Weak coherent backscattering makes the relative drive phase observable; a small phase difference from incident $k_y$, spot displacement, edges, or disorder then unfolds a Kerr directional instability and deterministically selects its sign.  The resulting population imbalance supplies the dynamical mirror-odd factor $\Ldyn=\lambda_\eta\eta$ in the structure-dependent $\Astr\Lamy\langle j_xs_y\rangle_t$ side-jump-like channel, which vanishes for the centered symmetric grating because $\Astr=0$.  We give the complete three-mode equations, the pump-only and saturated-state Bogoliubov problems, the imperfect-pitchfork reduction, and a validated normalized numerical realization.  The normalized equations exhibit a robust supercritical pitchfork with an approximately one-half critical exponent and phase-reversed branch selection.
}
\ifrevtex
  \begin{abstract}\manuscriptabstract\end{abstract}
  \maketitle
\else
  \maketitle
  \begin{abstract}\manuscriptabstract\end{abstract}
\fi

\section{Introduction}
\label{sec:introduction}

Optical rectification (OR) is the zero-frequency component of a nonlinear optical response.  For a monochromatic field it can be written schematically as
\begin{equation}
J_i^{(0)}=\sigma_{ijk}^{(2)}(0;\omega,-\omega)
E_j(\omega)E_k^*(\omega),
\label{eq:intro_or}
\end{equation}
and is therefore quadratic in the complex optical-field amplitude and linear in optical intensity~\cite{Bass1965,Boyd2020}.  This statement must be distinguished from the \emph{linear} electromagnetic process by which an incident photon or coherent optical field couples through a grating into an SPP.  In linear Maxwell theory every scattered-field amplitude, including the SPP amplitude, is proportional to the incident field, whereas reflected, transmitted, absorbed, and SPP powers are quadratic in that field.  The dc current in Eq.~\eqref{eq:intro_or} is consequently a nonlinear response even when the resonant local field entering it is obtained from a linear SPP-coupling calculation.  The grating redistributes the coherent optical energy among radiation, absorption, and surface-wave channels; it does not require a literal partition of one photon into two SPPs.  In the coherent-field, or semiclassical optical, regime relevant here, the experimentally observed power laws and phase-matching resonances are more naturally described by electromagnetic fields and nonlinear response functions than by counting individual pump photons.  This terminology concerns the optical-field description only and does not imply that the spin--orbit-coupled electronic response is classical.

Metallic gratings are particularly effective at converting free-space radiation into surface modes because their reciprocal lattice supplies the in-plane momentum missing from a smooth interface.  Deep and asymmetric gold gratings support several families of surface-plasmon-related resonances whose frequencies, linewidths, and field localization depend strongly on groove depth and unit-cell asymmetry~\cite{Kreiter2002}.  Their large near fields also enhance nonlinear polarization, while radiation leakage, Ohmic loss, surface response, and bulk response remain inseparably mixed in an open plasmonic structure~\cite{Kauranen2012,Sayer2025}.  This combination makes asymmetric Au gratings natural platforms for studying rectification, nonlinear surface transport, and driven plasmonic instabilities, but it also makes symmetry accounting essential: local enhancement determines the size of an allowed response, not whether that response is allowed at all.

The broader context includes electromagnetic ratchets, in which a spatially asymmetric periodic structure and a time-dependent drive produce a directed dc response.  Experiments and theory on semiconductor and graphene-based structures have established polarization-sensitive ratchet transport, classical and quantum contributions, and plasmon-enhanced photocurrents~\cite{Olbrich2009,Olbrich2011,Nalitov2012,Rozhansky2015}.  These works are important precedents for converting an ac field into directed transport in a patterned system.  They also clarify a point central to the present problem: broken inversion along the modulation direction can select a longitudinal ratchet current, but a current transverse to an otherwise one-dimensional modulation is constrained by the remaining transverse mirror symmetry and therefore requires an additional mirror-plane-odd factor.

A sequence of experiments coauthored by Osgood established the specific Au-grating motivation for this work.  Moroshkin, Osgood, Xu, and co-workers first reported longitudinal optical rectification in a one-dimensional asymmetric gold grating and interpreted the signal as ratchet transport enhanced at SPP resonances~\cite{Moroshkin2021}.  The measured dc response followed the grating asymmetry, illumination geometry, and resonant absorption even though the underlying optical excitation was described by linear grating electrodynamics.  In a subsequent experiment, the same collaboration electrically detected the transverse spin associated with SPP spin--momentum locking, providing direct evidence that the evanescent plasmonic field can imprint its optical spin on electronic transport in gold~\cite{Moroshkin2022}.  Finally, Plumitallo, Moroshkin, Osgood, and Xu reported a transverse, Hall-like OR current in a one-dimensional asymmetric Au grating; the signal was selected by $p$ polarization and appeared at the same incident-angle regions in which grating momentum matching excites the relevant SPPs~\cite{Plumitallo2023}.  These three experiments---longitudinal ratchet rectification, electrical detection of SPP spin--momentum locking, and transverse optical Hall rectification---constitute the complete experimental Au-grating sequence on which the present theory is based.

The transverse experiment poses a sharper symmetry question than the longitudinal effect.  An ideal profile $h(x)$ can break $x\to-x$ and thereby distinguish the two longitudinal directions, but it remains invariant under the reflection
$\Me:(x,y,z)\mapsto(x,-y,z)$.  A polar transverse current changes sign under this mirror, whereas the incident $p$-polarized field, its local components $E_x$ and $E_z$, and the quadratic combinations $|E_x|^2$, $|E_z|^2$, and $\operatorname{Re}(E_xE_z^*)$ are mirror even.  Thus the ideal one-dimensional geometry cannot by itself convert those bilinears into $J_y^{(0)}$.  The observed current is not called into question; rather, the fitted transverse susceptibility must implicitly contain an additional mirror-odd material, structural, or dynamical variable.  Identifying this missing variable is the organizing objective of the present paper.

Intrinsic nonlinear Hall mechanisms provide a useful comparison.  Berry-curvature dipoles and related quantum-geometric quantities generate second-order Hall currents in time-reversal-invariant crystals when the material symmetry permits the requisite pseudotensor~\cite{Sodemann2015,Parker2019,Xiao2019,Du2019,Du2021,Koenig2021,Jiang2025}.  Disorder, skew scattering, and side-jump terms can contribute at the same order and must be treated consistently~\cite{Berger1970,Sinitsyn2008,Nagaosa2010,Sinova2015,Atencia2022}.  For the ideal grating considered here, however, the remaining $xz$ mirror removes an $x$-directed Berry-curvature dipole capable of producing the measured $J_y$ from $|E_x|^2$.  Quantum geometry may still enter a microscopic surface response once a mirror-odd quantity is present, but it does not eliminate the need to specify that quantity.

Gold also supplies a natural spin--orbit conversion channel.  Au(111) supports spin-split surface states with a Rashba-like helical texture~\cite{Henk2003,Yan2015}, and an in-plane electric field can generate an Edelstein spin polarization perpendicular to the driven charge current~\cite{Edelstein1990,Johansson2021}.  The Rashba eigenstates and their momentum-locked spin expectation values are quantum-mechanical consequences of spin--orbit coupling.  The field-driven redistribution of these states, and hence the ac Edelstein polarization $s_y(t)$, is likewise a quantum electronic effect even though the optical drive is represented semiclassically.  In this materials-based sense, the patterned Au grating is a quantum-plasmonic device in which coherent plasmonic fields drive a quantum spin--orbit response.  We do not describe it as a quantum sensor, because no nonclassical light, entanglement, or quantum-enhanced detection protocol is assumed.  More broadly, surface-state engineering has been proposed as a route to nonlinear charge and spin photocurrents~\cite{Sivianes2025}.  The longitudinal pump produces an ac $s_y$; together with spin-dependent scattering this permits a quadratic side-jump-like correlator.  Because both $j_x$ and the axial component $s_y$ are even under $\Me$, their product requires a separate mirror-odd multiplier $\Lamy$.  The dynamical proposal below identifies the directional SPP imbalance as a possible realization $\Ldyn=\lambda_\eta\eta$, while the inversion-odd cell factor $\Astr$ enforces the observed distinction between asymmetric and centered symmetric gratings.

One possibility is that $\Lamy$ is static, arising from sidewall texture, deposition direction, Au(111) azimuthal texture, contacts, or an intentionally introduced transverse asymmetry.  The alternative developed here is dynamical.  Finite width, a finite laser spot, contacts, and radiative boundaries can support a mirror-related pair of outgoing SPP-like quasinormal modes with momenta $(K_x,+q_y)$ and $(K_x,-q_y)$.  The same third-order susceptibility produces self- and cross-Kerr shifts and degenerate four-wave mixing (FWM), so a pumped plasmonic mode can parametrically populate this pair while conserving both energy and transverse momentum.  Nonlinear SPP-mediated FWM and grating-enhanced FWM have been demonstrated in related plasmonic settings~\cite{Liu2011,Jin2016,Chakraborty2023}, and counterpropagating Kerr resonators exhibit spontaneous directional symmetry breaking, switching, and phase-controlled branch selection~\cite{DelBino2017,Hill2020,Woodley2021,Silver2021,Anashkina2025}.  These results motivate, but do not by themselves prove, the corresponding instability in the Au grating.

The central theoretical point is that pair creation alone cannot choose one transverse direction: the FWM vertex creates $+q_y$ and $-q_y$ excitations together and conserves their population difference.  Directional selection requires an additional mixing or dissipative channel.  In the model below, weak reciprocal backscattering couples the two traveling modes and makes their relative drive phase observable.  A small phase seed---from a controlled incident $k_y$, transverse spot displacement, edges, or disorder---then unfolds a Kerr instability and reproducibly selects the sign of the directional imbalance.  That imbalance supplies the sign of $\Ldyn$ and hence of the dynamical part of the structure-weighted transverse current.  We derive these statements from a three-mode pump--pair model and its Bogoliubov stability matrix, reduce it near threshold to an imperfect-pitchfork normal form, and keep all uncalculated coefficients explicit.

The paper is organized as follows.  Sections~\ref{sec:mirror_pathway}--\ref{sec:conclusion} give the symmetry argument, the FWM--Kerr--QNM model, its stability analysis, the structure-dependent electronic readout, momentum balance, and experimental tests.  The appendices are part of the present PRB article and contain the Maxwell--QNM projection, an analytically transparent two-mode limit, the microscopic Rashba response, a perturbative reduction of the SPP-dressed, structure-dependent scattering vertex, and a Keldysh analysis of the possible structure-independent directional-SPP channel.  They also give the gauge-invariant side-jump construction, the origin-independent structural invariant, and a precise separation between rigorous symmetry results and model-dependent hypotheses.  Appendix~\ref{app:roadmap_scope} provides a detailed map of these derivations, Appendix~\ref{app:direct_keldysh} separates virtual spectral dressing from real population injection in the direct channel, Appendix~\ref{app:rigorous_summary} separates symmetry-enforced conclusions from model-dependent assumptions, and Appendix~\ref{app:literature_remarks} summarizes the relationship to the relevant side-jump, nonlinear-Hall, QNM, and nonlinear-plasmonics literature.  This consolidated format keeps the assumptions and derivations in the archival article rather than relegating the theory required to assess the proposal to a separate online supplement.

\section{Mirror constraint and rectification pathway}
\label{sec:mirror_pathway}

We now apply the symmetry issue identified above directly to the idealized experimental geometry.  The grating is periodic along $x$, invariant along the grooves ($y$), and illuminated nominally in the $xz$ plane.  Its lithographic unit cell breaks $x\to-x$, but an ideal one-dimensional structure retains
\begin{equation}
\Me:(x,y,z)\longmapsto(x,-y,z).
\label{eq:My}
\end{equation}
Under this mirror a polar current obeys $J_y\to-J_y$, while $E_x$, $E_z$, the longitudinal current $j_x$, and the axial Edelstein spin component $s_y$ are even.  Consequently an ideal $h(x)$ model gives $J_y=0$ for $p$-polarized excitation unless a separate $\Me$-odd material, structural, or dynamical quantity is present.

The phenomenological susceptibility used to interpret the original optical Hall-rectification experiment~\cite{Plumitallo2023} parameterized
$P_y^{(2)}=\chi_{yxx}^{(2)}|E_x|^2+2\operatorname{Re}[\chi_{yxz}^{(2)}E_xE_z^*]+\chi_{yzz}^{(2)}|E_z|^2$.
Imposing Eq.~\eqref{eq:My} makes all three displayed coefficients vanish unless they contain an additional mirror-plane-odd factor: the left-hand side is odd and the optical bilinears are even.  This observation does not challenge the measured current or its SPP correlation.  It identifies the missing symmetry variable that any microscopic model must generate.

Intrinsic nonlinear Hall currents from quantum geometry provide a useful comparison~\cite{Sodemann2015,Parker2019,Jiang2025}.  For example, an $x$-directed Berry-curvature dipole would give $J_y^{\rm BCD}\propto D_x|E_x|^2$, but $\Me$ forces $D_x=0$ for the ideal device.  Here we retain an intraband Drude--Edelstein conversion on spin--orbit-coupled Au surface states and ask how a mirror-odd factor can arise in the driven open plasmonic system.

The binary profile supplies a distinct $x$-inversion-odd invariant.  Writing $h(x)=\sum_mh_m e^{\ii G_mx}$, with $G_m=2\pi m/D$, the lowest origin-independent handedness constructed from the first two spatial harmonics is
\begin{equation}
\Astr\equiv\operatorname{Im}(h_1^2h_2^*).
\label{eq:Astr}
\end{equation}
It is invariant under an origin shift, reverses under $x\to-x$, and vanishes for the centered symmetric cell.  Importantly, $\Astr$ is even under $\Me$ and therefore cannot replace $\Lamy$.
A complete derivation of this origin-independent structural invariant for the binary grating profile is given in Appendix~\ref{sec:structural_handedness_detailed}.

\begin{figure*}[t]
\centering
\includegraphics[width=0.985\linewidth]{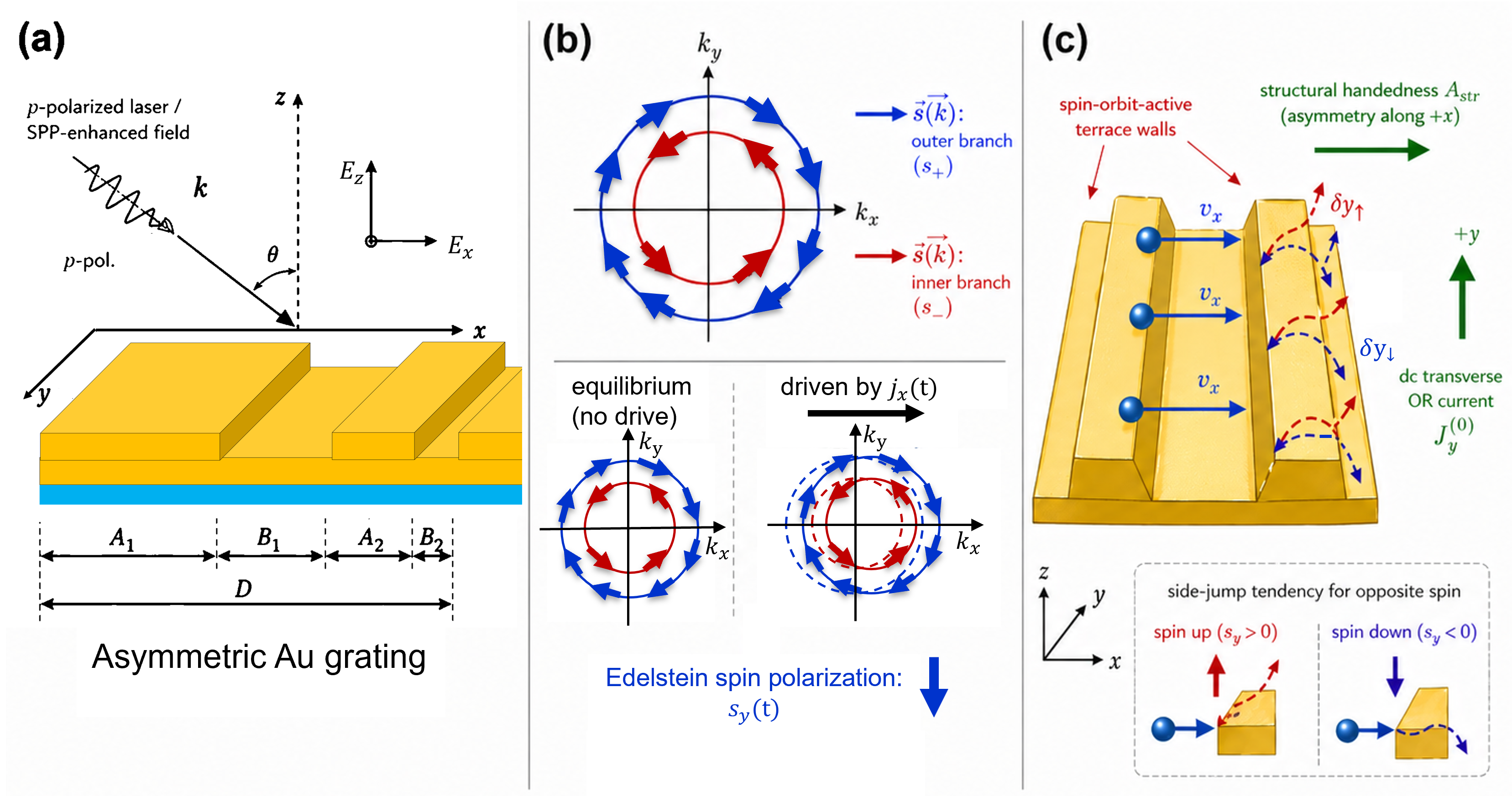}
\caption{$\Astr$-weighted Edelstein rectification pathway.  The asymmetric binary profile supplies the $x$-odd invariant $\Astr$ and enhances the local $p$-polarized field through SPP resonances.  Longitudinal drift produces an ac Edelstein spin density $s_y(t)$.  Spin--orbit-active conversion can produce this contribution to $J_y^{(0)}$ only in the presence of a separate $\Me$-odd factor $\Lamy$.  The sidewall sketch is a generic conversion vertex, not an assignment of the dominant microscopic mechanism.}
\label{fig:mechanism}
\end{figure*}

Figure~\ref{fig:mechanism} summarizes the $\Astr$-weighted pathway.  The asymmetric profile supplies the longitudinal handedness $\Astr$ and the SPP-enhanced local field, the longitudinal drive generates $j_x(t)$ and $s_y(t)$, and a distinct mirror-odd factor $\Lamy$ is required at the spin--orbit conversion step to obtain $J_y^{(0)}$.  The figure therefore separates plasmonic field enhancement and Edelstein polarization from the additional symmetry breaking required for this transverse dc channel.

For the electronic part, the leading surface Hamiltonian is
\begin{equation}
H_R(\bm k)=\frac{\hbar^2k^2}{2m}+\alpha_R(k_x\sigma_y-k_y\sigma_x),
\label{eq:Rashba}
\end{equation}
whose helical spin texture permits an $E_x$-driven Edelstein component $s_y$~\cite{Edelstein1990,Yan2015}.  A minimal moment reduction for the $\Astr$-weighted channel is
\begin{subequations}
\begin{align}
\dot j_x+\tau^{-1}j_x&=\frac{ne^2}{m}E_x^{\rm loc}(t),\\
\dot s_y+\tau_s^{-1}s_y&=\kappa_E j_x,\\
\dot J_{y,{\rm sj}}+\tau_y^{-1}J_{y,{\rm sj}}&=\Astr\Lamy j_xs_y .
\end{align}
\label{eq:moments}
\end{subequations}
The helicity-resolved equilibrium cancellation and the driven momentum-space displacement underlying the second line of Eq.~\eqref{eq:moments} are visualized in Appendix Fig.~\ref{fig:Supp_Rashba_Edelstein_visualization}.
The Rashba conventions and microscopic electronic model are defined in Appendix~\ref{app:rashba_model}.  Appendices~\ref{app:edelstein_derivation} and \ref{sec:Supp_firstharmonic} provide the semiclassical, Kubo, and first-harmonic derivations underlying Eq.~\eqref{eq:moments}, while Appendix~\ref{sec:Supp_frequency_magnitude} gives the corresponding frequency dependence and magnitude estimates.
An isotropic one-dimensional binary-profile vertex may produce spin-selective scattering, but it conserves $k_y$ and respects $\Me$; hence it cannot generate the last line with $\Lamy=0$.  The projected scalar and Hermitian terrace--sidewall Pauli matrix elements are derived in Appendix~\ref{sec:Supp_projected_matrix_elements}.  Appendix~\ref{app:microscopic_vertex} then shows how interference between the $G+G$ and $2G$ grating paths, one projected Pauli insertion, and the mirror-odd part of the SPP-dressed electronic self-energy produces the lowest longitudinal-to-transverse on-shell vertex proportional to $\Astr\lambda_\eta\eta$.  Because the same mirror-odd self-energy also exists outside the $\Astr$ projection, symmetry alone permits a structure-independent direct channel proportional to $\eta$.  Appendix~\ref{app:direct_keldysh} evaluates its Keldysh collision source and shows that its order-$g_{\rm SPP}^2$, $\eta$-linear one-SPP contribution vanishes in the clean, off-resonant Au(111) surface-state model: the $808$-nm photon energy cannot be reconciled with the electronic energy change generated by the available low-order in-plane momentum transfer.  This zero is kinematic rather than symmetry enforced; disorder-assisted, surface-to-bulk, interband, phonon-assisted, and finite-size channels can relax it.  Mode-off-diagonal coherences proportional to $a_+a_-^*$ instead produce a finite-wave-vector elastic dressing; Appendix~\ref{app:direct_keldysh} shows separately why this dressing is not a spatially uniform order-$g_{\rm SPP}^2$ population source.  The corresponding mirror-consistent collision and coordinate-shift projection of the retained structure-dependent channel is given in Appendix~\ref{sec:Supp_corrected_sidejump}.  Possible static mirror-odd realizations and their discriminating tests are discussed in Appendices~\ref{sec:Supp_mirror} and \ref{sec:Supp_tests}.  Solving the first two equations at frequency $\omega$ gives $j_x^{(1)}=\sigma_D E_x^{\rm loc}$ and $s_y^{(1)}=\chi_{EE}E_x^{\rm loc}$, so that
\begin{equation}
J_{y,{\rm sj}}^{(0)}=\frac{\tau_y\Astr\Lamy}{2}
\operatorname{Re}[\chi_{EE}(\omega)\sigma_D^*(\omega)]
|E_x^{\rm loc}|^2.
\label{eq:Jy_general}
\end{equation}

\section{Why a Kerr--FWM transverse pair is plausible}

The proposed optical order parameter concerns a \emph{finite-width} device.  It is not the unequal excitation of the opposite $K_x$ Floquet harmonics of an infinite one-dimensional calculation.  Those harmonics are not independent lateral output ports.  Instead, let a finite illuminated region support a pump-like plasmonic QNM $\Psi_p$ with $k_y\simeq0$ and a mirror pair $\Psi_\pm$ with
\begin{equation}
\bm K_\pm=(K_x,\pm q_y),\qquad \Me\Psi_+=\Psi_-.
\label{eq:mirror_pair}
\end{equation}
Such outgoing QNMs are the natural modal basis for an absorbing, radiating nanostructure~\cite{Sauvan2013,Lalanne2018}.  Finite edges, the finite optical spot, roughness, and contacts quantize and mix lateral surface waves.  Therefore the pair in Eq.~\eqref{eq:mirror_pair} is a physically testable finite-device hypothesis rather than an extra branch inserted into an infinite Bloch spectrum.

There is a concrete nonlinear-optical motivation.  Counterpropagating SPPs have been used to generate coherent surface-mediated four-wave-mixing (FWM) signals~\cite{Liu2011}; resonant plasmonic metasurfaces strongly enhance FWM~\cite{Jin2016}; and gold gratings exhibit FWM enhanced jointly by localized and surface-lattice plasmon resonances~\cite{Chakraborty2023}.  These works establish the ingredients---third-order mixing, surface-wave propagation, and grating-enhanced local fields---but do not demonstrate the specific $\pm q_y$ directional pitchfork proposed here.  The closest dynamical analogue is the Kerr competition of counterpropagating resonator modes~\cite{DelBino2017,Woodley2021,Silver2021}, particularly the demonstration that weak backscattering converts a relative pump phase into deterministic intensity asymmetry~\cite{Anashkina2025}.

For a monochromatic field, the third-order polarization is~\cite{Boyd2020,Kauranen2012}
\begin{equation}
P_i^{(3)}(\omega)=3\epsilon_0
\chi_{ijkl}^{(3)}(-\omega;\omega,\omega,-\omega)
E_jE_kE_l^*,
\label{eq:P3}
\end{equation}
where repeated Cartesian indices are summed.  Expanding the field as
\begin{align}
\bm E(\bm r,t)={}&p(t)\bm e_p(\bm r)e^{-\ii\omega t}
+a_+(t)\bm e_+(\bm r)e^{-\ii\omega t}\nn\\
&+a_-(t)\bm e_-(\bm r)e^{-\ii\omega t}+\text{c.c.},
\label{eq:modeexpansion}
\end{align}
and projecting Maxwell's equation onto the adjoint QNMs generates four classes of terms: pump self-phase modulation, pump--sideband and sideband cross-phase modulation, the phase-conjugate FWM term $p^2a_\mp^*$, and pump depletion $p^*a_+a_-$.  The same $\chi^{(3)}$ therefore produces both Kerr shifts and parametric pair creation; they are not independent phenomenological effects.  The QNM overlaps defining the coefficients are given in Appendix~\ref{app:projection}.

The degenerate pair process obeys
\begin{subequations}
\begin{align}
2\omega_p&=\omega_++\omega_-,\\
2\bm K_p&=\bm K_++\bm K_-+m\bm G,
\end{align}
\label{eq:matching}
\end{subequations}
where $\bm G=(2\pi/D)\hat{\bm x}$ is a grating reciprocal vector.  At $k_y^{\rm inc}=0$, the $y$ component is $q_y+(-q_y)=0$.  Thus the initial FWM event creates the two mirror partners together and conserves transverse momentum exactly.

For Au, ``Kerr'' must be understood as an effective third-order response, not automatically as an instantaneous, real refractive index.  Thin-metal $\chi^{(3)}$ can be delayed and thermally dominated~\cite{Conforti2012}, and nonlinear SPP propagation in Au strip waveguides exhibits pulse-duration-dependent nonlinear absorption~\cite{Lysenko2016}.  Consequently the modal coefficients below may be complex and frequency dependent.  A delayed reservoir should be included if the modulation period, pulse duration, or electron--lattice relaxation time makes it relevant.

\begin{figure*}[t]
\centering
\includegraphics[width=0.985\linewidth]{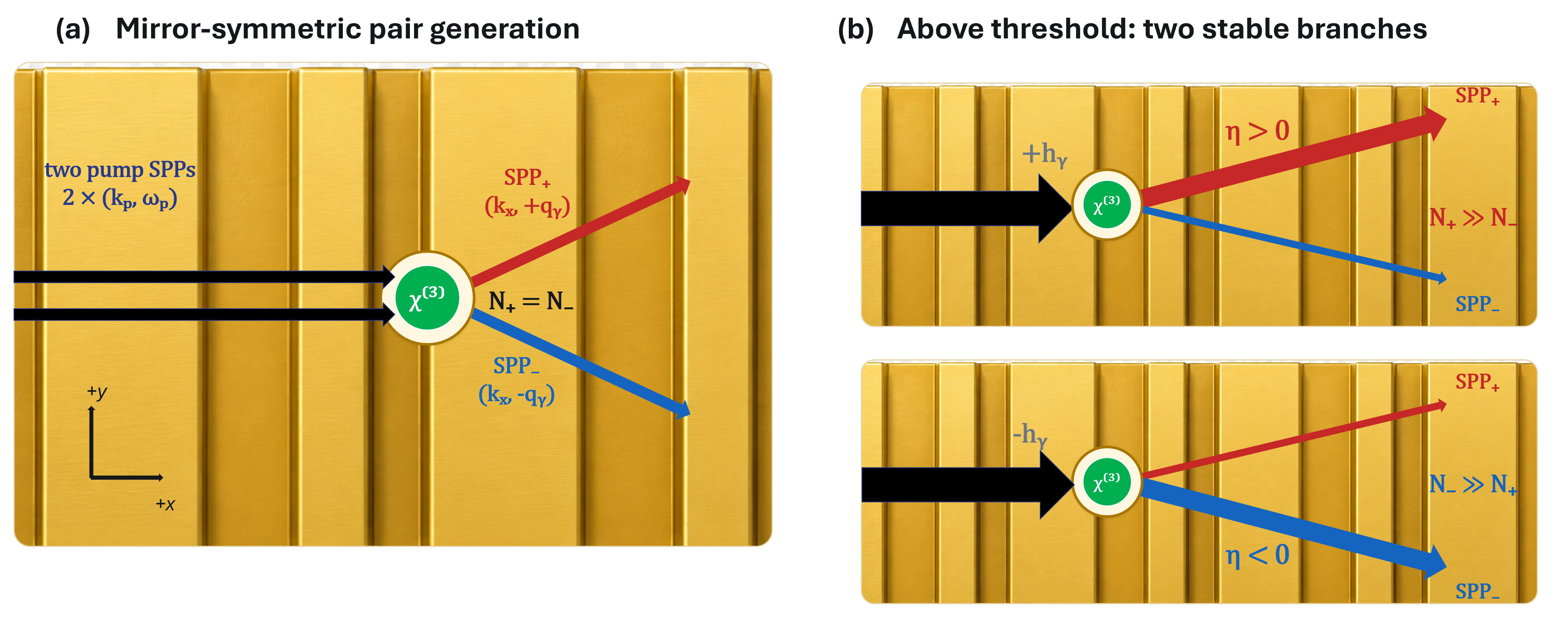}
\caption{\textbf{Kerr four-wave mixing and directional SPP bifurcation in the asymmetric Au grating.}
(a) Two pump SPP quanta with wave vector $\mathbf{k}_{p}$ and frequency $\omega_{p}$ drive the third-order interaction $\chi^{(3)}$ and generate the mirror-related outgoing modes $\mathrm{SPP}_{+}$ and $\mathrm{SPP}_{-}$ with transverse momenta $+q_y$ and $-q_y$, respectively. For $k_{p,y}=0$, four-wave mixing obeys $2\omega_p=\omega_{+}+\omega_{-}$ and $2k_{p,y}=k_{+,y}+k_{-,y}=0$. The mirror-symmetric source therefore produces equal modal populations, $N_{+}=N_{-}$, where $N_{\pm}=|a_{\pm}|^{2}$. 
(b) Above the nonlinear-instability threshold, Kerr--FWM feedback destabilizes the symmetric state and produces two stable directional branches. The normalized imbalance
$\eta=(N_{+}-N_{-})/(N_{+}+N_{-})$ is positive when $N_{+}\gg N_{-}$ and negative when $N_{-}\gg N_{+}$. A small effective mirror-odd seed $h_y$ unfolds the pitchfork: $+h_y$ selects the $\eta>0$ branch, whereas $-h_y$ selects the $\eta<0$ branch. Arrow widths schematically represent the modal populations $N_{\pm}$. The figure illustrates the mechanism of the normalized reduced three-mode model.}
\label{fig:kerr_fwm_spp_bifurcation}
\end{figure*}

Figure~\ref{fig:kerr_fwm_spp_bifurcation} visualizes the distinction between mirror-symmetric pair generation and directional branch selection.  In panel (a), the phase-conjugate $\chi^{(3)}$ interaction couples two pump excitations to the mirror-related $\mathrm{SPP}_{+}$ and $\mathrm{SPP}_{-}$ modes.  For $k_{p,y}=0$, their transverse momenta cancel and the elementary FWM source is balanced, so that $N_+=N_-$.  Panel (b) depicts the nonlinear state after the balanced solution loses stability: Kerr--FWM feedback permits two mirror-related steady branches with $\eta>0$ or $\eta<0$, and a small signed mirror-odd seed $h_y$ selects which branch is reached.  Reversing the seed interchanges $N_+$ and $N_-$, while the arrow widths represent their modal populations.  This schematic illustrates the normalized reduced-model mechanism and does not assign a dimensional instability threshold to the Au grating.

\section{Three-mode coupled equations}

\begin{figure*}[t]
\centering
\includegraphics[width=0.985\linewidth]{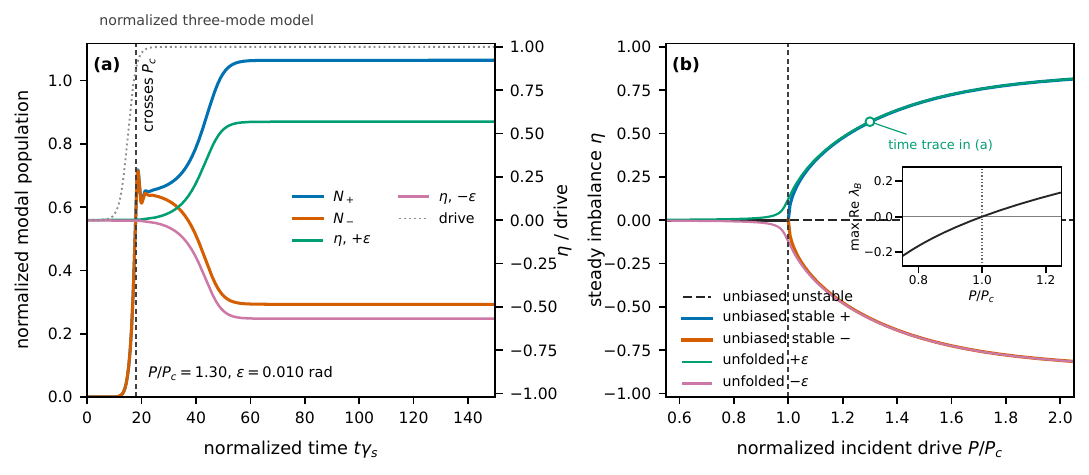}
\caption{Normalized three-mode Kerr--FWM dynamics. (a) Full stiff integration after a smooth pump turn-on at $P/P_c=1.30$. The modal populations $N_+$ and $N_-$ separate after the balanced state crosses its Bogoliubov instability. Reversing the small coherent phase bias, $\delta\phi=\pm0.01$ rad, reverses the selected imbalance $\eta$; time is shown only as $\gamma_s t$. (b) Steady imbalance versus normalized internal drive. Solid curves are stable solutions, the dashed balanced segment is unstable, and the two phase signs unfold the unbiased supercritical pitchfork in opposite directions. The inset shows $\max\operatorname{Re}\lambda_{\mathcal B}$ crossing zero. The marked point is used in panel (a). This is a normalized three-mode model, not a prediction of an Au-grating threshold or switching time.}
\label{fig:normalized_bifurcation}
\end{figure*}

Let $p$, $a_+$, and $a_-$ denote reduced modal amplitudes. Their squared moduli are normalized populations; they are not calibrated energies or photon numbers because the dispersive-QNM normalization has not converged. In a frame rotating at the laser frequency, the validated reduced equations are
\begin{subequations}
\begin{align}
\dot p={}&(\ii\Delta_p-\gamma_p)p+s_p
+\ii g_{pp}|p|^2p\nn\\
&+2\ii d_{\rm dep}\nu^*p^*a_+a_-,
\label{eq:three_p}\\
\dot a_+={}&(\ii\Delta_+-\gamma_+)a_++s_+\nn\\
&+\ii\kappa a_-+\ii\nu p^2a_-^*\nn\\
&+\ii\left(g_s|a_+|^2+g_c|a_-|^2
+2g_{p+}|p|^2\right)a_+,
\label{eq:three_plus}\\
\dot a_-={}&(\ii\Delta_--\gamma_-)a_-+s_-\nn\\
&+\ii\kappa^*a_++\ii\nu p^2a_+^*\nn\\
&+\ii\left(g_s|a_-|^2+g_c|a_+|^2
+2g_{p-}|p|^2\right)a_-.
\label{eq:three_minus}
\end{align}
\label{eq:three_mode}
\end{subequations}
Here $\Delta_j$ and $\gamma_j$ are detunings and total losses; $s_j$ are coherent drives; $\kappa$ is reciprocal linear mixing between the transverse modes; $g_{pp}$ and $g_s$ are self-Kerr coefficients; $g_c$ and $g_{p\pm}$ are cross-Kerr coefficients; and $\nu$ is the FWM overlap. The baseline uses the depletion multiplier $d_{\rm dep}=1$ and varies it only in the robustness audit. The last term in Eq.~\eqref{eq:three_p} is the retained pump-depletion channel. A more complete QNM truncation can add pump shifts proportional to the side-mode populations; they are absent from the specifically validated model in Fig.~\ref{fig:normalized_bifurcation}. In an open dispersive system all nonlinear coefficients are QNM overlaps and may be complex: their real parts in this convention produce frequency shifts, while positive imaginary parts produce nonlinear loss.

Every coefficient $g_{pp}$, $g_s$, $g_c$, $g_{p\pm}$, and $\nu$ is proportional to the same complex $\chi^{(3)}$ tensor contracted with a different four-field modal overlap. Thus $\chi^{(3)}$ supplies the dispersive Kerr shifts, cross-Kerr competition, nonlinear loss, FWM pair creation, and FWM depletion. No $\chi^{(2)}$ coefficient enters Eqs.~\eqref{eq:three_mode}. The dc OR readout instead requires the process-specific $\chi^{(2)}(0;\omega,-\omega)$ or an equivalent microscopic conductivity; a surface-SHG coefficient $\chi^{(2)}(2\omega;\omega,\omega)$ cannot replace it. A cascaded-$\chi^{(2)}$ correction to an effective Kerr coefficient would have to be derived separately with the intermediate-frequency Green function and is not included here.

Equations~\eqref{eq:three_mode} separate three roles that are easily conflated.  First, $\nu p^2a_\mp^*$ is the phase-sensitive parametric term that creates the pair.  Second, $g_s$ and $g_c$ shift the two resonances according to their populations and can destabilize a balanced state.  Third, $\kappa$ and the coherent phases of $s_\pm$ determine how the two propagation directions interfere.  None of these roles can generally be removed without changing the physical question.
The exact $a_+\leftrightarrow a_-$ symmetry of the complete unbiased equations is verified in Appendix~\ref{app:complete_symmetry}, while Appendix~\ref{sec:Supp_QNM} derives an analytically transparent two-mode limit and its instability criterion.

If the Au response is delayed, one may supplement Eqs.~\eqref{eq:three_mode} by
\begin{equation}
\tau_T\dot\Theta=-\Theta+\zeta_p|p|^2+
\zeta_s(|a_+|^2+|a_-|^2),
\label{eq:reservoir}
\end{equation}
with $\Delta_j\to\Delta_j+\xi_j\Theta$ and
$\gamma_j\to\gamma_j+\alpha_j\Theta$.  This reservoir is mirror even and cannot select $+q_y$ by itself, but it can shift a threshold, generate hysteresis, or change the response to a chopped beam.  It should therefore be added only when independently constrained by time-domain data.

\subsection{What FWM alone does not do}

Set $s_\pm=\kappa=0$ and retain only the FWM terms.  Directly from Eqs.~\eqref{eq:three_plus} and \eqref{eq:three_minus},
\begin{equation}
\left.\frac{d}{dt}(|a_+|^2-|a_-|^2)\right|_{\rm FWM}=0.
\label{eq:manleyrowe}
\end{equation}
This is the Manley--Rowe content of pair production: FWM changes $N_+$ and $N_-$ equally.  It locks the phase sum
$\phi_++\phi_--2\phi_p$ but leaves the relative phase $\phi_+-\phi_-$ free.  Therefore an isolated FWM process cannot by itself evolve from a balanced vacuum pair into a classical output containing only one transverse direction.  Directional selection requires coherent mixing, unequal driving or losses, nonlinear mode competition, or a finite-device momentum reservoir.

\subsection{How a tiny phase becomes a branch selector}

The essential result of Ref.~\cite{Anashkina2025} can be seen before solving the nonlinear problem.  Consider equal-frequency sideband drives with equal magnitude,
$s_+=s$ and $s_-=se^{\ii\psi}$, and choose a gauge in which $\kappa>0$.  In the linear limit with equal detuning and loss,
\begin{subequations}
\begin{align}
\dot a_+&=-L a_++\ii\kappa a_-+s,\\
\dot a_-&=-L a_-+\ii\kappa a_++se^{\ii\psi},
\\[-2pt]
L&=\gamma-\ii\Delta .
\end{align}
\label{eq:linear_phase}
\end{subequations}
The steady state is
\begin{subequations}
\begin{align}
a_+&=\frac{s(L+\ii\kappa e^{\ii\psi})}{L^2+\kappa^2},\\
a_-&=\frac{s(Le^{\ii\psi}+\ii\kappa)}{L^2+\kappa^2},
\end{align}
\label{eq:linear_solution}
\end{subequations}
and hence
\begin{equation}
|a_+|^2-|a_-|^2=
-\frac{4\gamma\kappa|s|^2\sin\psi}{|L^2+\kappa^2|^2}.
\label{eq:phase_bias}
\end{equation}
The sign in Eq.~\eqref{eq:phase_bias} depends on which mode is called $+$; its physical content does not.  No directional imbalance occurs if either $\kappa=0$ or $\psi=0,\pi$.  For complex $\kappa=|\kappa|e^{\ii\chi}$, the observable phase is the gauge-invariant combination
\begin{equation}
\Psi_{\rm eff}=\arg(s_-/s_+)+\arg\kappa,
\label{eq:gauge_phase}
\end{equation}
up to the sign convention used in Eq.~\eqref{eq:three_mode}.  It is $\sin\Psi_{\rm eff}$, not the phase of either drive alone, that controls the linear seed.

The same result is transparent in the standing-wave basis
\begin{equation}
a_s=\frac{a_++a_-}{\sqrt2},\qquad
a_a=\frac{a_+-a_-}{\sqrt2}.
\label{eq:standing}
\end{equation}
Backscattering splits their detunings by $2\kappa$.  Apart from an overall phase, the two drive amplitudes are proportional to $\cos(\psi/2)$ and $\sin(\psi/2)$.  At $\psi=0$ only one standing wave is driven; at $\psi=\pi$ only the other is driven; at intermediate phase both are driven and recombine into unequal traveling-wave powers.  The phase can simultaneously suppress or enhance the Kerr instability because the two standing waves have different effective detunings.  This is why weak backscattering does more than add a small intensity offset: it changes the nonlinear threshold landscape~\cite{Anashkina2025}.

For the Au grating, plausible contributions to $\Psi_{\rm eff}$ are a deliberately imposed small $k_y^{\rm inc}$, a transverse displacement of the beam center, unequal optical paths to the two sample edges, and coherent roughness or edge scattering.  A small incident transverse momentum produces approximately
\begin{equation}
\delta\psi\simeq 2k_y^{\rm inc}L_{\rm eff},
\label{eq:kyphase}
\end{equation}
where $L_{\rm eff}$ is an effective separation set by the illuminated width and the mode-overlap centroids.  Equation~\eqref{eq:kyphase} is a scaling relation, not a substitute for a three-dimensional calculation.  Reversing $k_y^{\rm inc}$ must reverse the corresponding selector.

Changing the polar angle of incidence while the beam remains in the $xz$ mirror plane preserves $\Me$ and therefore cannot by itself supply the mirror-odd seed $h_y$.  By contrast, $k_y^{\rm inc}\ne0$ represents an azimuthal tilt out of that plane; it is odd under $\Me$ and can provide the controlled selector used here.

\subsection{Kerr amplification of the seed}

The nonlinear differential detuning is
\begin{equation}
\delta\omega_+-\delta\omega_-=
(g_s-g_c)(|a_+|^2-|a_-|^2),
\label{eq:differential_kerr}
\end{equation}
for equal pump-induced shifts.  An initial imbalance therefore shifts the two modes in opposite directions relative to the drive.  On the appropriate side of a resonance, this shift increases the better-coupled population and decreases the other, producing positive feedback.  Counterpropagating Kerr resonators often have a cross-phase coefficient twice the self-phase coefficient in the simplest isotropic overlap model, a condition known to support mode competition and symmetry breaking~\cite{DelBino2017,Anashkina2025}.  The grating coefficients must not be assumed to have that ratio; $g_s$, $g_c$, and their dissipative parts must be calculated from the actual QNM fields.

Thus a phase difference can be tiny at low power but produce an order-one $\eta$ near a Kerr bifurcation.  The precise statement is important: the small phase is an \emph{explicit seed}, while the nonlinear instability is the amplifier.  For $\Psi_{\rm eff}\ne0,\pi$ the transition is an imperfect pitchfork and the direction is deterministic.  Only when every mirror-odd field vanishes can the two signs be selected spontaneously by noise.

\section{Bogoliubov stability problem}

The first instability is found by solving for the pumped steady state $\bar p$ with $a_\pm=0$ and linearizing the transverse fields.  Define
\begin{equation}
\bm v=(\delta a_+,\delta a_-,\delta a_+^*,\delta a_-^*)^{\mathsf T}.
\end{equation}
For equal sideband loss and detuning, and after gauging $\kappa$ real, the pump-only Bogoliubov equation is
\begin{equation}
\dot{\bm v}=\mathcal B_4\bm v,
\quad
\mathcal B_4=
\begin{pmatrix}
A&\ii\kappa&0&\ii G\\
\ii\kappa&A&\ii G&0\\
0&-\ii G^*&A^*&-\ii\kappa\\
-\ii G^*&0&-\ii\kappa&A^*
\end{pmatrix},
\label{eq:B4}
\end{equation}
where
\begin{equation}
A=-\gamma+\ii\Omega,\qquad
\Omega=\Delta_s+g_p|\bar p|^2,\qquad
G=\nu\bar p^2.
\label{eq:AG}
\end{equation}
The pump-only state becomes unstable when
\begin{equation}
\max_n\operatorname{Re}\lambda_n(\mathcal B_4)=0
\label{eq:threshold}
\end{equation}
is crossed from below.  For $\kappa=0$ and real coefficients this reduces to the familiar parametric condition
\begin{equation}
|G|^2>\gamma^2+\Omega^2.
\label{eq:parametric_threshold}
\end{equation}
Finite $\kappa$ splits the standing-wave detunings, schematically $\Omega\to\Omega\pm\kappa$, so the relative phase determines which split channel is populated and can raise or lower the observed onset.

Equation~\eqref{eq:B4} describes pair creation but not its saturated directional state.  Above onset, pump depletion, self- and cross-Kerr shifts, direct sideband drives, unequal losses, and the phase seed must all be retained.  Let
\begin{equation}
\bm u=(p,a_+,a_-)^{\mathsf T},\qquad \dot{\bm u}=\bm F(\bm u,\bm u^*),
\end{equation}
where $\bm F$ is the right-hand side of Eqs.~\eqref{eq:three_mode}.  Around any nonlinear steady state $\bar{\bm u}$, define
\begin{equation}
\delta\bm U=(\delta\bm u,\delta\bm u^*)^{\mathsf T}.
\end{equation}
The complete three-mode Bogoliubov problem is
\begin{equation}
\delta\dot{\bm U}=\mathcal B_6\delta\bm U,
\qquad
\mathcal B_6=
\begin{pmatrix}
J&K\\ K^*&J^*
\end{pmatrix}_{\bar{\bm u}},
\label{eq:B6}
\end{equation}
with
\begin{equation}
J_{jk}=\pdv{F_j}{u_k},\qquad
K_{jk}=\pdv{F_j}{u_k^*}.
\label{eq:Wirtinger}
\end{equation}
Equation~\eqref{eq:B6}, rather than a two-mode Kerr matrix with an undepleted pump, is the required stability test for the three-mode model.  It contains pump amplitude and phase fluctuations, FWM pair correlations, linear backscattering, self/cross-mode competition, dissipative nonlinearities, and explicit bias.  Each steady solution is stable only if every eigenvalue has negative real part.  The parity and port flux of the critical eigenvector identify whether an instability is predominantly pair-generating, standing-wave-like, or directional.

If the reservoir in Eq.~\eqref{eq:reservoir} is retained, $\delta\Theta$ adds one real dimension (or an equivalent pair in a complex implementation) to Eq.~\eqref{eq:B6}.  That enlarged problem can distinguish a fast electronic Kerr instability from a slower thermo-optic or hot-electron instability.

\section{Reduction to the directional order parameter}

Define the normalized transverse flux imbalance
\begin{equation}
\eta=\frac{|a_+|^2-|a_-|^2}{|a_+|^2+|a_-|^2}.
\label{eq:eta}
\end{equation}
It is odd under $\Me$ and supplies the dynamical factor
\begin{equation}
\Lamy=\Lstat+\Ldyn,\qquad \Ldyn=\lambda_\eta\eta.
\label{eq:Lambda_split}
\end{equation}
Here $\eta$ fixes only the sign and fractional imbalance.  The coefficient $\lambda_\eta$ must also contain the total directional-mode population, the electron--SPP matrix elements, the transverse momentum $q_y$, and the relevant electronic energy denominators.  Appendix~\ref{app:microscopic_vertex} obtains this structure by eliminating the optical Floquet sidebands and expanding the retarded scattering $T$ matrix.  The resulting coefficient is left as an explicit sum over unknown matrix elements; no absolute Au value is inferred.  Because this microscopic dynamical contribution contains the directional-mode population in addition to the two ac factors in $\langle j_xs_y\rangle$, it is an intensity-dependent readout of a nonlinear optical state rather than a strict perturbative $\chi^{(2)}$ coefficient.  The static-$\Lamy$ term remains quadratic in the local field.

Equation~\eqref{eq:Lambda_split} also makes an important distinction: $\eta=0$ implies only $\Ldyn=0$.  It does not require $\chi^{(3)}=0$; mirror-even self- and cross-Kerr shifts and balanced FWM with $N_+=N_-$ may remain even when the dynamical mirror-odd factor vanishes.  If $\eta=0$ but $\Lamy\ne0$, the entire mirror-plane-odd factor must instead reside in $\Lstat$.  Such a static contribution requires a signed transverse structural, material, or externally imposed term that is odd under the $xz$ mirror plane; mirror-even scalars such as the average grain size, or a change of polar incidence angle that keeps the beam in that mirror plane, cannot supply it.

Near a simple directional instability, elimination of the stable pump, total-intensity, and phase-sum variables gives the center-manifold normal form
\begin{equation}
\begin{split}
\tau_\eta\dot\eta={}&
\mu(I,\Delta,|\kappa|,|G|)\eta-\beta_\eta\eta^3+h_y\\
&+O(\eta^5,h_y\eta^2),\qquad \beta_\eta>0.
\end{split}
\label{eq:normalform}
\end{equation}
The explicit field has the symmetry-allowed expansion
\begin{align}
h_y={}&C_\phi|\kappa||s_+s_-|\sin\Psi_{\rm eff}
+C_k k_y^{\rm inc}+C_r(y_0-y_c)
\nn\\
&+C_\Delta(\Delta_+-\Delta_-)
+C_\gamma(\gamma_--\gamma_+)+\cdots .
\label{eq:hy}
\end{align}
All $C_j$ are reduced coefficients, not universal constants.  The first term is the nonlinear continuation of Eq.~\eqref{eq:phase_bias}; the remaining terms represent other mirror-odd perturbations.

For $h_y=0$, $\eta=0$ is stable for $\mu<0$, and the two branches $\eta_\pm=\pm\sqrt{\mu/\beta_\eta}$ are stable for $\mu>0$.  For $h_y\ne0$, no exact pitchfork remains: one branch is favored before the nominal threshold and evolves into the high-contrast state.  A subcritical coefficient set can instead produce bistability and hysteresis.  The phase-sensitive resonator results~\cite{Anashkina2025} demonstrate both the strength of this unfolding and its threshold dependence: weak backscattering can suppress symmetry breaking for one phase and enhance it for another.  Translating that result to the grating is a hypothesis to be tested, not evidence that the experimental sample already crosses the Kerr threshold.

Figure~\ref{fig:normalized_bifurcation} is a normalized solution of Eqs.~\eqref{eq:three_mode} and is intended to demonstrate the dynamical mechanism permitted by the Kerr--FWM equations. All rates are divided by $\gamma_s$, time is $\gamma_s t$, and modal amplitudes absorb an arbitrary common QNM scale. Drive amplitudes consequently absorb both that modal scale and the uncalibrated external-coupling coefficient. The internal value $P_c=0.9553947$ is the square of a reduced drive amplitude; it is not optical power in watts. Determining incident power or physical response time requires converged dispersive-QNM normalization, nonlinear modal overlaps, external-coupling efficiencies, physical linewidths, and the appropriate complex Au susceptibility at the experimental pulse duration.

The symmetric steady branch was continued in the mirror-even subspace, while the asymmetric and phase-unfolded solutions were solved in all six real modal variables. Stability was determined from the largest real part of the analytic $6\times6$ Nambu matrix, independently checked against a fifth-order finite-difference Jacobian. Full time traces used an implicit BDF integrator. At the displayed parameter set the eigenvalue and time-domain growth rates agree to $1.49\times10^{-7}$ relatively, the analytic Jacobian agrees with finite differences to $1.63\times10^{-10}$ in maximum entry, and the fitted critical exponent is $0.4898$. Reversing $\delta\phi$ gives equal and opposite steady imbalances, while the bias magnitude $0.01$ is only $1.76\%$ of the saturated imbalance at $P/P_c=1.30$.

The robustness audit varied every displayed parameter separately by at least $\pm20\%$ and then sampled joint local and broader ranges. All 16 individual local cases retained a real mirror-odd crossing; 15 were direct supercritical pitchforks and one side-detuning case acquired a preceding saddle-node. All 24 joint local cases retained the pitchfork, although eight acquired a preceding multistable jump. Across the resolved direct pitchforks the critical exponent remained between $0.46$ and $0.50$, and phase reversal remained exact to the declared numerical tolerance. The broad scan also found regimes in which no crossing occurred and regimes dominated first by a saddle-node; these cases are reported in Appendix~\ref{app:numerical_bogoliubov} and summarized in Fig.~\ref{fig:app_robustness} rather than averaged away.

Substituting Eq.~\eqref{eq:Lambda_split} into Eq.~\eqref{eq:Jy_general} yields the $\Astr$-weighted side-jump-like contribution
\begin{equation}
J_{y,{\rm sj}}^{(0)}=\frac{\tau_y\Astr}{2}
(\Lstat+\lambda_\eta\eta)
\operatorname{Re}[\chi_{EE}\sigma_D^*]|E_x^{\rm loc}|^2.
\label{eq:Jy_final}
\end{equation}
This is the electronic readout retained in the present interpretation.  The factor $\Astr$ changes sign under reversal of the unit-cell handedness, the dynamical part of $\Lamy$ changes sign under the $y$ mirror or reversal of the phase seed, and the optical intensity is even.  The centered symmetric cell therefore suppresses the resonant structure-dependent contribution through $\Astr=0$, even if its optical subsystem develops $\eta\ne0$.

The experimental control sharpens this choice.  The asymmetric grating exhibits pronounced SPP-correlated transverse-current peaks, whereas the symmetric control retains only a weak, broad, peakless background of comparable scale to the nonresonant background in the asymmetric sample~\cite{Plumitallo2023}.  At the same time, the COMSOL calculation in Ref.~\onlinecite{Plumitallo2023} shows sharp $p$-polarized absorbance resonances for both gratings, including the centered symmetric control; no experimental absorbance trace was reported.  The modeled symmetric grating can therefore couple resonantly to the optical/SPP field even though its measured $J_y$ has no corresponding resonant peaks.  This comparison indicates that resonant absorption and local-field enhancement are not sufficient to produce the transverse electrical signal: the structure-dependent readout is suppressed through $\Astr=0$.  A direct $\Astr$-independent readout of the directional SPP imbalance is allowed by the $yz$ mirror and, once $\eta\ne0$, also by the broken $xz$ mirror.  It is therefore not symmetry-forced to vanish in the centered symmetric grating.  Appendix~\ref{app:direct_keldysh} nevertheless shows that its leading one-SPP population-injection term vanishes in the clean low-energy surface-state model at $808~\mathrm{nm}$ because the required transition is off shell.  More general absorptive channels are not excluded, but the absence of resonant current peaks in the symmetric control gives no experimental reason to invoke them as the origin of the measured resonant contrast.  We write the measured current schematically as
\begin{equation}
J_y^{\rm meas}=J_y^{\rm bg}+J_{y,{\rm sj}}^{(0)},
\label{eq:Jy_measured}
\end{equation}
where $J_y^{\rm bg}$ denotes the smooth contribution common to the two samples.  Its microscopic origin is not assigned here.  Reversing the $x$-handedness and reversing the phase seed remain complementary tests because they reverse $\Astr$ and $\eta$, respectively.

\section{Momentum conservation and the meaning of a directional output}

There is no conflict between branch selection and momentum conservation.  At onset, FWM creates a $+q_y,-q_y$ pair and Eq.~\eqref{eq:matching} conserves momentum.  Linear backscattering $\kappa$ exists only because the finite device, edges, or disorder break continuous $y$ translation; these objects can exchange transverse momentum with the field.  A nonzero $k_y^{\rm inc}$ also injects a known momentum.  Radiation, the substrate, and the mount carry the remaining Maxwell stress.  The correct conservation law applies to the total field--matter--structure system, not to the two-mode subsystem alone.

Nor does the model predict that every microscopic FWM event emits two SPPs into one direction.  The elementary parametric vertex remains pairwise.  The classical steady state becomes directional because coherent mode mixing, drive, loss, and Kerr shifts continually redistribute the pair-generated population.  A large $|\eta|$ therefore means that the measured outgoing SPP flux is dominated by one port, not that the Manley--Rowe pair-production rule has been violated.

\section{Most likely selectors and decisive experiments}

Table~\ref{tab:selectors} separates mechanisms that merely seed a sign from those that can alter the instability itself.  In the present geometry, a deliberately imposed small $k_y^{\rm inc}$ is the cleanest controllable selector.  Finite edges and coherent roughness are the most natural source of $\kappa$.  Unequal losses or detunings are possible but should be diagnosed independently because they generate a static imbalance even without Kerr amplification.

\begin{table*}[t]
\caption{Directional selectors and their discriminating signatures.}
\label{tab:selectors}
\ifrevtex\begin{ruledtabular}\else\centering\small\fi
\begin{tabular}{p{0.19\linewidth}p{0.34\linewidth}p{0.37\linewidth}}
\ifrevtex\else\toprule\fi
Mechanism & Role in Eqs.~\eqref{eq:three_mode}--\eqref{eq:hy} & Experimental signature \\
\hline
Incident $k_y^{\rm inc}$ or beam displacement & Changes $s_-/s_+$ and therefore $\Psi_{\rm eff}$; supplies a controlled $h_y$ & $\eta$ and its contribution to $J_y$ reverse when $k_y^{\rm inc}$ or centered displacement reverses \\
Coherent edge/roughness scattering & Produces reciprocal $\kappa$; makes relative phase observable and splits standing modes & Phase dependence disappears as $\kappa\to0$; threshold and selected mode depend on edge condition \\
Kerr self/cross competition & Produces differential shift $(g_s-g_c)(N_+-N_-)$ and nonlinear amplification & Rapid crossover or threshold, critical slowing, bistability, or hysteresis in a power scan \\
Complex nonlinear loss & Imaginary parts of $g_s,g_c,g_p$ give population-dependent damping & Pulse-duration and temperature dependence; different up/down sweeps; altered linewidths \\
Unequal $\Delta_\pm$ or $\gamma_\pm$ & Direct static imperfection in $h_y$ and in the Bogoliubov matrix & Imbalance persists in the linear low-power limit and follows the perturbation rather than a threshold \\
Delayed Au reservoir & Shifts detuning/loss through Eq.~\eqref{eq:reservoir}; not a mirror-odd selector by itself & Strong dependence on chopper frequency, duty cycle, pulse duration, and thermal boundary conditions \\
\ifrevtex\else\bottomrule\fi
\end{tabular}
\ifrevtex\end{ruledtabular}\fi
\end{table*}

The decisive measurement is simultaneous detection of $J_y$ and the two lateral SPP powers while scanning incident power and the controlled seed.  The model predicts: (i) an odd response under $k_y^{\rm inc}\to-k_y^{\rm inc}$ after laboratory-fixed offsets are removed; (ii) a one-to-one correlation between the signs of the resonant part of $J_y$ and $\eta$; (iii) strong susceptibility to the seed near threshold; (iv) possible hysteresis or critical slowing; and (v) a threshold shift when spot size, position, or edge condition is changed.  A centered symmetric lithographic control suppresses the $\Astr$-weighted contribution in Eq.~\eqref{eq:Jy_final} even if it develops the directional optical state.  Reversing the $x$-handedness should reverse $\Astr$, whereas reversing $k_y^{\rm inc}$ should reverse $\eta$ and the dynamical contribution to Eq.~\eqref{eq:Jy_final}.  These operations therefore test the structural and dynamical factors independently, while the smooth background in Eq.~\eqref{eq:Jy_measured} need not follow either reversal.

The optical spin of an SPP provides a useful port label because its transverse spin reverses with propagation direction~\cite{Moroshkin2022,RodriguezFortuno2013}.  It does not, however, supply a mirror-odd selector in an unbiased reciprocal structure.  Spin--momentum locking correlates the chosen port with optical transverse spin after selection; it does not by itself choose $+q_y$ over $-q_y$.

\section{Static bias and magnitude}

Static contributions remain symmetry allowed but are not assigned a privileged role.  We write
\begin{equation}
\Lstat=\lambda_{\rm tex}T_y+\lambda_{\rm dep}\Ffab+
\lambda_3\Mbar+\lambda_g\mathcal A_y+\cdots,
\label{eq:static}
\end{equation}
where $T_y$ is a transverse sidewall-texture moment, $\Ffab$ a laboratory-fixed deposition direction, $\Mbar=\langle\sin3\phi_c\rangle$ a weighted Au(111) texture moment, and $\mathcal A_y$ a deliberately introduced transverse geometry.  These are neutral, nonexclusive realizations of a symmetry-allowed bias.  In the dynamical model they can simply select a nonlinear branch rather than generate the full response.

Here $T_y$ denotes a signed transverse texture moment, not the scalar average grain size.  Average grain size is mirror-plane even and cannot by itself contribute to $\Lstat$; only a $y$-odd difference or gradient in grain morphology or crystalline texture can do so.  Likewise, changing the polar angle while keeping the incident wave vector in the $xz$ mirror plane is mirror-plane even.  An imposed $k_y^{\rm inc}$ or transverse beam displacement is mirror-plane odd and, in the dynamical model above, is treated primarily as a selector of $\eta$.

The measured magnitude is modest.  The reported $I_y^{\rm pk}\simeq300~\mathrm{pA}$ over an approximately $150~\mu\mathrm m$ illuminated width corresponds to $J_y^{\rm pk}\simeq2.0~\mu\mathrm{A\,m^{-1}}$~\cite{Plumitallo2023}.  Using the complex Au optical conductivity near $808~\mathrm{nm}$~\cite{Johnson1972} and the incident-field estimate in Appendix~\ref{sec:Supp_frequency_magnitude} gives an estimated oscillatory longitudinal optical sheet-current amplitude $K_x^{\rm ac}$ of order $10^3~\mathrm{A\,m^{-1}}$.  Thus $J_y^{\rm pk}/K_x^{\rm ac}\sim10^{-9}$ compares a measured transverse dc sheet current with an estimated longitudinal ac optical sheet-current amplitude; it is not a comparison with ordinary dc conductivity, a Hall angle, or the ratio of experimental voltage responsivities.  The reported dimensionful voltage responsivities (termed efficiencies in the original reports) are instead $\mathcal R_y=1.25\times10^{-9}~\mathrm{V}/(\mathrm{W\,cm^{-2}})$ for the transverse experiment~\cite{Plumitallo2023} and $\mathcal R_x=12\times10^{-9}~\mathrm{V}/(\mathrm{W\,cm^{-2}})$ for the longitudinal experiment~\cite{Moroshkin2021}, giving $\mathcal R_y/\mathcal R_x\simeq0.104$, or a transverse value about $9.6$ times smaller.  Because these numbers were obtained in different device configurations and depend on sample resistance, contact geometry, illuminated area, and local-field enhancement, they provide a useful device-level comparison but not the microscopic conversion factor entering Eq.~\eqref{eq:Jy_final}.  Only weak conversion is required, but neither comparison determines $\nu$, $g_s$, $g_c$, $\lambda_\eta$, the process-specific side-jump coefficient, or the threshold.

\section{Conclusion}
\label{sec:conclusion}

The $xz$ mirror supplies a strict organizing principle for optical Hall rectification in a one-dimensional grating.  The $x$-handedness $\Astr$ and transverse mirror-odd factor $\Lamy$ have different symmetry roles.  The dynamical proposal has a definite causal sequence: a pumped plasmonic resonance drives FWM pair creation into mirror-related $(K_x,\pm q_y)$ QNMs; weak reciprocal backscattering makes their relative drive phase observable; self/cross-Kerr competition destabilizes or sharply amplifies the balanced state; and a small phase seed, most cleanly supplied by controlled $k_y^{\rm inc}$, deterministically selects the sign.  The resulting population imbalance supplies $\Ldyn\propto\eta$ in the structure-dependent $\Astr\Lamy\langle j_xs_y\rangle_t$ side-jump-like channel.  If $\eta=0$, this dynamical part vanishes and any nonzero $\Lamy$ must come from a separate static term that is odd under the $xz$ mirror plane.  The factor $\Astr$ makes the resonant contribution vanish for the centered symmetric grating, in accord with the absence of corresponding resonant current peaks even though the calculated absorbance of that control remains resonant.  A separate $\Astr$-independent current proportional to $\eta$ is symmetry allowed, but its leading $\eta$-linear one-SPP on-shell contribution vanishes in the clean, off-resonant Au(111) surface-state reduction developed in Appendix~\ref{app:direct_keldysh}; realistic momentum-relaxing or interband channels remain outside that controlled statement.  The distinct $a_+a_-^*$ coherence produces a transverse standing-wave dressing rather than a uniform order-$g_{\rm SPP}^2$ population source; a net contribution from it requires a finite-device phase reference or additional momentum mixing.  Four-wave mixing alone neither selects a direction nor violates momentum conservation.

The normalized calculation now demonstrates that Eqs.~\eqref{eq:three_mode} possess a robust supercritical pitchfork, an approximately one-half critical exponent, and opposite branch selection under reversal of a small coherent phase bias. The numerical continuation, analytic Bogoliubov matrix, finite-difference Jacobian, and time integration agree.  The perturbative scattering analysis additionally identifies an $O(g_{\rm SPP}^2\lambda_Ph^3)$ route to the symmetry-allowed $\Astr\lambda_\eta\eta\sigma_y$ vertex, but it does not establish its magnitude or even exclude helicity cancellation in the fabricated sample. What remains unambiguous but uncompleted is the dimensional Au calculation: one must obtain the three finite-device QNMs, evaluate converged complex $\chi^{(3)}$ overlaps, determine physical linewidths and external coupling, calculate the electron--SPP matrix elements entering $\lambda_\eta$, and evaluate the process-specific side-jump readout. The output must be converged lateral SPP port powers, not individual evanescent Fourier harmonics. Until that program and the corresponding power/phase experiments are completed, the Kerr--FWM mechanism and its structure-dependent electronic readout remain falsifiable, literature-motivated hypotheses rather than a microscopic calculation of the measured current.

\ifrevtex
  \begin{acknowledgments}
  M.N.L. acknowledges support by the ORISE program at DEVCOM SC in 2022, 2023, 2024, 2025, and 2026 and funding by DEVCOM SC SPD.  M.N.L. also acknowledges support by the Army Research Office under Grant No.~W911NF-24-1-0163.
  \end{acknowledgments}
\else
  \section*{Acknowledgments}
  M.N.L. acknowledges support by the ORISE program at DEVCOM SC in 2022, 2023, 2024, 2025, and 2026 and funding by DEVCOM SC SPD.  M.N.L. also acknowledges support by the Army Research Office under Grant No.~W911NF-24-1-0163.
\fi

\appendix

\section{Normalized three-mode numerics and robustness}
\label{app:numerical_bogoliubov}

\subsection{Normalization and parameter provenance}

Choose the side-mode loss as the reference rate and define $t'=\gamma_s t$. If the dimensional QNM amplitudes are $A_j=A_0a_j$, then the normalized drives are $s_j'=s_j/(\gamma_sA_0)$ and a generic nonlinear coefficient is $g'=gA_0^2/\gamma_s$. The unspecified amplitude $A_0$ contains the dispersive biorthogonal QNM normalization; the dimensional drive additionally contains the external coupling. These absorbed quantities are precisely why neither $|a_j|^2$ nor $|s_j|^2$ can be labeled as energy, photon number, or incident watts. In the numerical files $P/P_c$ is a ratio of squared internal drive amplitudes and the stored $P_c=0.9553947$ is dimensionless.

The linear losses, detunings, and reciprocal mixing $\kappa$ are illustrative normalized QNM parameters. The coherent $s_\pm$ encode equal-magnitude side drives and the gauge-invariant phase loop $\delta\phi$. A single common complex third-order scale $c=1+0.01\ii$ multiplies every Kerr and FWM overlap; positive $\operatorname{Im}c$ gives nonlinear loss for the $e^{-\ii\omega t}$ convention. The optional reservoir is a mirror-even phenomenological hot-electron diagnostic. It is disabled in Fig.~\ref{fig:normalized_bifurcation}; when enabled, its strength and relaxation time are varied independently.

The complete normalized parameter set used for Fig.~\ref{fig:normalized_bifurcation}, together with the provenance and role of each parameter, is listed in Table~\ref{tab:normalized_parameters}.

\begin{table*}[t]
\caption{Displayed normalized parameter set. Values are reduced rates or overlap ratios, not fitted Au coefficients.}
\label{tab:normalized_parameters}
\ifrevtex\begin{ruledtabular}\else\centering\small\fi
\begin{tabular}{llll}
Quantity & Value & Provenance & Role \\
\hline
$\gamma_p/\gamma_s,\gamma_\pm/\gamma_s$ & $2,1$ & normalized linear model & pump and side losses \\
$\Delta_p/\gamma_s,\Delta_\pm/\gamma_s$ & $0,-1.75$ & normalized linear model & rotating-frame detunings \\
$\kappa/\gamma_s$ & $0.10$ & illustrative reciprocal mixing & closes the coherent selector loop \\
$g_{pp},g_s,g_c$ & $(0.05,1,3)c$ & common $\chi^{(3)}$ times overlap ratios & self- and cross-Kerr shifts/loss \\
$g_{p+},g_{p-},\nu$ & $(0.02,0.02,0.05)c$ & common $\chi^{(3)}$ times overlap ratios & pump cross-Kerr and FWM \\
$d_{\rm dep}$ & $1$ & FWM depletion normalization & pump backaction \\
$s_p,s_+,s_-$ & $A,Ae^{+\ii\delta\phi/2},Ae^{-\ii\delta\phi/2}$ & normalized coherent drives & turn-on and phase selector \\
$\delta\phi$ & $0,\pm0.01$ rad & small coherent bias & unfolds the pitchfork \\
Reservoir & off in Fig.~\ref{fig:normalized_bifurcation} & optional mirror-even diagnostic & common detuning/loss shift \\
\end{tabular}
\ifrevtex\end{ruledtabular}\fi
\end{table*}

\subsection{Full Bogoliubov matrix}

Let $\bm z=(p,u,v)=(p,a_+,a_-)$ and write $\dot{\bm z}=\bm f(\bm z,\bm z^*)$. With the Wirtinger blocks $A=\partial\bm f/\partial\bm z$ and $B=\partial\bm f/\partial\bm z^*$, the Nambu matrix in $(\delta\bm z,\delta\bm z^*)$ ordering is
\begin{equation}
\mathcal B_6=\begin{pmatrix}A&B\\B^*&A^*\end{pmatrix}.
\label{eq:app_B6_explicit}
\end{equation}
For $P=|p|^2$, $U=|u|^2$, and $V=|v|^2$, the nonzero entries are as follows. We write $d=d_{\rm dep}$ and $k_\pm=\kappa e^{\pm\ii\phi_\kappa}$.
\ifrevtex\begin{widetext}\fi
\begingroup\small
\begin{align}
A_{pp}&=\ii\Delta_p-\gamma_p+2\ii g_{pp}P,&
A_{pu}&=2\ii d\nu^*p^*v,&
A_{pv}&=2\ii d\nu^*p^*u,\nn\\
B_{pp}&=\ii g_{pp}p^2+2\ii d\nu^*uv,&&&\nn\\
A_{up}&=2\ii\nu pv^*+2\ii g_{p+}p^*u,&
A_{uu}&=\ii\Delta_+-\gamma_++\ii(2g_sU+g_cV+2g_{p+}P),&
A_{uv}&=\ii k_++\ii g_cv^*u,\nn\\
B_{up}&=2\ii g_{p+}pu,&B_{uu}&=\ii g_su^2,&
B_{uv}&=\ii\nu p^2+\ii g_cvu,\nn\\
A_{vp}&=2\ii\nu pu^*+2\ii g_{p-}p^*v,&
A_{vu}&=\ii k_-+\ii g_cu^*v,&
A_{vv}&=\ii\Delta_- -\gamma_-+\ii(2g_sV+g_cU+2g_{p-}P),\nn\\
B_{vp}&=2\ii g_{p-}pv,&B_{vu}&=\ii\nu p^2+\ii g_cuv,&B_{vv}&=\ii g_sv^2.
\label{eq:app_AB_entries}
\end{align}
\endgroup
\ifrevtex\end{widetext}\fi
The equivalent real Jacobian in $[\operatorname{Re}\bm z,\operatorname{Im}\bm z]$ ordering is
\begin{equation}
J_R=\begin{pmatrix}
\operatorname{Re}(A+B)&-\operatorname{Im}(A-B)\\
\operatorname{Im}(A+B)&\operatorname{Re}(A-B)
\end{pmatrix}.
\label{eq:app_real_jacobian}
\end{equation}
A steady state is stable only when every eigenvalue satisfies $\operatorname{Re}\lambda<0$; the threshold is the first continuous crossing $\max\operatorname{Re}\lambda=0$. A discontinuous branch jump is recorded as a saddle-node/multistable onset and is never forced into this definition of $P_c$.

\subsection{Numerical methods and robustness}

The mirror-even branch was solved in four real variables and continued in drive. If ordinary parameter continuation encountered a fold, constrained BDF integration located the dynamically adjacent symmetric branch and the jump amplitude was recorded. Asymmetric branches were solved in six real variables. The analytic matrix in Eq.~\eqref{eq:app_B6_explicit} was checked entry by entry against an independent fifth-order central finite difference. Time-domain integration used BDF with relative and absolute tolerances $10^{-9}$ and $10^{-11}$; tightening both tolerances and halving the maximum step changed the displayed imbalance by $7.57\times10^{-9}$.

The audit contains 118 cases. In the 16 one-at-a-time local variations, every case retained a real mirror-odd crossing, 15 remained a direct supercritical pitchfork, and one had a preceding multistable jump. In 24 joint local variations, all retained the crossing, 16 were direct supercritical pitchforks, and eight had a preceding multistable jump. Sixteen additional joint local cases with the delayed reservoir all retained the crossing; ten were direct and six had a preceding jump. Across direct supercritical solutions the exponent stayed close to one half and the signed phase bias remained much smaller than the selected saturated imbalance. The broad scan intentionally included limits where the crossing disappeared or a discontinuous symmetric jump occurred first. No oscillatory Hopf or resolved subcritical pitchfork was found within the declared ranges, but neither is excluded outside them.

\begin{figure*}[t]
\centering
\includegraphics[width=0.98\linewidth]{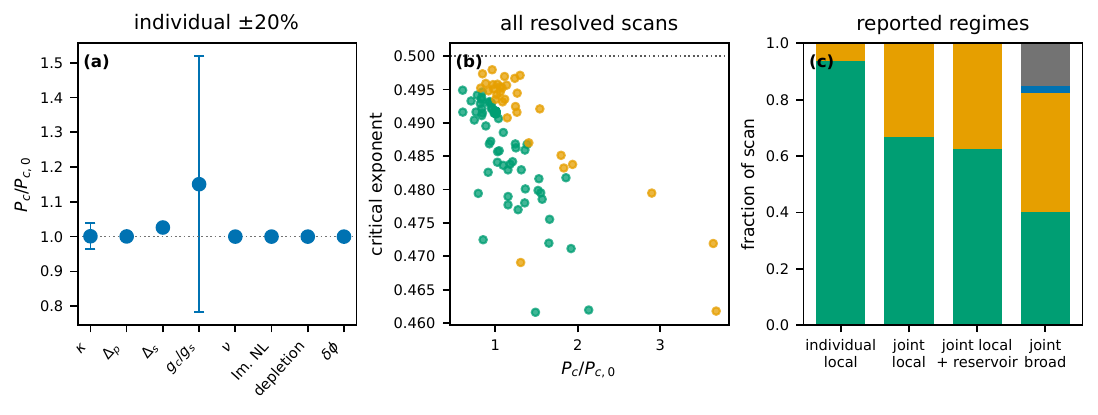}
\caption{Robustness of the normalized three-mode calculation. (a) Mean and full range of $P_c/P_{c,0}$ for each pair of individual $\pm20\%$ parameter variations; the dotted line marks the baseline threshold. (b) Fitted critical exponent versus normalized threshold for all resolved scans. Green circles denote direct supercritical pitchforks, orange circles denote cases in which a saddle-node or multistability occurs before the pitchfork, and the dotted line marks the mean-field value $1/2$. (c) Stacked fractions of the reported regimes for, from left to right, the individual-local, joint-local, joint-local-plus-reservoir, and joint-broad scans. Green denotes a direct supercritical pitchfork, orange denotes multistability before the pitchfork, blue denotes a discontinuous symmetric jump, and gray denotes no crossing within the scanned range. All axes are dimensionless; the diagram contains no Au threshold or dimensional time.}
\label{fig:app_robustness}
\end{figure*}

Figure~\ref{fig:app_robustness} summarizes the audit.  Panel (a) shows the threshold variation under one-at-a-time parameter changes, panel (b) compares the fitted critical exponent with the shifted normalized threshold for the resolved direct pitchforks, and panel (c) reports the fractions of direct, multistable-before-pitchfork, discontinuous-jump, and no-crossing regimes across the scan classes.  The figure therefore displays both the local robustness of the pitchfork used in the main text and the broader parameter regions in which the onset changes character or disappears.

\section{Projection of the Kerr Maxwell equation}
\label{app:projection}

With no free charge, Maxwell's equation in a dispersive medium is
\begin{equation}
\nabla\times\nabla\times\bm E+
\mu_0\epsilon_0\epsilon_r(\bm r,\ii\partial_t)\partial_t^2\bm E
=-\mu_0\partial_t^2\bm P^{(3)}.
\label{eq:maxwell_kerr}
\end{equation}
Insert Eq.~\eqref{eq:modeexpansion}, retain resonant terms under the slowly varying envelope approximation, and project with the adjoint QNM $\widetilde{\bm e}_j^{L}$.  A generic coefficient has the form
\begin{equation}
\Gamma_{jklm}=\frac{3\omega_j}{2\mathcal N_j}
\int_{V_{\rm nl}}d^3r\,
\widetilde e_{j,i}^{L}
\epsilon_0\chi_{iabc}^{(3)}
e_{k,a}e_{l,b}e_{m,c}^*,
\label{eq:overlap}
\end{equation}
where $\mathcal N_j$ is the dispersive QNM normalization and $V_{\rm nl}$ is the nonlinear Au or nonlinear dielectric volume.  Numerical prefactors can move between $\Gamma$ and the amplitude normalization; the invariant content is the field overlap and its phase.

Terms with indices $(jklm)=(++++),(+--+)$ generate $g_s$ and $g_c$.  The overlap $(+pp-)$ generates $\nu p^2a_-^*$, while its reciprocal partner generates $\nu p^2a_+^*$.  The adjoint pump projection produces $2\nu^*p^*a_+a_-$, including the factor of two associated with annihilating either pump quantum.  The mirror relations among $\bm e_+$ and $\bm e_-$ enforce equal coefficients in the ideal limit.  A complex $\chi^{(3)}$ or complex QNM normalization makes the effective coefficients complex, naturally incorporating nonlinear loss.

The resonant FWM interaction may equivalently be written
\begin{equation}
H_{\rm FWM}=-\hbar\left(\nu p^2a_+^\dagger a_-^\dagger+
\nu^*p^{\dagger2}a_+a_-\right),
\label{eq:Hfwm}
\end{equation}
which immediately shows pair creation and the conservation of $N_+-N_-$ by this vertex.

\section{Symmetry of the complete equations}
\label{app:complete_symmetry}

In the ideal limit,
\begin{equation}
\Delta_+=\Delta_-,\quad \gamma_+=\gamma_-,\quad
s_+=s_-,\quad \kappa\in\mathbb R,
\label{eq:ideal}
\end{equation}
Eqs.~\eqref{eq:three_mode} are invariant under
$(a_+,a_-)\mapsto(a_-,a_+)$.  The order parameter $\eta$ changes sign.  A phase $\Psi_{\rm eff}\ne0,\pi$, unequal loss, unequal detuning, nonzero $k_y^{\rm inc}$, or transverse beam displacement breaks this exchange symmetry explicitly.  At $\Psi_{\rm eff}=0$ or $\pi$, backscattering can still change the threshold through standing-wave splitting without itself choosing a traveling direction.  This distinction explains why phase-zero and phase-$\pi$ pumping can show different instability thresholds while intermediate phases deterministically control the sign~\cite{Anashkina2025}.

\section{Roadmap and scope}
\label{app:roadmap_scope}

These appendices separate rigorous symmetry statements from proposed microscopic realizations and collect the derivations needed to assess the model without consulting a separate online supplement. The ideal one-dimensional profile $h(x)$ retains the reflection $\mathsf M_y:(x,y,z)\mapsto(x,-y,z)$. The transverse current $J_y$ is odd under $\mathsf M_y$, while the $p$-polarized bilinears $|E_x|^2$, $|E_z|^2$, and $\mathrm{Re}(E_xE_z^*)$ are even. Therefore a response $J_y^{(0)}$ requires a separate mirror-odd factor $\Lamy$.

This observation also qualifies the effective-susceptibility interpretation in Ref.~\onlinecite{Plumitallo2023}. Its phenomenological coefficients $\chi_{yxx}^{(2)}$, $\chi_{yxz}^{(2)}$, and $\chi_{yzz}^{(2)}$ were not constrained by the remaining $xz$ mirror. In the ideal one-dimensional model that mirror sets all three to zero. The measured current is not in question; rather, a microscopic explanation must identify the additional mirror-odd material, structural, or dynamical quantity hidden in the fitted coefficients.

The edge-position invariant
\begin{equation}
\Astr\equiv \Im\!\left[h_1^2 h_2^*\right]
\label{eq:Supp_Astr}
\end{equation}
is odd under $x\to-x$ but even under $y\to-y$. It must consequently multiply, rather than replace, $\Lamy$ in the explicitly projected side-jump-like channel,
\begin{equation}
J_{y,{\rm sj}}^{(0)}
=
\frac{\tau_y\Astr}{2}\Lamy\,
\Re\!\left[\chi_{EE}(\omega)\sigma_D^*(\omega)\right]
\left|E_x^{\rm loc}\right|^2 .
\label{eq:Supp_Jy_scope}
\end{equation}
The main text proposes that $\Lamy$ can be generated dynamically by a Kerr instability of a mirror-related directional-SPP quasinormal-mode pair.  The next section gives an analytically transparent two-mode limit of the full three-mode pump--pair equations, Appendix~\ref{sec:Supp_projected_matrix_elements} obtains the bare grating edge vertices, and Appendix~\ref{app:microscopic_vertex} derives the SPP-dressed, $\Astr\eta$-odd projected $T$-matrix component.  Appendix~\ref{app:direct_keldysh} then tests the distinct, symmetry-allowed $\Astr$-independent channel by deriving its retarded, lesser, and greater self-energies and the associated on-shell collision source.  In the controlled clean Au(111) surface-band model that direct order-$g_{\rm SPP}^2$ source vanishes at $808~\mathrm{nm}$ by energy--momentum mismatch, although more general absorptive channels are not ruled out.  Appendix~\ref{sec:Supp_corrected_sidejump} connects the retained structure-dependent component and the possible static anisotropies to the collision and coordinate-shift formulation.  Sidewall texture, deposition direction, and Au(111) texture are retained only as nonexclusive static contributions or weak branch-selection biases. No such realization is assigned as a fabrication error or as the established origin of the experiment.

\section{Two-mode analytic limit of the pump--pair QNM model}
\label{sec:Supp_QNM}

\subsection{Assumptions and mirror-related open-system modes}

The main text retains the pumped mode explicitly, includes the phase-conjugate FWM terms, and tests the resulting three-mode steady states with a $6\times6$ Bogoliubov matrix.  A useful limiting case is obtained when the pump is undepleted and the FWM-generated seed has been absorbed into effective coherent drives $s_\pm$.  Eliminating the clamped pump then leaves two mirror-related Kerr modes.  This reduction does not replace the full calculation; it isolates the self/cross-phase competition and yields a closed-form threshold against which numerical implementations can be checked.

Metal loss and radiation make the grating an open non-Hermitian resonator. Near an isolated spectral window, its field may be expanded in quasinormal modes (QNMs) with complex eigenfrequencies $\widetilde\omega_\nu=\omega_\nu-i\gamma_\nu$~\cite{Sauvan2013,Lalanne2018}. We consider a pair
\begin{equation}
\widetilde{\bm E}(\bm r,t)
\simeq
a_+(t)\widetilde{\bm E}_+(\bm r)
+a_-(t)\widetilde{\bm E}_-(\bm r)
+\bm E_{\rm bg}(\bm r,t)
\label{eq:Supp_QNM_expansion}
\end{equation}
whose members are related by
\begin{equation}
\mathsf M_y\widetilde{\bm E}_+
=
\widetilde{\bm E}_-,
\qquad
\mathsf M_y:\ a_+\longleftrightarrow a_- .
\label{eq:Supp_QNM_mirror}
\end{equation}
The pair is assumed to carry opposite transverse SPP flux, schematically $(K_x,+q_y)$ and $(K_x,-q_y)$. Directional excitation of guided and surface modes is known in other near-field settings~\cite{RodriguezFortuno2013}; its occurrence here remains to be demonstrated. Such a pair requires the finite illuminated width, line ends, contacts, or a fully three-dimensional resonance. It is not the pair of opposite-$K_x$ SPPs obtained from a strictly one-dimensional linear Bloch problem. Linear reciprocal calculations of opposite-$K_x$ launching therefore neither establish nor rule out the transverse nonlinear order parameter introduced here.

The QNM description is appropriate because it treats Ohmic loss and radiation on the same footing. Its use does not imply gain, an exceptional point, or nonreciprocity. The proposed instability is instead a nonlinear loss-driven bifurcation of a reciprocal system.

\subsection{Reduced two-mode Kerr equations}

In a frame rotating at the laser frequency, a minimal reciprocal model is
\begin{subequations}
\begingroup\small
\begin{align}
\dot a_+&=(i\Delta_+-\gamma_+)a_+
+i\varkappa a_-+s_+
+i\left(g_s|a_+|^2+g_c|a_-|^2\right)a_+,
\label{eq:Supp_QNM_plus}\\
\dot a_-&=(i\Delta_--\gamma_-)a_-
+i\varkappa a_++s_-
+i\left(g_s|a_-|^2+g_c|a_+|^2\right)a_- .
\label{eq:Supp_QNM_minus}
\end{align}
\endgroup
\label{eq:Supp_QNM_CMT}
\end{subequations}
The detunings $\Delta_\pm=\omega-\omega_\pm$ and linewidths $\gamma_\pm$ include the linear QNM spectrum; $\varkappa$ mixes the two modes; $s_\pm$ are drive amplitudes; and $g_s$ and $g_c$ are effective self- and cross-Kerr coefficients. A microscopic projection can include the nonlinear response of the metal, surface states, dielectric environment, and hot-electron sector~\cite{Kauranen2012}. Because the resonator is open and absorptive, the general QNM-projected coefficients are complex. The real-coefficient limit below is sufficient to display the instability analytically.

In the unbiased mirror-symmetric case,
\begin{equation}
\Delta_+=\Delta_-=\Delta,\quad
\gamma_+=\gamma_-=\gamma,\quad
s_+=s_-=s,
\label{eq:Supp_QNM_symmetry}
\end{equation}
Eqs.~\eqref{eq:Supp_QNM_CMT} are invariant under $a_+\leftrightarrow a_-$. A symmetric steady state $a_+=a_-=A$ obeys
\begin{equation}
\left[
-\gamma+i\left\{
\Delta+\varkappa+(g_s+g_c)|A|^2
\right\}
\right]A+s=0,
\label{eq:Supp_QNM_symmetric}
\end{equation}
and hence
\begin{equation}
|s|^2
=
I_A\left[
\gamma^2+
\left\{\Delta+\varkappa+(g_s+g_c)I_A\right\}^2
\right],
\qquad I_A=|A|^2.
\label{eq:Supp_QNM_input_output}
\end{equation}
Equation~\eqref{eq:Supp_QNM_input_output} already allows ordinary Kerr multistability. The transverse mirror-breaking instability is identified by the antisymmetric perturbation
\begin{equation}
a_+=A+\delta,\qquad a_-=A-\delta .
\end{equation}
For real $\varkappa$, $g_s$, and $g_c$, linearization gives
\begin{equation}
\frac{d}{dt}
\begin{pmatrix}
\delta\\ \delta^*
\end{pmatrix}
=
\begin{pmatrix}
-\gamma+i\Omega_a & iG_a\\
-iG_a^* & -\gamma-i\Omega_a
\end{pmatrix}
\begin{pmatrix}
\delta\\ \delta^*
\end{pmatrix},
\label{eq:Supp_QNM_Jacobian}
\end{equation}
where
\begin{equation}
\Omega_a=\Delta-\varkappa+2g_sI_A,
\qquad
G_a=(g_s-g_c)A^2 .
\label{eq:Supp_QNM_Omega_G}
\end{equation}
The two antisymmetric eigenvalues are
\begin{equation}
\lambda_\pm=-\gamma
\pm\sqrt{|G_a|^2-\Omega_a^2}.
\label{eq:Supp_QNM_eigenvalues}
\end{equation}
The symmetric state loses stability when
\begin{equation}
|g_s-g_c|^2I_A^2
>
\gamma^2+\left(\Delta-\varkappa+2g_sI_A\right)^2 .
\label{eq:Supp_QNM_threshold}
\end{equation}
Thus a difference between self- and cross-phase modulation can overcome loss in an appropriate detuning window, consistent with symmetry breaking observed in other counterpropagating Kerr resonators~\cite{DelBino2017,Hill2020}. For complex QNM coefficients the threshold is obtained from the largest real part of the full four-dimensional real Jacobian; Eq.~\eqref{eq:Supp_QNM_threshold} is the transparent conservative-Kerr limit, not a numerical threshold for the Au grating.

\subsection{Directional order parameter and pitchfork normal form}

The normalized directional imbalance
\begin{equation}
\eta
=
\frac{|a_+|^2-|a_-|^2}
{|a_+|^2+|a_-|^2}
\label{eq:Supp_eta}
\end{equation}
satisfies $\mathsf M_y:\eta\to-\eta$. Near a simple instability, symmetry requires the center-manifold equation
\begin{equation}
\tau_\eta\dot\eta
=
\mu(I,\Delta)\eta-\beta\eta^3+h_y
+\mathcal O(\eta^5,h_y\eta^2),
\label{eq:Supp_normal_form}
\end{equation}
where $\mu=0$ is the linear-stability boundary and $h_y$ is a weak explicit mirror-odd bias. For a supercritical pitchfork ($\beta>0$) and $h_y=0$,
\begin{equation}
\eta=0\quad(\mu<0),\qquad
\eta_\pm=\pm\sqrt{\mu/\beta}\quad(\mu>0).
\label{eq:Supp_pitchfork_branches}
\end{equation}
The two nonzero states are exchanged by $\mathsf M_y$. An unbiased ensemble has zero average, while a single driven realization can occupy either branch. Higher-order terms can make the transition subcritical and hysteretic. A small $h_y$ unfolds the pitchfork, selecting one sign and smoothing the ideal threshold.

The dynamical mirror factor is
\begin{equation}
\Ldyn=\lambda_\eta\eta,\qquad
\Lamy=\Lstat+\Ldyn.
\label{eq:Supp_Lambda_dynamic}
\end{equation}
Substitution in Eq.~\eqref{eq:Supp_Jy_scope} yields
\begin{equation}
J_{y,{\rm sj}}^{(0)}
=
\frac{\tau_y\Astr}{2}
\left(\Lstat+\lambda_\eta\eta\right)
\Re\!\left[\chi_{EE}(\omega)\sigma_D^*(\omega)\right]
\left|E_x^{\rm loc}\right|^2 .
\label{eq:Supp_Jy_dynamic}
\end{equation}
This equation keeps the $x$-handedness and $y$-mirror breaking conceptually distinct. The symmetric lithographic control suppresses this coupled contribution through $\Astr=0$ even if the optical subsystem develops $\eta\ne0$.

\subsection{Static bias, tests, and limits of the proposal}

A convenient neutral decomposition of the static term is
\begin{equation}
\Lstat
=
\lambda_{\rm tex}T_y
+\lambda_{\rm dep}\Ffab
+\lambda_3\Mbar
+\lambda_g\mathcal A_y+\cdots .
\label{eq:Supp_Lambda_static}
\end{equation}
Here $T_y$ is a transverse sidewall-texture moment, $\Ffab$ is a laboratory-fixed deposition-direction factor, $\Mbar$ is a weighted Au(111) azimuthal texture, and $\mathcal A_y$ is an intentionally introduced transverse geometric asymmetry. These are ordinary sample-dependent symmetry fields, not evidence of fabrication mistakes. None is established as the dominant source by the present characterization. In the nonlinear model their most economical role is to provide $h_y$ and select a directional branch.

The model is falsifiable. It predicts a threshold or rapid power crossover, possible two-sign switching near threshold, hysteresis for a subcritical branch, and correlation of $J_y$ with an independently measured $+q_y/-q_y$ leakage or near-field imbalance. Finite-width QNMs further predict spot-size and spot-position dependence. No Kerr coefficient, QNM pair, or threshold has yet been extracted for the reported sample, so the present work does not claim numerical corroboration of Eqs.~\eqref{eq:Supp_QNM_CMT}--\eqref{eq:Supp_QNM_threshold}. Quantitative validation requires three-dimensional nonlinear Maxwell--material calculations and microscopic evaluation of the Au surface-state response.

\section{Mirror constraint and nonexclusive static realizations}
\label{sec:Supp_mirror}

Let $\mathsf M_y$ denote the reflection $(x,y,z)\mapsto(x,-y,z)$. A polar vector transforms as
\begin{equation}
(J_x,J_y,J_z)\stackrel{\mathsf M_y}{\longrightarrow}(J_x,-J_y,J_z),
\end{equation}
whereas an axial vector transforms as
\begin{equation}
(s_x,s_y,s_z)\stackrel{\mathsf M_y}{\longrightarrow}(-s_x,s_y,-s_z).
\end{equation}
Thus $J_y$ is odd, while $|E_x|^2$, $|E_z|^2$, $\Re(E_xE_z^*)$, $j_x$, and the Edelstein component $s_y$ are even. The binary profile $h(x)$ and every invariant constructed only from its $x$-edge positions are also even under $\mathsf M_y$. Therefore
\begin{equation}
J_y^{(0)}\propto \Astr j_xs_y
\label{eq:Supp_forbidden_form}
\end{equation}
is forbidden in the ideal one-dimensional isotropic model.

\subsection{Directional deposition as one possible static bias}

Physical vapor deposition in high vacuum is directional. If the substrate is off axis or tilted and is not continuously rotated, the arriving flux has a fixed in-plane azimuth. Inclined evaporation through a lithographic resist mask is known to generate direction-dependent shadowing, sidewall coverage, edge fences, and tilted or porous microstructure~\cite{Robbie1998,Eschimese2019,Liu2025}. These observations motivate the dimensionless fabrication factor
\begin{equation}
\Ffab=\hat{\bm y}\cdot\hat{\bm q},
\qquad
\mathsf M_y:\ \Ffab\longrightarrow-\Ffab,
\label{eq:Supp_Ffab}
\end{equation}
where $\hat{\bm q}$ is the unit propagation direction of the arriving Au flux. Normal-incidence deposition has $\Ffab=0$, and continuous rotation averages $\Ffab$ toward zero. A flux tilted only in the $xz$ plane can make the two $x$-facing walls different but does not break the $xz$ mirror; a nonzero transverse component $q_y$ is essential.

For an edge whose sidewall normal is
\begin{equation}
\hat{\bm n}_i=\nu_i\hat{\bm x},
\qquad \nu_i=\pm1,
\end{equation}
the lowest local deposition handedness is
\begin{equation}
\zeta_i^{({\rm ev})}
=\hat{\bm z}\cdot
\left(\hat{\bm n}_i\times\hat{\bm q}\right)
=\nu_i\Ffab.
\label{eq:Supp_local_evap_handedness}
\end{equation}
Opposite walls therefore acquire opposite handedness. If $W_i$ denotes the optical and spin-dependent scattering weight of wall $i$, their evaporation-induced contribution has the schematic form
\begin{equation}
\mathcal C_{\rm ev}
=\sum_i W_i\zeta_i^{({\rm ev})}
=\Ffab\sum_i\nu_iW_i.
\label{eq:Supp_edge_pair_sum}
\end{equation}
For a symmetric unit cell, opposite walls are equivalent at the leading paired-edge order, $W_+=W_-$, and Eq.~\eqref{eq:Supp_edge_pair_sum} cancels. In the asymmetric binary cell the edge positions and SPP near fields make the weights inequivalent. The leading origin-independent quantity that changes sign under reversal of this $x$-asymmetry is $\Astr$, so the paired-edge contribution is parameterized as
\begin{equation}
\begin{aligned}
\sum_i\nu_iW_i&=C_{\rm ev}\Astr+
\text{higher structural invariants},\\
\mathcal C_{\rm ev}&=C_{\rm ev}\Astr\Ffab+\cdots .
\end{aligned}
\label{eq:Supp_pair_reduction}
\end{equation}
This provides the product symmetry needed by the experiment: the fabrication direction breaks $y\to-y$, while the binary cell prevents cancellation of the oppositely handed walls.

If $\Ffab\neq0$, symmetry also permits a background term proportional to $\Ffab|E_x|^2$ that need not contain $\Astr$. Its absence is not guaranteed by point-group symmetry; equivalent opposite walls cancel it in the minimal paired-edge model. The coupled $\Astr\Ffab$ term is retained as one possible static channel, not identified as the origin of the larger SPP-resonant signal.

The available sample report does not specify the source-to-sample azimuth, substrate rotation, or cross-sectional sidewall morphology. Consequently, the factor $\Ffab$ is a symmetry-allowed example and a possible branch-selection bias, not an established characterization of the measured device.

\subsection{Au(111) texture as another possible static bias}

For an Au(111) domain the physical surface group is $C_{3v}$ rather than the artificial $C_{\infty v}$ symmetry of the isotropic Rashba Hamiltonian. If $\phi_c$ is measured from a crystalline mirror line to the grating $x$ axis, threefold equivalence requires periodicity under $\phi_c\to\phi_c+2\pi/3$, and $\mathsf M_y$ sends $\phi_c\to-\phi_c$. The lowest mirror-odd harmonic is consequently
\begin{equation}
\mathcal M_3(\phi_c)=\sin3\phi_c.
\label{eq:Supp_M3}
\end{equation}
It vanishes whenever the grating axis is aligned with a crystalline mirror line. For a domain distribution $w(\phi_c)$, the macroscopic factor is
\begin{equation}
\Mbar=
\frac{\int d\phi_c\,w(\phi_c)W_{\rm opt}(\phi_c)W_{\rm tr}(\phi_c)\sin3\phi_c}
{\int d\phi_c\,w(\phi_c)W_{\rm opt}(\phi_c)W_{\rm tr}(\phi_c)}.
\label{eq:Supp_Mbar}
\end{equation}
A uniform or otherwise mirror-symmetric azimuthal distribution has $\Mbar=0$. Polycrystallinity removes a global crystallographic mirror from any one microscopic realization, but it does not guarantee a nonzero macroscopic coefficient; a weighted orientation imbalance must be measured or inferred.

\section{Model and notation}
\label{app:rashba_model}

The top terrace is described by the Rashba Hamiltonian
\begin{equation}
H_R(\bm k)=\frac{\hbar^2k^2}{2m}+\alpha_R(k_x\sigma_y-k_y\sigma_x)
=\frac{\hbar^2k^2}{2m}+\alpha_R(\hat{\bm z}\times\bm k)\cdot\bm\sigma,
\label{eq:Supp_HR}
\end{equation}
where we adopt the standard Rashba convention, with $\hat{\bm z}$ pointing from the crystal toward vacuum. For Au(111), we take $\alpha_R>0$ in this convention.
Its helicity eigenvalues are
\begin{equation}
\epsilon_{\lambda k}=\frac{\hbar^2k^2}{2m}+\lambda\alpha_R k,
\qquad \lambda=\pm,
\label{eq:Supp_epslambda}
\end{equation}
and we choose the normalized spinors
\begin{equation}
\ket{u_{\lambda,\varphi}}
=
\frac{1}{\sqrt2}
\begin{pmatrix}
1\\
i\lambda e^{i\varphi}
\end{pmatrix},
\label{eq:Supp_spinor}
\end{equation}
for which
\begin{equation}
\ev{\bm \sigma}_{\lambda,\varphi}
=
\lambda(-\sin\varphi,\cos\varphi,0),
\qquad
\ev{\sigma_y}_{\lambda,\varphi}
=
\lambda\cos\varphi.
\label{eq:Supp_spintexture}
\end{equation}

The one-dimensional grating profile is periodic along $x$ with period $D$,
\begin{equation}
h(x)=\sum_{m\neq 0} h_m e^{iG_m x},
\qquad
G_m=\frac{2\pi m}{D},
\label{eq:Supp_hx}
\end{equation}
and we denote by $\mathcal A_{\rm str}$ the translation-invariant inversion-odd structural invariant defined in Eq.~(\ref{eq:Supp_Astr}).

\section{Projected binary-profile scattering matrix elements}
\label{sec:Supp_projected_matrix_elements}

In this section we derive in detail the scalar and spin-orbit scattering matrix elements
used in the main text. The important point is that the Born amplitude is the matrix
element of the perturbation operator between the unperturbed surface states. In the
present problem, the unperturbed states are not three-dimensional plane waves, but
Rashba surface states with a plane-wave dependence in the $(x,y)$ plane, a bound
surface envelope $\phi(z)$ in the normal direction, and a spinor structure.

\subsection{Surface-confined basis states}

We take the surface-state basis in factorized form,
\begin{equation}
\Psi_{\bm k s}(\bm r,z)
=
\frac{1}{\sqrt A}\,
e^{i\bm k\cdot \bm r}\,
\phi(z)\,
\chi_s,
\qquad
\bm r=(x,y),
\label{eq:Supp_basis_states_PM}
\end{equation}
where $A$ is the in-plane normalization area, $\phi(z)$ is the bound top-surface envelope,
and $\chi_s$ is a spin basis state.  The grating is one dimensional, so its profile depends
only on $x$ and has period $D$.

For the binary unit cell used below, let the two raised terraces in each period occupy
$x_1<x<x_2$ and $x_3<x<x_4$.  A sharp periodic indicator of the raised regions is
\begin{equation}
\begin{split}
\Pi(x)=\sum_{n\in\mathbb Z}\big[&\Theta(x-x_1-nD)-\Theta(x-x_2-nD)
\\
&+\Theta(x-x_3-nD)-\Theta(x-x_4-nD)\big].
\end{split}
\label{eq:Supp_binary_indicator_PM}
\end{equation}
Thus $\Pi=1$ on a raised terrace and $\Pi=0$ in a lower region, and the physical height is
$h(x)=H_1\Pi(x)$.  Its average only shifts the reference surface and is absorbed into
$U_0$.  The nonzero Fourier components are therefore written as
\begin{equation}
h(x)-h_0=\sum_{m\neq0}h_m e^{iG_mx},
\qquad
G_m=\frac{2\pi m}{D}.
\label{eq:Supp_hx_PM}
\end{equation}
Introduce the signed edge structure factor
\begin{equation}
\begin{aligned}
S_m&=\sum_{j=1}^{4}s_j e^{-iG_mx_j},\\
(s_1,s_2,s_3,s_4)&=(+1,-1,+1,-1).
\end{aligned}
\label{eq:Supp_Sm_PM}
\end{equation}
For a sharp profile, direct integration of Eq.~\eqref{eq:Supp_binary_indicator_PM} gives
$\Pi_m=S_m/(iG_mD)$.  Convolution with a symmetric, normalized edge-rounding kernel
multiplies each component by its Fourier transform $F_w(G_m)$.  In a notation covering
both cases,
\begin{equation}
\Pi_m=\frac{F_w(G_m)}{iG_mD}S_m,
\qquad
h_m=H_1\Pi_m.
\label{eq:Supp_hm_step_PM}
\end{equation}
Here $F_w(G)$ is the Fourier transform of a normalized edge-rounding kernel.  A perfectly
sharp step has $F_w(G)=1$; a finite edge width suppresses the large-$|G|$ components.
Equivalently, the derivative is the rounded edge comb
\begin{equation}
\partial_x h(x)
=\frac{H_1}{D}\sum_m F_w(G_m)S_m e^{iG_mx},
\label{eq:Supp_dhdx_step_PM}
\end{equation}
whose sharp-edge real-space limit is
\begin{equation}
\partial_xh(x)=H_1\sum_{n\in\mathbb Z}\sum_{j=1}^{4}
s_j\delta(x-x_j-nD).
\label{eq:Supp_edge_comb_real_PM}
\end{equation}

The confining potential may be written as $U(x,z)=U_0[z-H_1\Pi(x)]$.  Because the sharp
binary indicator takes only the values zero and one, the following decomposition is exact:
\begin{equation}
\begin{aligned}
U(x,z)&=U_0(z)+\Pi(x)\Delta U_H(z),\\
\Delta U_H(z)&\equiv U_0(z-H_1)-U_0(z).
\end{aligned}
\label{eq:Supp_Uxz_PM}
\end{equation}
The first term defines the reference surface band, while
\begin{equation}
\delta U(x,z)=\Pi(x)\Delta U_H(z)
\label{eq:Supp_deltaU_PM}
\end{equation}
is the binary-profile perturbation.  Equation~\eqref{eq:Supp_Uxz_PM} is exact with respect
to the two height levels; the use of a single reference envelope $\phi(z)$ remains a
surface-projection approximation.  In the shallow-height limit,
$\Delta U_H(z)=-H_1U_0'(z)+O(H_1^2)$, recovering the usual displaced-interface expansion
$U_0[z-h(x)]\simeq U_0(z)-U_0'(z)h(x)$.

\subsection{Scalar Born matrix element}

The first-Born matrix element of the scalar perturbation is
\begin{equation}
\mel{\bm k' s'}{\delta U}{\bm k s}
=
\int d^2r\,dz\,
\Psi_{\bm k' s'}^*(\bm r,z)\,
\delta U(x,z)\,
\Psi_{\bm k s}(\bm r,z).
\label{eq:Supp_scalar_Born_start}
\end{equation}
Substituting Eqs.~\eqref{eq:Supp_basis_states_PM} and
\eqref{eq:Supp_deltaU_PM}, and writing
$\Pi(x)=\sum_m\Pi_m e^{iG_mx}$, gives
\begin{equation}
\resizebox{\columnwidth}{!}{$\displaystyle
\begin{aligned}
\mel{\bm k' s'}{\delta U}{\bm k s}
&=
\frac{\delta_{s's}}{A}
\sum_m\Pi_m
\int dz\,|\phi(z)|^2\Delta U_H(z)
\int d^2r\,
e^{-i\bm k'\cdot\bm r}e^{iG_mx}e^{i\bm k\cdot\bm r}.
\end{aligned}$}
\label{eq:Supp_scalar_Born_step1}
\end{equation}
The in-plane integral factorizes:
\begin{equation}
\resizebox{\columnwidth}{!}{$\displaystyle
\begin{aligned}
\int d^2r\,
e^{-i\bm k'\cdot \bm r}
e^{iG_m x}
e^{i\bm k\cdot \bm r}
&=
\int dx\,e^{-ik_x' x}e^{iG_m x}e^{ik_x x}
\int dy\,e^{-ik_y' y}e^{ik_y y}
\\
&=
A\,\delta_{k_x',\,k_x+G_m}\,\delta(k_y'-k_y).
\end{aligned}$}
\label{eq:Supp_xy_integral_PM}
\end{equation}
Define the finite-height surface projection
\begin{equation}
\mathcal U_H\equiv
\int dz\,|\phi(z)|^2\Delta U_H(z).
\label{eq:Supp_UH_PM}
\end{equation}
The scalar Born matrix element is therefore
\begin{equation}
\resizebox{\columnwidth}{!}{$\displaystyle
\mel{\bm k' s'}{\delta U}{\bm k s}
=
u_m\,
\delta_{s's}\,
\delta_{k_x',\,k_x+G_m}\,
\delta(k_y'-k_y),
\qquad
u_m\equiv\mathcal U_H\Pi_m
=\frac{\mathcal U_H}{H_1}h_m.$}
\label{eq:Supp_scalar_Born_final}
\end{equation}
This is the Fourier coefficient of the \emph{surface-projected} scalar perturbation, not
of the bare three-dimensional potential alone.  In the shallow-height limit,
$\mathcal U_H\simeq-H_1u_z$, where
$u_z=\int dz\,|\phi(z)|^2U_0'(z)$, and Eq.~\eqref{eq:Supp_scalar_Born_final}
reduces to $u_m=-u_zh_m$.

\subsection{Projected terrace-and-edge spin-orbit matrix element}

The spin-orbit scattering comes from the Pauli term
\begin{equation}
H_{\rm SO}
=
\lambda_P(\nabla U\times \bm p)\cdot \bm \sigma,
\qquad
\lambda_P=\frac{\hbar}{4m_0^2c^2}.
\label{eq:Supp_PauliSOC_PM}
\end{equation}
Because the grating is one dimensional, $\partial_y U=0$, so
\begin{equation}
\resizebox{\columnwidth}{!}{$\displaystyle
H_{\rm SO}
=
\lambda_P\Big[
-(\partial_z U)p_y\sigma_x
+\big((\partial_z U)p_x-(\partial_x U)p_z\big)\sigma_y
+(\partial_x U)p_y\sigma_z
\Big].$}
\label{eq:Supp_Pauli_components_PM}
\end{equation}
The flat-surface part
$\lambda_PU_0'(z)(p_x\sigma_y-p_y\sigma_x)$ of
Eq.~\eqref{eq:Supp_Pauli_components_PM} defines the reference Rashba coupling already
included in the unperturbed surface Hamiltonian.  From Eq.~\eqref{eq:Supp_Uxz_PM}, the
complete grating-dependent Pauli perturbation is
\begin{equation}
\resizebox{\columnwidth}{!}{$\displaystyle
\delta H_{\rm SO}^{\rm gr}
=\lambda_P\Big[
-\Pi\Delta U_H' p_y\sigma_x
+\big(\Pi\Delta U_H'p_x-\Pi'\Delta U_Hp_z\big)\sigma_y
+\Pi'\Delta U_Hp_y\sigma_z
\Big].$}
\label{eq:Supp_Pauli_grating_full_PM}
\end{equation}
Primes on $\Pi$ and $\Delta U_H$ denote derivatives with respect to $x$ and $z$,
respectively.  The $\sigma_y$ component selected by the longitudinal Edelstein spin is
therefore
\begin{equation}
\delta H_{{\rm SO},y}^{\rm gr}
=\lambda_P\big[\Pi(x)\Delta U_H'(z)p_x
-\Pi'(x)\Delta U_H(z)p_z\big]\sigma_y.
\label{eq:Supp_SOC_y_complete_PM}
\end{equation}
The first contribution is the modulation of the horizontal-interface Rashba term; the
second is localized at the sidewalls through the edge comb $\Pi'(x)$.  Keeping only the
second term would not constitute a complete Hermitian first-order Pauli perturbation.

Define the two surface projections
\begin{align}
\mathcal J_H&\equiv
\int dz\,|\phi(z)|^2\Delta U_H'(z),
\label{eq:Supp_JH_PM}\\
\mathcal P_H&\equiv
\int dz\,\phi^*(z)\Delta U_H(z)p_z\phi(z).
\label{eq:Supp_PH_PM}
\end{align}
For a real bound-state envelope, with vanishing boundary terms and
$p_z=-i\hbar\partial_z$, integration by parts gives
\begin{equation}
\mathcal P_H=\frac{i\hbar}{2}\mathcal J_H.
\label{eq:Supp_PH_JH_relation_PM}
\end{equation}
Consequently, projection of Eq.~\eqref{eq:Supp_SOC_y_complete_PM} onto $\phi(z)$ combines
the terrace and sidewall pieces into the explicitly Hermitian operator
\begin{equation}
\delta H_{{\rm SO},y}^{\rm proj}
=\frac{\lambda_P\mathcal J_H}{2}
\left\{\Pi(x),p_x\right\}\sigma_y,
\label{eq:Supp_SOC_projected_anticommutator_PM}
\end{equation}
where $\{A,B\}=AB+BA$.  Its matrix element is
\begin{equation}
\resizebox{\columnwidth}{!}{$\displaystyle
\begin{aligned}
\mel{\bm k's'}{\delta H_{{\rm SO},y}^{\rm proj}}{\bm ks}
&=\zeta_m(\bm k',\bm k)(\sigma_y)_{s's}
\delta_{k_x',\,k_x+G_m}\delta(k_y'-k_y),\\
\zeta_m(\bm k',\bm k)
&\equiv\lambda_P\hbar\mathcal J_H
\frac{k_x+k_x'}{2}\Pi_m.
\end{aligned}$}
\label{eq:Supp_SOC_Born_final_PM}
\end{equation}
The sidewall part alone is
\begin{equation}
\zeta_m^{\rm edge}
=-\lambda_P(iG_m\Pi_m)\mathcal P_H
=-\frac{\lambda_P}{D}F_w(G_m)S_m\mathcal P_H,
\label{eq:Supp_SOC_edge_Sm_PM}
\end{equation}
which displays explicitly how the signed edge sum enters.  Equation~\eqref{eq:Supp_SOC_Born_final_PM},
rather than Eq.~\eqref{eq:Supp_SOC_edge_Sm_PM} by itself, is used below as the Hermitian
projected Pauli vertex.  This formulation also removes the otherwise ambiguous extra
factor of $i$ generated when the derivative factor $iG_m$ is combined with the imaginary
$p_z$ matrix element.

\subsection{Total projected amplitude}

Combining Eqs.~\eqref{eq:Supp_scalar_Born_final} and
\eqref{eq:Supp_SOC_Born_final_PM}, the projected Born vertex for the $m$th channel is
\begin{equation}
\hat M_m(\bm k',\bm k)
=u_m\1+\zeta_m(\bm k',\bm k)\sigma_y.
\label{eq:Supp_Mm_PM}
\end{equation}
For a real profile and real confining potential,
\begin{equation}
u_{-m}=u_m^*,
\qquad
\zeta_{-m}(\bm k,\bm k')=\zeta_m^*(\bm k',\bm k),
\label{eq:Supp_Mm_hermiticity_PM}
\end{equation}
so the full projected grating operator is Hermitian.  Both $u_m$ and $\zeta_m$ are
proportional to the same profile harmonic $\Pi_m=h_m/H_1$; the sidewall component within
$\zeta_m$ is separately proportional to the edge factor $F_w(G_m)S_m$.

\subsection{Matrix element in the Rashba helicity basis}

We now pass from the spin basis \(\chi_s\) to the Rashba helicity basis
\begin{equation}
\ket{\lambda,\varphi}
=
\sum_s [u_{\lambda\varphi}]_s \ket{s},
\qquad
\ket{u_{\lambda,\varphi}}
=
\frac{1}{\sqrt2}
\begin{pmatrix}
1\\
i\lambda e^{i\varphi}
\end{pmatrix}.
\label{eq:Supp_helicity_basis_PM}
\end{equation}
Then
\begin{equation}
\resizebox{\columnwidth}{!}{$\displaystyle
\begin{aligned}
\mel{\lambda',\varphi';\bm k'}{\hat M_m}{\lambda,\varphi;\bm k}
&=
\sum_{s',s}
[u_{\lambda'\varphi'}^\dagger]_{s'}
\mel{\bm k' s'}{\hat M_m}{\bm k s}
[u_{\lambda\varphi}]_s
\\
&=
\Big[
u_m\mel{u_{\lambda'\varphi'}}{\1}{u_{\lambda\varphi}}
+\zeta_m(\bm k',\bm k)
\mel{u_{\lambda'\varphi'}}{\sigma_y}{u_{\lambda\varphi}}
\Big]
\delta_{k_x',\,k_x+G_m}\,
\delta(k_y'-k_y).
\end{aligned}$}
\label{eq:Supp_Mm_helicity_PM}
\end{equation}
with
\begin{equation}
\mel{u_{\lambda'\varphi'}}{\1}{u_{\lambda\varphi}}
=
\frac12\Big(1+\lambda\lambda' e^{i(\varphi-\varphi')}\Big).
\label{eq:Supp_spinor_overlap_PM}
\end{equation}
The needed spinor overlaps are
\begin{equation}
\mel{u_{\lambda'\varphi'}}{\sigma_y}{u_{\lambda\varphi}}
=
\frac12\Big(\lambda e^{i\varphi}+\lambda' e^{-i\varphi'}\Big).
\label{eq:Supp_sigma_y_overlap_PM}
\end{equation}
Hence the exact helicity-basis amplitude is
\begin{equation}
\resizebox{\columnwidth}{!}{$\displaystyle
\begin{aligned}
\mel{\lambda',\varphi';\bm k'}{\hat M_m}{\lambda,\varphi;\bm k}
&=
\Bigg[
\frac{u_m}{2}
\Big(1+\lambda\lambda' e^{i(\varphi-\varphi')}\Big)
+\frac{\zeta_m(\bm k',\bm k)}{2}
\Big(\lambda e^{i\varphi}+\lambda' e^{-i\varphi'}\Big)
\Bigg]
\delta_{k_x',\,k_x+G_m}\,
\delta(k_y'-k_y).
\end{aligned}$}
\label{eq:Supp_Mm_helicity_exact_PM}
\end{equation}

For the dominant intraband nearly resonant channel, \(\lambda'=\lambda\) and
\(\varphi'\approx\varphi\), so
\begin{equation}
\innerproduct{u_{\lambda\varphi'}}{u_{\lambda\varphi}}\approx 1,
\qquad
\mel{u_{\lambda\varphi'}}{\sigma_y}{u_{\lambda\varphi}}
\approx \lambda\cos\varphi.
\label{eq:Supp_intraband_approx_PM}
\end{equation}
Therefore
\begin{equation}
\resizebox{\columnwidth}{!}{$\displaystyle
\mel{\lambda,\varphi';\bm k'}{\hat M_m}{\lambda,\varphi;\bm k}
\approx
\Big[
u_m+\zeta_m(\bm k',\bm k)
\mel{u_{\lambda\varphi'}}{\sigma_y}{u_{\lambda\varphi}}
\Big]
\delta_{k_x',\,k_x+G_m}\,
\delta(k_y'-k_y),$}
\label{eq:Supp_Mmmatrix_PM}
\end{equation}
and, if one further sets \(\varphi'=\varphi\),
\begin{equation}
\resizebox{\columnwidth}{!}{$\displaystyle
\mel{\lambda,\varphi;\bm k'}{\hat M_m}{\lambda,\varphi;\bm k}
\approx
\left(
u_m+\lambda\zeta_m(\bm k',\bm k)\cos\varphi
\right)
\delta_{k_x',\,k_x+G_m}\,
\delta(k_y'-k_y).$}
\label{eq:Supp_Mm_intraband_final_PM}
\end{equation}

\subsection{Implication of the one-dimensional grating geometry}

All projected matrix elements contain
\begin{equation}
\delta(k_y'-k_y),
\label{eq:Supp_ky_conservation_PM}
\end{equation}
which expresses the fact that the ideal grating is translationally invariant along \(y\).
Thus the scattering channels do not generate ordinary skew deflection into inequivalent
\(+k_y'\) and \(-k_y'\) channels. 

More strongly, the binary-profile amplitude in Eq.~\eqref{eq:Supp_Mm_PM} respects the $xz$ mirror and cannot generate any net transverse current from $E_x$ alone. The intuitive Pauli side-jump direction reaches the same conclusion: for an Edelstein spin along $y$ and momentum transfer in the $xz$ plane, $\bm\sigma\times\Delta\bm p$ lies in the $xz$ plane. The projected $\sigma_y$ term contains both the horizontal-interface modulation and its sidewall companion, but it does not by itself provide a $y$-directed displacement. A transverse side jump requires an additional scattering anisotropy that is odd under the $xz$ mirror. In the present model it is supplied primarily by the evaporation factor $\Ffab$ and secondarily by the crystalline factor $\mathcal M_3(\phi_c)$. If $y$-dependent disorder or a deliberately $y$-asymmetric unit cell is present, further geometric skew or side-jump channels can also occur.


\section{Microscopic reduction of the SPP-dressed structure-dependent scattering vertex}
\label{app:microscopic_vertex}

This section derives the lowest projected scattering vertex containing the three ingredients required by the proposed dynamical readout: the inversion-odd unit-cell invariant $\Astr$, the directional SPP imbalance $\eta$, and a spin--orbit operator.  The derivation is microscopic in the sense that it starts from Hamiltonian vertices and the retarded Born series, while leaving the electron--SPP matrix elements and intermediate-state linewidths unevaluated.  It therefore fixes the perturbative order, symmetry, angular projection, and energy denominators, but not an absolute coefficient for the experimental sample.

The object denoted by $\delta T$ below is an \emph{on-shell, Floquet-reduced, first-angular-harmonic component} of the scattering $T$ matrix.  It is not a new bare term postulated in the static one-electron Hamiltonian.  This distinction is essential because the ideal $y$-independent grating vertex derived in Appendix~\ref{sec:Supp_projected_matrix_elements} conserves $k_y$ and cannot produce the required transverse harmonic by itself.

\subsection{Hamiltonian vertices}

We take
\begin{equation}
H(t)=H_R+V_{\rm gr}+V_{\rm SPP}(t),
\label{eq:Supp_vertex_H}
\end{equation}
where $H_R$ is Eq.~\eqref{eq:Supp_HR}.  In terms of the surface-projected vertices of
Appendix~\ref{sec:Supp_projected_matrix_elements}, the grating perturbation is defined by
\begin{equation}
\resizebox{\columnwidth}{!}{$\displaystyle
\mel{\bm k'}{V_{\rm gr}}{\bm k}
=\sum_{m\ne0}
\hat M_m(\bm k',\bm k)
\delta_{k_x',\,k_x+G_m}\delta(k_y'-k_y),
\qquad
\hat M_m=u_m\1+\zeta_m(\bm k',\bm k)\sigma_y.$}
\label{eq:Supp_vertex_Vgr}
\end{equation}
Equations~\eqref{eq:Supp_scalar_Born_final} and
\eqref{eq:Supp_SOC_Born_final_PM} show that both $u_m$ and $\zeta_m$ are proportional to
$\Pi_m=h_m/H_1$; the sidewall part of $\zeta_m$ is proportional to
$iG_m\Pi_m=F_w(G_m)S_m/D$.  For a real profile,
\begin{equation}
\hat M_{-m}(\bm k,\bm k')
=\hat M_m^\dagger(\bm k',\bm k),
\label{eq:Supp_vertex_hermitian_grating}
\end{equation}
so Eq.~\eqref{eq:Supp_vertex_Vgr} is Hermitian.  In the operator products below the
intermediate momenta are implicit.  We write the scalar and spin--orbit pieces as
$\hat M_m^{(0)}=u_m\1$ and
$\hat M_m^{({\rm so})}=\zeta_m(\bm k',\bm k)\sigma_y$.

The two oblique SPP modes are retained through the most general linear electron--field coupling
\begin{equation}
V_{\rm SPP}(t)=\sum_{\mu=\pm}
\left[
a_\mu e^{-i\omega t}\hat G_\mu
+a_\mu^*e^{i\omega t}\hat G_\mu^\dagger
\right].
\label{eq:Supp_vertex_Vspp}
\end{equation}
Here $\hat G_\mu$ may contain scalar-potential, vector-potential, and surface-field contributions.  Its matrix elements
\begin{equation}
g^\mu_{\ell\ell'}
=\mel{\ell}{\hat G_\mu}{\ell'},
\qquad \ell=(\lambda,\bm k),
\label{eq:Supp_vertex_gdef}
\end{equation}
are deliberately left unspecified.  They contain the finite-device QNM field, surface-state envelope, grating Umklapp components, and screening.  Before a directional branch is selected, mirror symmetry relates the two operators by
\begin{equation}
\mathsf M_y\hat G_+\mathsf M_y^{-1}=\hat G_-.
\label{eq:Supp_vertex_Gmirror}
\end{equation}

\subsection{Elimination of the optical sidebands}

Let
\begin{equation}
G_0^R(E)=\left(E-H_R+i\Gamma\right)^{-1}
\label{eq:Supp_vertex_G0}
\end{equation}
denote the retarded electronic propagator, with $\Gamma$ standing for the appropriate intermediate-state broadening.  Eliminating the Floquet sidebands to second order in Eq.~\eqref{eq:Supp_vertex_Vspp} gives the elastic optical self-energy
\begin{equation}
\Sigma_{\rm SPP}^R(E)=\sum_{\mu=\pm}N_\mu\mathcal S_\mu^R(E),
\label{eq:Supp_vertex_Sigma}
\end{equation}
where, for a classical coherent field,
\begin{equation}
\begin{split}
\mathcal S_\mu^R(E)=&\,
\hat G_\mu G_0^R(E+\hbar\omega)\hat G_\mu^\dagger
\\
&+\hat G_\mu^\dagger G_0^R(E-\hbar\omega)\hat G_\mu .
\end{split}
\label{eq:Supp_vertex_Smu}
\end{equation}
Number-independent vacuum terms in a quantized description are mirror even and do not affect the following odd projection.  With $\overline N=(N_++N_-)/2$, Eq.~\eqref{eq:Supp_vertex_Sigma} separates into
\begin{align}
\Sigma_{\rm even}^R
&=\overline N\left(\mathcal S_+^R+\mathcal S_-^R\right),
\nn\\
\Sigma_{\rm odd}^R
&=\overline N\eta\left(\mathcal S_+^R-\mathcal S_-^R\right)
\equiv\lambda_\eta\eta\,\widehat{\mathcal S}_y^R.
\label{eq:Supp_vertex_Sigma_odd}
\end{align}
The normalization of $\widehat{\mathcal S}_y^R$ is arbitrary; Eq.~\eqref{eq:Supp_vertex_Sigma_odd} defines $\lambda_\eta$.  Thus $\lambda_\eta$ contains the total side-mode population as well as the electron--SPP matrix elements and energy denominators.  The fractional imbalance $\eta$ alone is not a scattering strength.

Equation~\eqref{eq:Supp_vertex_Gmirror} implies
\begin{equation}
\mathsf M_y\widehat{\mathcal S}_y^R\mathsf M_y^{-1}
=-\widehat{\mathcal S}_y^R.
\label{eq:Supp_vertex_Soddmirror}
\end{equation}
For mirror-related wave vectors $(K_x,\pm q_y)$ and small $q_y$, the odd self-energy begins as
\begin{equation}
\Sigma_{\rm odd}^R
=2\overline N\eta q_y
\left.\pdv{\mathcal S_{\bm Q}^R}{Q_y}\right|_{Q_y=0}
+O(q_y^3).
\label{eq:Supp_vertex_smallqy}
\end{equation}
This equation identifies the microscopic origin of the missing $y$-odd quantity: it is the difference of two virtual absorption--emission processes carrying opposite transverse SPP momenta.  It vanishes for $q_y=0$, $N_+=N_-$, or zero electron--SPP coupling.

\subsection{Interference of the \texorpdfstring{$G+G$ and $2G$}{G+G and 2G} grating paths}

The retarded scattering operator generated by the effective elastic perturbation is
\begin{equation}
T^R(E)=V_{\rm eff}
+V_{\rm eff}G_0^RV_{\rm eff}
+V_{\rm eff}G_0^RV_{\rm eff}G_0^RV_{\rm eff}+\cdots,
\label{eq:Supp_vertex_Tseries}
\end{equation}
with $V_{\rm eff}=V_{\rm gr}+\Sigma_{\rm even}^R+\Sigma_{\rm odd}^R$.  The lowest translation-invariant product that distinguishes a binary unit cell from its $x$-reversed partner is the closed reciprocal-lattice path
\begin{equation}
(+G)+(+G)+(-2G)=0.
\label{eq:Supp_vertex_loopmomentum}
\end{equation}
For example,
\begin{equation}
\hat M_{-2}G_0^R\hat M_1G_0^R\hat M_1
\propto h_1^2h_2^*.
\label{eq:Supp_vertex_loop}
\end{equation}
The same invariant appears directly in the transition probability as the interference term
\begin{equation}
\begin{aligned}
\delta W_{112}&\propto
2\operatorname{Re}\!\left[
(T_{2G}^{(1)})^\dagger T_{G+G}^{(2)}
\right],\\
T_{2G}^{(1)}&=\hat M_2,
&
T_{G+G}^{(2)}&=\hat M_1G_0^R\hat M_1.
\end{aligned}
\label{eq:Supp_vertex_path_interference}
\end{equation}
The $x$-reversed loop is proportional to $(h_1^2h_2^*)^*$.  If $D^R$ denotes the remaining product of propagators and matrix-element prefactors, their interference obeys
\begin{equation}
2\operatorname{Re}(D^Rh_1^2h_2^*)
-2\operatorname{Re}[D^R(h_1^2h_2^*)^*]
=-4\operatorname{Im}D^R\,\Astr.
\label{eq:Supp_vertex_interference_identity}
\end{equation}
Thus $\Astr$ is not inserted by hand: it is the handedness-odd part of the interference between the one-step $2G$ path and the two-step $G+G$ path.  A nonzero handedness-odd \emph{rate} also requires an absorptive or on-shell phase, $\operatorname{Im}D^R\ne0$.  Such a phase can be supplied by the retarded $i\Gamma$ denominators, Ohmic or radiative damping, or complex screened electron--SPP matrix elements.  In a strictly lossless, wholly off-resonant model with real denominators, this contribution vanishes at this order.

Retaining one Pauli spin--orbit insertion in the loop gives
\begin{equation}
\begin{split}
\mathcal L_{112}^{(y)}=&\,
\hat M_{-2}^{({\rm so})}G_0^R\hat M_1^{(0)}G_0^R\hat M_1^{(0)}
\\
&+\hat M_{-2}^{(0)}G_0^R\hat M_1^{({\rm so})}G_0^R\hat M_1^{(0)}
\\
&+\hat M_{-2}^{(0)}G_0^R\hat M_1^{(0)}G_0^R\hat M_1^{({\rm so})}
+\text{orderings}.
\end{split}
\label{eq:Supp_vertex_L112}
\end{equation}
Every displayed term is proportional to $h_1^2h_2^*\sigma_y$.  This is the microscopic origin of the $\sigma_y$ factor: it comes from the Hermitian projected Pauli corrugation operator in Eq.~\eqref{eq:Supp_SOC_projected_anticommutator_PM}, whose sidewall component is $-(\partial_xU)p_z\sigma_y$, rather than from the scalar grating profile.

\subsection{Combined handedness-odd and SPP-odd vertex}

The required term is obtained by inserting Eq.~\eqref{eq:Supp_vertex_Sigma_odd} at every distinct position in the loop of Eq.~\eqref{eq:Supp_vertex_L112}, adding the conjugate reciprocal-lattice paths, and retaining terms linear in both $\Astr$ and $\eta$.  A compact operator expression is
\begin{equation}
\delta T_{A\eta}^R
=\left.
\mathcal P_A\mathcal P_\eta
\sum_{\pi}
X_{\pi_1}G_0^RX_{\pi_2}G_0^RX_{\pi_3}G_0^RX_{\pi_4}
\right|_{\text{one SO}},
\label{eq:Supp_vertex_compact}
\end{equation}
where $(X_1,X_2,X_3,X_4)=(\hat M_1,\hat M_1,\hat M_{-2},\Sigma_{\rm odd}^R)$, the sum contains the distinct orderings, and ``one SO'' means that exactly one of the three grating matrices is replaced by $\hat M_m^{({\rm so})}$.  The projectors are
\begin{align}
\mathcal P_A F
&=\frac12\left[F(h_1,h_2)-F(h_1^*,h_2^*)\right],
\nn\\
\mathcal P_\eta F
&=\frac12\left[F(N_+,N_-)-F(N_-,N_+)\right].
\label{eq:Supp_vertex_projectors}
\end{align}
Equation~\eqref{eq:Supp_vertex_compact} is of order
\begin{equation}
\delta T_{A\eta}^R
=O\!\left(g_{\rm SPP}^2
[V_{\rm gr}^{(0)}]^2V_{\rm gr}^{({\rm so})}\right),
\label{eq:Supp_vertex_order}
\end{equation}
namely second order in electron--SPP coupling, first order in Pauli spin--orbit coupling,
and three grating vertices.  In the shallow-height limit this reduces to
$O(g_{\rm SPP}^2\lambda_Ph^3)$.  Within the minimal vertices of
Eqs.~\eqref{eq:Supp_vertex_Vgr} and \eqref{eq:Supp_vertex_Vspp}, with handedness carried
by the first two profile harmonics and spin dependence supplied by one complete projected
Pauli insertion, this is the lowest order that can contain $N_+-N_-$, $\sigma_y$, and
$\Im(h_1^2h_2^*)$ simultaneously.  Additional atomistic or intrinsically spin-dependent
electron--SPP vertices could generate other terms and are absorbed into the general
matrix elements in Eq.~\eqref{eq:Supp_vertex_gdef}.

For clarity, insertion of complete intermediate electronic states converts Eq.~\eqref{eq:Supp_vertex_compact} into
\begin{equation}
\resizebox{\columnwidth}{!}{$\displaystyle
\begin{aligned}
\mel{\ell}{\delta T_{A\eta}^R(E)}{\ell'}
&=\Astr\lambda_\eta\eta
\sum_{\pi}\sum_{\alpha\beta\gamma}
\frac{
\mathcal V^{\pi_1}_{\ell\alpha}
\mathcal V^{\pi_2}_{\alpha\beta}
\mathcal V^{\pi_3}_{\beta\gamma}
\mathcal V^{\pi_4}_{\gamma\ell'}
}{
(E-\epsilon_\alpha+i\Gamma_\alpha)
(E-\epsilon_\beta+i\Gamma_\beta)
(E-\epsilon_\gamma+i\Gamma_\gamma)
}
+\cdots .
\end{aligned}$}
\label{eq:Supp_vertex_spectral}
\end{equation}
After the factors $\Astr\lambda_\eta\eta$ have been extracted, the $\mathcal V^{\pi_j}$ are the reduced scalar, spin--orbit, and SPP-odd matrix elements specified by Eq.~\eqref{eq:Supp_vertex_compact}.  The ellipsis includes the absorption--emission denominators already contained in Eq.~\eqref{eq:Supp_vertex_Smu}, the conjugate loops, and permutations in which the external states lie between different vertices.  Equation~\eqref{eq:Supp_vertex_spectral}, rather than a fitted scalar, is the microscopic definition of the unknown coefficient.

\subsection{First-angular-harmonic projection}

The complete $\delta T_{A\eta}^R(\varphi,\varphi')$ contains several angular harmonics.  The measured channel retains only an incoming longitudinal distortion and an outgoing transverse current.  The corresponding coefficient is defined without approximation by
\begin{equation}
\begin{split}
C_{\lambda\lambda'}^{yx}(E)
=4\int_0^{2\pi}\frac{d\varphi}{2\pi}
\int_0^{2\pi}\frac{d\varphi'}{2\pi}
\sin\varphi\cos\varphi'\qquad\qquad\\[-6pt]
\times\frac12\Tr\!\left[
\sigma_y
\frac{\mel{\bm k_\lambda(\varphi)}{\delta T_{A\eta}^R(E)}{\bm k_{\lambda'}(\varphi')}}
{\Astr\lambda_\eta\eta}
\right].
\end{split}
\label{eq:Supp_vertex_Cprojection}
\end{equation}
In Eq.~\eqref{eq:Supp_vertex_Cprojection}, the kets specify the orbital momentum direction and the radial momenta $k_\lambda$ and $k_{\lambda'}$ of the two Rashba contours, while the spin indices remain open for $\Tr_s$.  The subsequent helicity-spinor overlaps are retained in the spin trace entering $\gamma_{\lambda\lambda'}$ below.
The factor four compensates $\langle\sin^2\varphi\rangle=\langle\cos^2\varphi'\rangle=1/2$.  By definition, the $y\leftarrow x$ rank-one projection is therefore
\begin{equation}
\boxed{
\left[\delta T_{\lambda\lambda'}^R\right]_{yx}
=C_{\lambda\lambda'}^{yx}
\Astr\lambda_\eta\eta\,
\sigma_y\sin\varphi\cos\varphi'.}
\label{eq:Supp_vertex_target}
\end{equation}
The factors $\sin\varphi=k_y/k$ and $\cos\varphi'=k_x'/k'$ are consequently not independent microscopic potentials.  They are the lowest angular projectors onto the transverse output and longitudinal input sectors.  The source of transverse momentum in Eq.~\eqref{eq:Supp_vertex_target} is Eq.~\eqref{eq:Supp_vertex_smallqy} or a finite-width boundary contained in $g^\mu_{\ell\ell'}$; the ideal static grating alone still conserves $k_y$.

Equation~\eqref{eq:Supp_vertex_target} should not be mistaken for the complete two-angle kernel.  A Hermitian effective Hamiltonian has the first-harmonic completion
\begin{equation}
\begin{split}
\delta T_{\lambda\lambda'}^{(1)}
=\Astr\lambda_\eta\eta\,\sigma_y
\Big\{&C_{\lambda\lambda'}^{s}
[\sin\varphi\cos\varphi'+\sin\varphi'\cos\varphi]
\\
&+iC_{\lambda\lambda'}^{a}
[\sin\varphi\cos\varphi'-\sin\varphi'\cos\varphi]
\Big\},
\end{split}
\label{eq:Supp_vertex_reciprocal_completion}
\end{equation}
with the corresponding conjugation relations under $\ell\leftrightarrow\ell'$.  For the retarded scattering operator the exact relation is $T^A_{\ell'\ell}=(T^R_{\ell\ell'})^\dagger$.  On contraction with an incoming $\cos\varphi'$ distribution and an outgoing $\sin\varphi$ current, the exchanged term integrates to zero because $\langle\sin\varphi'\cos\varphi'\rangle=0$.  Equation~\eqref{eq:Supp_vertex_target} is therefore the relevant projected component without asserting that the omitted reciprocal partner is absent.

\subsection{From the vertex to the quadratic collision source}

Let $T_0^R$ denote the mirror-even background scattering amplitude.  To first order in Eq.~\eqref{eq:Supp_vertex_target}, Fermi's golden rule gives
\begin{equation}
\delta W_{\ell\ell'}
=\frac{4\pi}{\hbar}
\delta(\epsilon_\ell-\epsilon_{\ell'})
\operatorname{Re}\Tr_s
\left[(T_{0,\ell\ell'}^R)^\dagger
\delta T_{\ell\ell'}^R\rho_s\right],
\label{eq:Supp_vertex_deltaW}
\end{equation}
where $\rho_s$ is the spin-density matrix on the spin leg.  In the second-order kinetic or Kubo response, one optical insertion produces the longitudinal distribution $\delta f_x^{(1)}(\varphi',t)\propto j_x^{(1)}(t)\cos\varphi'$, while the other produces the Edelstein spin insertion $\rho_s^{(1)}(t)\propto s_y^{(1)}(t)\sigma_y$.  Contracting the latter with the $\sigma_y$ in Eq.~\eqref{eq:Supp_vertex_target} gives the reduced rate
\begin{equation}
\mathcal W_{\lambda\lambda'}^{s}
(\varphi,\varphi';t)
=2\gamma_{\lambda\lambda'}
\Astr\lambda_\eta\eta\,
s_y^{(1)}(t)\sin\varphi\cos\varphi',
\label{eq:Supp_vertex_Wresult}
\end{equation}
where $\gamma_{\lambda\lambda'}$ contains the density of final states and $\operatorname{Re}[(T_0^R)^*C_{\lambda\lambda'}^{yx}]$, together with the helicity spin traces.  Inserting Eq.~\eqref{eq:Supp_vertex_Wresult} into the collision integral and using $2\langle\cos^2\varphi'\rangle=1$ yields
\begin{equation}
I_{\lambda,y}^{(2)}(\varphi,t)
=-\partial_\epsilon f_\lambda^{(0)}
\Gamma_\lambda\Astr\lambda_\eta\eta\,
j_x^{(1)}(t)s_y^{(1)}(t)\sin\varphi,
\label{eq:Supp_vertex_collision_result}
\end{equation}
and hence
\begin{equation}
J_{y,{\rm str}}^{(0)}
=\mathcal C_{\rm str}\Astr\lambda_\eta\eta
\ev{j_x^{(1)}(t)s_y^{(1)}(t)}_t.
\label{eq:Supp_vertex_Jresult}
\end{equation}

Equation~\eqref{eq:Supp_vertex_deltaW} is a compact representation of a second-order quantum-kinetic contraction; it is not an equilibrium scattering rate that somehow contains two independent first-order distribution functions.  Equivalently, in a self-consistent kinetic formulation one may regard Eq.~\eqref{eq:Supp_vertex_target} as the vertex per unit Edelstein spin and write
\begin{equation}
\delta T^R(t)
=s_y^{(1)}(t)
\left.\pdv{T^R}{s_y}\right|_{s_y=0}.
\label{eq:Supp_vertex_per_spin}
\end{equation}
Such a formulation requires exchange, screening, or a two-particle collision kernel.  If the scattering operator is assumed to be strictly static, single-particle, and independent of the driven density matrix, its linear collision integral cannot by itself generate the product $j_xs_y$.

There is also an important optical-power distinction.  Equation~\eqref{eq:Supp_vertex_Sigma_odd} gives $\lambda_\eta\eta\propto N_+-N_-$.  Since $j_x^{(1)}s_y^{(1)}$ is already second order in the local ac field, the SPP-dressed term in Eq.~\eqref{eq:Supp_vertex_Jresult} is generally of higher optical order and may be nonanalytic at the bifurcation.  It should not be identified with a power-independent bulk $\chi^{(2)}(0;\omega,-\omega)$.  In a restricted operating range it can instead be viewed as an intensity-dependent effective rectification coefficient.  Measuring $J_y$ and $N_+-N_-$ simultaneously as functions of incident power would test this distinction.

Finally, a spin-dependent rate asymmetry and a side jump are not identical.  A genuine side-jump contribution must be evaluated with the gauge-invariant coordinate shift
\begin{equation}
\delta\bm r_{\ell\ell'}
=\bm{\mathcal A}_\ell-\bm{\mathcal A}_{\ell'}
-\left(\nabla_{\bm k}+\nabla_{\bm k'}\right)
\arg T_{\ell\ell'},
\label{eq:Supp_vertex_coordinate_shift}
\end{equation}
and with both the side-jump velocity and anomalous-distribution terms~\cite{Sinitsyn2008,Nagaosa2010,Atencia2022}.  The phase of Eq.~\eqref{eq:Supp_vertex_target} contributes to this coordinate shift.  The part arising solely from $\delta|T|^2$ is more properly called skew-scattering-like.  We therefore retain the term ``side-jump-like'' for Eq.~\eqref{eq:Supp_vertex_Jresult} until the full finite-device $T$ matrix, its phase derivatives, and the companion anomalous distribution have been evaluated.

The derivation establishes that the proposed vertex is allowed and identifies a concrete perturbative path to it.  It does not guarantee a nonzero value: cancellations between Rashba helicities, grating paths, or absorption and emission denominators can make $C_{\lambda\lambda'}^{yx}$ small or zero.  Determining its magnitude and sign requires the finite-device QNMs, $g^\mu_{\ell\ell'}$, the surface and edge potential, electronic broadening, and the actual Au(111) domain distribution.

\section{Keldysh analysis of the structure-independent directional-SPP channel}
\label{app:direct_keldysh}

\subsection{Question, symmetry, and controlled scope}

The projection in Appendix~\ref{app:microscopic_vertex} isolates the term that is odd in both the longitudinal unit-cell handedness $\Astr$ and the directional SPP imbalance $\eta$.  It does not remove the terms that are linear in $\eta$ but independent of $\Astr$ from the full $T$ matrix.  Consequently, one must separately test a possible direct contribution of the form
\begin{equation}
J_{y,{\rm dir}}^{(0)}=\mathcal C_{\rm dir}\lambda_\eta\eta+\cdots .
\label{eq:Supp_direct_candidate}
\end{equation}
Such a term is symmetry allowed.  Under the $xz$ mirror $\mathsf M_y$, $\eta$, the axial spin $S_x$, and the polar current $J_y$ are all odd.  Under the $yz$ mirror $\mathsf M_x:(x,y,z)\mapsto(-x,y,z)$, $S_x$ and $J_y$ are both even.  Thus the $yz$ mirror of a centered symmetric grating cannot force Eq.~\eqref{eq:Supp_direct_candidate} to zero once the optical state itself has $\eta\ne0$.

We now calculate this channel in a deliberately controlled limit.  The assumptions are: (i) one clean Au(111) Rashba surface band described by Eq.~\eqref{eq:Supp_HR}; (ii) a spatially homogeneous steady state coupled adiabatically to an equilibrium reservoir; (iii) electron--SPP coupling retained to second order; (iv) the low reciprocal harmonics of the experimental grating and the two directional QNMs; and (v) no surface-to-bulk, remote-interband, disorder-assisted, phonon-assisted, or contact-assisted transitions.  These assumptions make the calculation closed and expose the relevant energy--momentum constraint.  They do not constitute a complete model of real Au at an optical photon energy.

A further scope restriction is important.  For a coherently driven mode pair, the second-order self-energy is a matrix in the optical-mode indices.  The calculation in the next nine subsections retains its mode-diagonal entries, weighted by $N_\mu=|a_\mu|^2$, and projects their part linear in the population imbalance $\eta$.  Mode-off-diagonal entries proportional to $a_+a_-^*$ instead generate electron coherences at a nonzero transverse wave vector and are not part of this $\eta$-linear projection.  Their elastic kinematics, transport order, and finite-device interpretation are derived separately in the final subsection of this appendix.  Thus the clean-limit zero obtained below bounds the $\eta$-linear direct population-injection channel, not every possible coherence-dependent finite-device response.

\subsection{Contour Green functions and electron--SPP vertices}

Let $\ell=(\lambda,\bm k)$ denote a Rashba helicity and in-plane momentum.  Equation~\eqref{eq:Supp_vertex_Vspp} may be written in second-quantized form as
\begin{equation}
H_{e{\text{-}}{\rm SPP}}(t)=
\sum_{\mu=\pm}\sum_{\ell\ell'}
\left[a_\mu e^{-i\omega t}g^\mu_{\ell\ell'}
c_\ell^\dagger c_{\ell'}+{\rm H.c.}\right].
\label{eq:Supp_direct_Hespp}
\end{equation}
For a translationally invariant surface component, absorption from mode $\mu$ changes the electron momentum by
\begin{equation}
\bm Q_{\mu m}=(K_x+mG,\mu q_y),
\label{eq:Supp_direct_Qmum}
\end{equation}
where $m$ labels a grating Umklapp harmonic.  In a finite device this selection rule is broadened by the QNM envelope; the calculation below keeps the low-order components for which Eq.~\eqref{eq:Supp_direct_Qmum} is the central momentum.

The helicity projectors and equilibrium Green functions are
\begin{align}
P_\lambda(\bm k)
&=\frac12\left[\1+\lambda(\hat{\bm z}\times\hat{\bm k})\cdot\bm\sigma\right],
\label{eq:Supp_direct_projector}\\
G_0^{R,A}(\bm k,E)
&=\sum_\lambda
\frac{P_\lambda(\bm k)}{E-\epsilon_{\lambda k}\pm i0^+},
\label{eq:Supp_direct_GRA}\\
G_0^<(\bm k,E)
&=2\pi i f(E)\sum_\lambda P_\lambda(\bm k)
\delta(E-\epsilon_{\lambda k}),
\label{eq:Supp_direct_Gless}\\
G_0^>(\bm k,E)
&=-2\pi i[1-f(E)]\sum_\lambda P_\lambda(\bm k)
\delta(E-\epsilon_{\lambda k}).
\label{eq:Supp_direct_Ggreater}
\end{align}

\subsection{Retarded, lesser, and greater self-energies}

The second-order Born self-energy on the Keldysh contour is
\begin{equation}
\Sigma(1,2)=V_{e{\text{-}}{\rm SPP}}(1)G_0(1,2)V_{e{\text{-}}{\rm SPP}}(2).
\label{eq:Supp_direct_contourSigma}
\end{equation}
After averaging over the optical cycle, its retarded component in the mode-diagonal population sector is precisely the sideband self-energy of Eqs.~\eqref{eq:Supp_vertex_Sigma} and \eqref{eq:Supp_vertex_Smu}.  The corresponding lesser and greater components in this sector are
\begin{equation}
\begin{split}
\Sigma_\mu^{\lessgtr}(E)=N_\mu\big[&
\hat G_\mu G_0^{\lessgtr}(E+\hbar\omega)\hat G_\mu^\dagger
{}+\hat G_\mu^\dagger G_0^{\lessgtr}(E-\hbar\omega)\hat G_\mu\big].
\end{split}
\label{eq:Supp_direct_Sigmalg}
\end{equation}
Momentum arguments of the intermediate propagators are shifted according to Eq.~\eqref{eq:Supp_direct_Qmum}.  The first and second terms describe the two Floquet directions; in a quantized-field notation they become the stimulated absorption and stimulated-emission terms, with the usual additional vacuum factor in spontaneous emission.  The vacuum part is identical for the two mirror-related modes and therefore cancels from the $\eta$-odd projection.

For every Keldysh component $X=R,A,<,>$, define
\begin{equation}
\Sigma_\eta^X
=\overline N\eta\left(\mathcal S_+^X-\mathcal S_-^X\right),
\qquad
\overline N=\frac{N_++N_-}{2}.
\label{eq:Supp_direct_Sigmaeta}
\end{equation}
Mirror covariance gives
\begin{equation}
\mathsf M_y\Sigma_\eta^X(\bm k,E)\mathsf M_y^{-1}
=-\Sigma_\eta^X(k_x,-k_y,E).
\label{eq:Supp_direct_Sigma_mirror}
\end{equation}
The retarded quantity in Eq.~\eqref{eq:Supp_direct_Sigmaeta} is therefore the same mirror-odd virtual self-energy used inside the mixed vertex of Appendix~\ref{app:microscopic_vertex}.  A retarded self-energy alone, however, does not yet specify a nonequilibrium occupation.

\subsection{Separation of spectral dressing and population injection}

To first order in the $\eta$-odd self-energy, the Langreth rules give
\begin{equation}
\begin{split}
\delta G_\eta^<=&
G_0^R\Sigma_\eta^<G_0^A
+G_0^R\Sigma_\eta^RG_0^<
+G_0^<\Sigma_\eta^AG_0^A .
\end{split}
\label{eq:Supp_direct_dGless_raw}
\end{equation}
Using $G_0^<=f(E)(G_0^A-G_0^R)$ and defining
\begin{align}
\delta G_\eta^{R,A}&=G_0^{R,A}\Sigma_\eta^{R,A}G_0^{R,A},
\label{eq:Supp_direct_dGRA}\\
\mathcal I_\eta(E)&=
\Sigma_\eta^<(E)-f(E)\left[\Sigma_\eta^A(E)-\Sigma_\eta^R(E)\right],
\label{eq:Supp_direct_Ieta}
\end{align}
one obtains the exact first-order rearrangement
\begin{equation}
\boxed{
\delta G_\eta^<
=f(E)\left(\delta G_\eta^A-\delta G_\eta^R\right)
+G_0^R\mathcal I_\eta G_0^A .}
\label{eq:Supp_direct_dG_decomp}
\end{equation}
The first term changes the spectral function.  It is the reactive, Stark-like or Floquet dressing produced by virtual absorption and emission.  The second term changes the occupation and is the genuine Keldysh collision source.  This distinction is central: a nonzero $\operatorname{Re}\Sigma_\eta^R$ does not imply $\mathcal I_\eta\ne0$.

There is a useful equilibrium check.  If the driven optical field merely produces a static cycle-averaged quasienergy $\widetilde\epsilon_{n\bm k}$ and the reservoir populates it with $f(\widetilde\epsilon_{n\bm k})$, the gauge-complete band current is
\begin{equation}
\begin{split}
J_{y,{\rm spec}}
&=-\frac{e}{\hbar}\sum_n\int_{\bm k}
\pdv{\widetilde\epsilon_{n\bm k}}{k_y}
f(\widetilde\epsilon_{n\bm k})
\\
&=-\frac{e}{\hbar}\sum_n\int_{\bm k}
\pdv{\mathcal F(\widetilde\epsilon_{n\bm k})}{k_y}=0,
\qquad \mathcal F'(\epsilon)=f(\epsilon),
\end{split}
\label{eq:Supp_direct_total_derivative}
\end{equation}
where $\int_{\bm k}\equiv\int d^2k/(2\pi)^2$.  The last equality follows from Brillouin-zone periodicity, or from a vanishing boundary term in the continuum model.  Equation~\eqref{eq:Supp_direct_total_derivative} is the equilibrium no-current check that is missed if the paramagnetic current is evaluated without the contact, diamagnetic, and vertex terms required by the chosen electromagnetic gauge.  A direct dc transport current therefore requires the nonequilibrium part $G_0^R\mathcal I_\eta G_0^A$, or an explicitly open boundary flux.

\subsection{On-shell reduction of the collision source}

Inserting Eqs.~\eqref{eq:Supp_direct_Gless} and \eqref{eq:Supp_direct_Ggreater} into Eq.~\eqref{eq:Supp_direct_Sigmalg} produces the Golden-rule delta functions explicitly.  With $s=+1$ for absorption and $s=-1$ for stimulated emission, define the detuning for the transition $\ell=(\lambda,\bm k)\to\ell'=(\lambda',\bm k+s\bm Q_{\mu m})$ by
\begin{equation}
\Delta_{\mu m,s}^{\lambda'\lambda}(\bm k)
=\epsilon_{\lambda',|\bm k+s\bm Q_{\mu m}|}
-\epsilon_{\lambda k}-s\hbar\omega.
\label{eq:Supp_direct_transition_detuning}
\end{equation}
The corresponding Golden-rule rate is
\begin{equation}
W_{\mu m,s}(\ell'\leftarrow\ell)
=\frac{2\pi}{\hbar}N_\mu
\left|g_{\mu m,s}^{\lambda'\lambda}(\bm k)\right|^2
\delta\!\left(\Delta_{\mu m,s}^{\lambda'\lambda}\right).
\label{eq:Supp_direct_Wgolden}
\end{equation}
The diagonal kinetic equation follows either from Eq.~\eqref{eq:Supp_direct_Ieta} or from the gain--loss form
\begin{equation}
\begin{split}
\left.\dot f_\ell\right|_\eta
=\mathcal P_\eta\sum_{\mu m s\ell'}\big[&
W_{\mu m,-s}(\ell\leftarrow\ell')f_{\ell'}(1-f_\ell)
\\[-2pt]
&-W_{\mu m,s}(\ell'\leftarrow\ell)f_\ell(1-f_{\ell'})\big],
\end{split}
\label{eq:Supp_direct_collision}
\end{equation}
where $\mathcal P_\eta F=[F(N_+,N_-)-F(N_-,N_+)]/2$.  Equations~\eqref{eq:Supp_direct_Wgolden} and \eqref{eq:Supp_direct_collision} show precisely where a signed rate can arise: unequal $N_+$ and $N_-$ weight transitions carrying opposite $q_y$, but only if the corresponding energy-conserving delta function has support.  Virtual paths contribute to $\operatorname{Re}\Sigma_\eta^R$ even when this on-shell support is absent.

\subsection{Kinematic test at the experimental wavelength}

For the Rashba dispersion in Eq.~\eqref{eq:Supp_epslambda}, the energy change associated with a momentum transfer $\bm Q$ is
\begin{equation}
\begin{split}
\Delta\epsilon_{\lambda'\lambda}(\bm k,\bm Q)
=&\frac{\hbar^2}{2m}
\left(2\bm k\cdot\bm Q+Q^2\right)
\\
&+\alpha_R\left(\lambda'|\bm k+\bm Q|-\lambda k\right).
\end{split}
\label{eq:Supp_direct_Deltaeps}
\end{equation}
The experimental wavelength gives
\begin{equation}
\hbar\omega=\frac{hc}{808~{\rm nm}}\simeq1.535~{\rm eV}.
\label{eq:Supp_direct_photonenergy}
\end{equation}
For the period $D=1488~{\rm nm}$ of Ref.~\onlinecite{Plumitallo2023},
\begin{equation}
G=\frac{2\pi}{D}=4.22\times10^6~{\rm m}^{-1}.
\label{eq:Supp_direct_Gnumber}
\end{equation}
The propagating SPP wave vector and the low grating harmonics therefore give representative transfers $Q\sim10^7~{\rm m}^{-1}$.  With a representative Au(111) surface-state velocity $v_F\sim8\times10^5~{\rm m\,s}^{-1}$,
\begin{equation}
\hbar v_FQ\sim5~{\rm meV},
\qquad
\frac{\hbar^2Q^2}{2m}\ll1~{\rm meV}.
\label{eq:Supp_direct_energy_scale}
\end{equation}
Even allowing for an inter-helicity Rashba splitting of order $0.1~{\rm eV}$ on Au(111)~\cite{Henk2003,Yan2015}, the largest energy change available within the low-energy surface-state sector and the retained low-order momentum transfers remains far below $1.535~{\rm eV}$:
\begin{equation}
\left|\Delta\epsilon_{\lambda'\lambda}\right|
\lesssim O(10^{-1}~{\rm eV})
\ll\hbar\omega.
\label{eq:Supp_direct_mismatch}
\end{equation}

The size of the missing momentum can be stated quantitatively.  Retain the parabolic part of Eq.~\eqref{eq:Supp_direct_Deltaeps}, take the most favorable intraband geometry $\bm Q\parallel\bm k$ with $k=k_F$, and define
\begin{equation}
E_0=\frac{\hbar^2k_F^2}{2m},
\qquad
u=\frac{Q}{k_F}.
\label{eq:Supp_direct_E0u}
\end{equation}
Then $\Delta\epsilon=E_0(2u+u^2)$, and the momentum $Q_*$ required to absorb one $808$-nm quantum obeys
\begin{equation}
\boxed{
\frac{Q_*}{k_F}
=-1+\sqrt{1+\frac{\hbar\omega}{E_0}}.}
\label{eq:Supp_direct_Qstar}
\end{equation}
Using representative Au(111) surface-state parameters $m\simeq0.26m_e$ and $E_0\simeq0.42~{\rm eV}$, which give $k_F\simeq0.169~\text{\AA}^{-1}$~\cite{Henk2003,Yan2015}, Eq.~\eqref{eq:Supp_direct_Qstar} yields
\begin{equation}
Q_*\simeq1.16k_F\simeq0.196~\text{\AA}^{-1}
\simeq1.96\times10^9~{\rm m}^{-1}.
\label{eq:Supp_direct_Qstar_number}
\end{equation}
This is approximately $247$ times the propagating SPP wave vector and $464$ reciprocal vectors of the $1488$-nm grating.  Equivalently, it corresponds to the few-nanometer spatial period $2\pi/Q_*\simeq3.2~{\rm nm}$, rather than to the lithographic period.  Rashba and inter-helicity corrections can change the numerical prefactor, but they do not close this two-orders-of-magnitude momentum gap.  The very high Umklapp components capable of doing so lie outside the low-harmonic reduction and depend on unmodeled nanometer-scale edge or disorder structure.

Accordingly, the delta function in Eq.~\eqref{eq:Supp_direct_Wgolden} has no support in this controlled model.  To order $g_{\rm SPP}^2$ and in the clean limit,
\begin{equation}
\boxed{
\mathcal I_\eta=0,
\qquad
\delta G_{\eta,{\rm neq}}^<=0,
\qquad
S_{x,{\rm dir}}=J_{y,{\rm dir}}^{(0)}=0.}
\label{eq:Supp_direct_clean_zero}
\end{equation}
This result is an energy--momentum, or on-shell, selection rule for the $\eta$-linear one-SPP channel.  It is not a mirror-symmetry prohibition.

\subsection{Explicit spin and current generated when an on-shell channel is opened}

The vanishing result in Eq.~\eqref{eq:Supp_direct_clean_zero} can be checked by deriving the response that would occur if an additional process opened the delta function.  The nonequilibrium spin and current are
\begin{align}
S_{x,{\rm dir}}
&=-\frac{i\hbar}{2}\int\frac{dE}{2\pi}\int_{\bm k}
\Tr\!\left[\sigma_xG_0^R\mathcal I_\eta G_0^A\right],
\label{eq:Supp_direct_SxKeldysh}\\
J_{y,{\rm dir}}
&=ie\int\frac{dE}{2\pi}\int_{\bm k}
\Tr\!\left[v_yG_0^R\mathcal I_\eta G_0^A\right]
+J_y^{\rm contact}+J_y^{\rm vert},
\label{eq:Supp_direct_JyKeldysh}
\end{align}
where
\begin{equation}
v_y=\frac{\hbar k_y}{m}-\frac{\alpha_R}{\hbar}\sigma_x.
\label{eq:Supp_direct_vy}
\end{equation}
The contact and vertex terms in Eq.~\eqref{eq:Supp_direct_JyKeldysh} are required when $\hat G_\mu$ contains a vector-potential coupling.  They implement the relevant Ward identity and prevent a purely reactive gauge-dependent term from being misidentified as transport.

For an intuitive closed-form check, suppose a real injection process produces a small rigid displacement $\delta k_y$ of both occupied helicity contours.  At zero temperature,
\begin{equation}
\delta f_\lambda(\bm k)
=-\delta k_y\pdv{f_\lambda^{(0)}}{k_y}
=\delta k_y\sin\varphi\,
\delta(k-k_{F\lambda}).
\label{eq:Supp_direct_shifted_distribution}
\end{equation}
Using $\langle\sigma_x\rangle_{\lambda\varphi}=-\lambda\sin\varphi$ from Eq.~\eqref{eq:Supp_spintexture}, one finds, when both Rashba contours are occupied,
\begin{align}
S_{x,{\rm dir}}
&=\frac{\hbar}{2}\sum_\lambda\int_{\bm k}
\langle\sigma_x\rangle_{\lambda\varphi}\delta f_\lambda
\nn\\
&=-\frac{\hbar\delta k_y}{8\pi}
(k_{F+}-k_{F-})
=\frac{\hbar k_R}{4\pi}\delta k_y,
\qquad
k_R=\frac{m\alpha_R}{\hbar^2}.
\label{eq:Supp_direct_Sxshift}
\end{align}
The spin-dependent part of Eq.~\eqref{eq:Supp_direct_vy} then gives the inverse-Edelstein contribution
\begin{equation}
J_{y,{\rm IEE}}^{\rm dir}
=\frac{2e\alpha_R}{\hbar^2}S_{x,{\rm dir}}
=\frac{e\alpha_R k_R}{2\pi\hbar}\delta k_y.
\label{eq:Supp_direct_IEEshift}
\end{equation}
Equations~\eqref{eq:Supp_direct_Sxshift} and \eqref{eq:Supp_direct_IEEshift} show that Rashba spin--momentum locking does convert an actual $y$-directed occupation shift into $S_x$ and a transverse current.  In the clean off-resonant calculation, however, Eq.~\eqref{eq:Supp_direct_clean_zero} gives $\delta k_y=0$ because no real one-SPP transition injects the required momentum.

\subsection{Momentum-balance check}

The same conclusion follows without evaluating spinor traces.  Let $R_\mu$ be the net rate per unit area for electronic absorption minus stimulated emission from directional mode $\mu$.  Each event transfers transverse momentum $\mu\hbar q_y$, so
\begin{equation}
\dot{\mathcal P}_y
=\hbar q_y(R_+-R_-).
\label{eq:Supp_direct_momentum_injection}
\end{equation}
Including a momentum-relaxation time $\tau_p$ gives
\begin{equation}
\dot{\mathcal P}_y+\frac{\mathcal P_y}{\tau_p}
=\hbar q_y(R_+-R_-),
\qquad
J_{y,{\rm drag}}^{\rm dir}=-\frac{e}{m}\mathcal P_y.
\label{eq:Supp_direct_momentum_balance}
\end{equation}
For mirror-related matrix elements, $R_\pm=N_\pm r_0$ and
\begin{equation}
R_+-R_-=2\overline N\eta r_0.
\label{eq:Supp_direct_Rdifference}
\end{equation}
Thus a direct $\eta$-odd drag and Edelstein response would indeed be present if $r_0\ne0$.  In the clean low-energy model, Eq.~\eqref{eq:Supp_direct_Wgolden} gives $r_0=0$, reproducing Eq.~\eqref{eq:Supp_direct_clean_zero}.  This makes explicit why the direct term is absent from the calculated collision kernel but remains symmetry allowed in a more complete material model.

\subsection{Finite broadening and channels outside the reduction}

A phenomenological electronic linewidth replaces the clean delta function by
\begin{equation}
\delta_\Gamma(x)=\frac{1}{\pi}\frac{\Gamma}{x^2+\Gamma^2}.
\label{eq:Supp_direct_lorentzian}
\end{equation}
For a detuning $|x|\simeq\hbar\omega\gg\Gamma$, its absolute tail scales as $\Gamma/[\pi(\hbar\omega)^2]$ and its value relative to the resonant peak scales as $(\Gamma/\hbar\omega)^2$.  Therefore a constant-broadening surface-band model gives a small, not mathematically exact, direct term.  A numerical value cannot be assigned consistently without the energy dependence of the spectral functions and the microscopic mechanism that produces the broadening.

Several real processes can remove the clean-limit restriction: (i) optical transitions from the surface state into bulk Au bands; (ii) interband transitions within the full Au band structure; (iii) disorder- or edge-assisted transitions that supply additional momentum; (iv) phonon-assisted absorption or emission; (v) finite-spot and contact momentum broadening; and (vi) higher grating Umklapp components.  The prominent bulk Au $5d\to6sp$ interband onset lies near $2.4$--$2.5~{\rm eV}$~\cite{Johnson1972}, above $\hbar\omega=1.535~{\rm eV}$.  This reduces the phase space for that particular channel at $808~{\rm nm}$ but does not make the real device rigorously transparent: lifetime broadening, surface-to-bulk transitions, and nonvertical or disorder-assisted processes can leave subthreshold absorption.  These mechanisms contribute new matrix elements and new intermediate spectral densities to $\Sigma^{<,>}$.  The mode-off-diagonal coherence terms discussed below are not included in this list because they are elastic two-vertex dressing rather than real one-SPP absorption or emission.  The present calculation therefore does not prove that $\mathcal C_{\rm dir}$ in Eq.~\eqref{eq:Supp_direct_candidate} is exactly zero in the fabricated device.  It proves the narrower and useful statement that its $\eta$-linear one-SPP contribution vanishes at order $g_{\rm SPP}^2$ in the clean, low-harmonic, single-surface-band model.

\subsection{Relation to the retained structure-dependent vertex}

The direct term was not removed algebraically from the unprojected $T$ matrix.  Its $\eta$-linear part begins as
\begin{equation}
\begin{split}
T_\eta^R=&\,\Sigma_\eta^R
+V_{\rm gr}G_0^R\Sigma_\eta^R
+\Sigma_\eta^RG_0^RV_{\rm gr}
+\cdots .
\end{split}
\label{eq:Supp_direct_Teta}
\end{equation}
Appendix~\ref{app:microscopic_vertex} instead applies $\mathcal P_A\mathcal P_\eta$ and retains the mixed closed reciprocal-lattice loop containing $\Astr\eta$.  The two mechanisms use different components of the same optical dressing:
\begin{description}
\item[Direct $\eta$ channel.] The required Keldysh object is $\Sigma_\eta^{<,>}$, or equivalently $\mathcal I_\eta$.  Its physical role is real one-SPP population and momentum injection.
\item[Mixed $\Astr\eta$ channel.] The required object is $\Sigma_\eta^R$ inside $\mathcal P_A\mathcal P_\eta T^R$.  Its physical role is virtual SPP dressing of an elastic, handedness-sensitive grating-scattering vertex.
\end{description}
Consequently, the absence of an on-shell one-SPP transition does not set $\operatorname{Re}\Sigma_\eta^R$ to zero and does not remove the mixed virtual path.  The latter can modulate the elastic grating vertex and rectify the already driven correlator $\langle j_x^{(1)}s_y^{(1)}\rangle_t$.  It still requires the absorptive phase discussed below Eq.~\eqref{eq:Supp_vertex_interference_identity}, and its coefficient can cancel after the helicity and path sums; the Keldysh result here does not establish its magnitude.

The conclusion for the population-imbalance channel is therefore twofold.  First, a structure-independent current $J_y\propto\eta$ is allowed by symmetry and must not be declared identically forbidden by the $yz$ mirror.  Second, its leading $\eta$-linear one-SPP collision source is zero in the controlled clean, off-resonant Au(111) surface-state model at the experimental wavelength.  This kinematic result, together with the absence of resonant $J_y$ peaks in the symmetric control, supports retaining the $\Astr$-weighted pathway as the working explanation of the measured resonant contrast while leaving more general direct absorptive channels as a testable correction.

\subsection{Mode-off-diagonal coherence and the spatially averaged current}

For completeness, we now restore the optical-mode coherences omitted from Eq.~\eqref{eq:Supp_direct_Sigmalg}.  In a coherent classical two-mode field, the complete cycle-averaged second-order object has the mode-matrix form
\begin{equation}
\begin{split}
\Sigma^X(E)=\sum_{\mu,\nu=\pm}\big[&
a_\mu a_\nu^*\hat G_\mu G_0^X(E+\hbar\omega)\hat G_\nu^\dagger
\\
&+a_\mu^*a_\nu\hat G_\mu^\dagger G_0^X(E-\hbar\omega)\hat G_\nu\big],
\end{split}
\label{eq:Supp_direct_full_mode_sigma}
\end{equation}
where $X=R,A,<,>$ and the momentum labels are implicit.  The $\mu=\nu$ entries reduce to Eq.~\eqref{eq:Supp_direct_Sigmalg}.  The $\mu\ne\nu$ entries contain $a_+a_-^*$ and its conjugate.  They describe a virtual Raman-like sequence in which the electron absorbs from one member of the degenerate pair and emits into the other.  For matched longitudinal Umklapp components, the net energy and momentum delivered to the electron are
\begin{equation}
\begin{aligned}
\Delta E_{+-}&=\hbar\omega_+-\hbar\omega_-=0,\\
\Delta\bm Q_{+-}&=\bm Q_{+m}-\bm Q_{-m}=(0,2q_y).
\end{aligned}
\label{eq:Supp_direct_coh_transfer}
\end{equation}
The intermediate electronic state in Eq.~\eqref{eq:Supp_direct_full_mode_sigma} remains virtual and carries the denominator at $E\pm\hbar\omega$, but the external electron scattering is elastic.  Its intraband on-shell condition is
\begin{equation}
\epsilon_{\lambda,|\bm k+\Delta\bm Q_{+-}|}
=\epsilon_{\lambda,k}.
\label{eq:Supp_direct_coh_onshell}
\end{equation}
For the isotropic Rashba dispersion, $|\bm k+\Delta\bm Q_{+-}|=|\bm k|$ satisfies Eq.~\eqref{eq:Supp_direct_coh_onshell} and gives $k_y=-q_y$.  Thus this external elastic amplitude is not subject to the one-photon mismatch in Eq.~\eqref{eq:Supp_direct_mismatch}.

This observation does not, however, produce a missing uniform order-$g_{\rm SPP}^2$ current.  Let $\bm k'=\bm k+\Delta\bm Q$.  At that order the off-diagonal self-energy generates an electron density-matrix coherence between different momenta,
\begin{equation}
\begin{split}
\delta G_{{\rm coh};\bm k',\bm k}^<={}&
\left[G_0^R\Sigma_{\rm coh}^<G_0^A\right]_{\bm k',\bm k}
\\
&+\left[G_0^R\Sigma_{\rm coh}^RG_0^<
+G_0^<\Sigma_{\rm coh}^AG_0^A\right]_{\bm k',\bm k},
\end{split}
\label{eq:Supp_direct_coh_dG}
\end{equation}
rather than a diagonal population $\delta G^<_{\bm k,\bm k}$.  A spatially averaged current contracts only the diagonal momentum component,
\begin{equation}
\overline J_y
=ie\int\frac{dE}{2\pi}\int_{\bm k}
\Tr\!\left[v_y(\bm k)\delta G^<_{\bm k,\bm k}(E)\right],
\label{eq:Supp_direct_uniform_current}
\end{equation}
and therefore receives no direct contribution from Eq.~\eqref{eq:Supp_direct_coh_dG} in the translationally invariant bulk reduction.  The coherence instead produces a finite-wave-vector density or current modulation.  A diagonal population correction requires either a second insertion of the effective coherence vertex, giving a rate of order $g_{\rm SPP}^4|a_+a_-^*|^2$, or interference with a static scattering or boundary matrix element carrying $-\Delta\bm Q$, giving a mixed correction of order $U_{-\Delta Q}g_{\rm SPP}^2a_+a_-^*$.  The ideal infinitely extended one-dimensional grating supplies no such transverse matrix element because it conserves $k_y$; edges, contacts, finite illumination, or $y$-dependent disorder can supply it.

The real-space picture makes the same result transparent.  With $a_\pm=|a_\pm|e^{i\phi_\pm}$, the cycle-averaged interference pattern contains
\begin{equation}
\delta I(y)\propto
2|a_+||a_-|\cos\!\left(2q_yy+\Delta\phi\right),
\qquad
\Delta\phi=\phi_+-\phi_-.
\label{eq:Supp_direct_coh_standingwave}
\end{equation}
At a fixed origin, the quadrature
\begin{equation}
\mathcal P_{\rm coh}
=\Im(a_+a_-^*)
=\overline N\sqrt{1-\eta^2}\sin\Delta\phi
\label{eq:Supp_direct_Pcoh}
\end{equation}
is odd under $\mathsf M_y$.  It is nevertheless not an origin-independent bulk invariant: under $y\to y+y_0$, $\Delta\phi\to\Delta\phi-2q_yy_0$.  A single standing-wave harmonic always has a mirror plane displaced to $y=-\Delta\phi/(2q_y)$, just as a single spatial harmonic of the grating cannot define the translation-invariant handedness derived in Appendix~\ref{sec:structural_handedness_detailed}.  Consequently, averaging Eq.~\eqref{eq:Supp_direct_coh_standingwave} over a translationally invariant surface or over complete periods yields no signed bulk response.

A finite device provides a physical phase reference through the beam envelope, line ends, contacts, or transverse disorder.  For a measurement window $W(y-y_c)$, the relevant quantity is an overlap such as
\begin{equation}
\mathcal P_W=
\int dy\,W(y-y_c)
\sin\!\left(2q_yy+\Delta\phi\right),
\label{eq:Supp_direct_coh_overlap}
\end{equation}
which depends only on the displacement of the standing-wave mirror plane from the window or device centroid.  This belongs to the same boundary-referenced class as the $C_r(y_0-y_c)$ term in Eq.~\eqref{eq:hy}; its magnitude is an uncalculated finite-device overlap and need not be parametrically small when only a few fringes span the illuminated region.  It is not an independent, translation-invariant bulk coefficient.

The coherent and population-imbalance channels can be distinguished experimentally.  The former scales with $\mathcal P_{\rm coh}\propto\sqrt{1-\eta^2}\sin\Delta\phi$ and can be largest at $\eta=0$, whereas both $\eta$-linear channels correlate with $N_+-N_-$.  Displacing the spot transversely changes the overlap in Eq.~\eqref{eq:Supp_direct_coh_overlap}; reversing the population imbalance instead tests the $\eta$ channels.  Because the coherence term is not proportional to $\Astr$, a finite-device contribution is not symmetry-forbidden in the centered symmetric grating, but the absence of resonant transverse-current peaks in that control empirically bounds it.  The precise conclusion of this appendix is therefore that the leading $\eta$-linear one-SPP population source vanishes in the stated clean model, while mode coherence gives a finite-$2q_y$ elastic dressing whose conversion into a spatially averaged current requires an additional boundary, disorder, or momentum-mixing process.

\section{Mirror-consistent dynamical, fabrication-, and crystalline-anisotropic side-jump projection}
\label{sec:Supp_corrected_sidejump}

\subsection{Coordinate-shift constraint}

For a scattering event $\ell=(\lambda,\bm k)\to\ell'=(\lambda',\bm k')$, the gauge-invariant side-jump displacement is~\cite{Berger1970,Sinitsyn2008,Nagaosa2010,Atencia2022}
\begin{equation}
\delta\bm r_{\ell'\ell}
=\bm{\mathcal A}_{\ell'}-\bm{\mathcal A}_{\ell}
-\left(\nabla_{\bm k}+\nabla_{\bm k'}\right)
\arg T_{\ell'\ell}.
\label{eq:Supp_sidejump_correct}
\end{equation}
For an $xz$-mirror-symmetric scatterer, the $y$ component satisfies
\begin{equation}
\delta y_{\lambda'\lambda}(k_x',-k_y';k_x,-k_y)
=-\delta y_{\lambda'\lambda}(k_x',k_y';k_x,k_y).
\label{eq:Supp_deltay_mirror}
\end{equation}
The $E_x$-driven first-order occupation is even under $(k_y,k_y')\to(-k_y,-k_y')$. Therefore the angularly integrated transverse side-jump current from the mirror-even binary-profile amplitude $u_m\1+\zeta_m\sigma_y$ vanishes pairwise. Conservation of $k_y$ does not evade this cancellation.

For a sidewall morphology created with fabrication factor $\Ffab$, reflection maps the device to one fabricated with the reversed transverse evaporation azimuth. The corresponding constraint is
\begin{align}
&\delta y_{\lambda'\lambda}(k_x',-k_y';k_x,-k_y;-\Ffab)
\nn\\
&\hspace{2cm}=-\delta y_{\lambda'\lambda}(k_x',k_y';k_x,k_y;\Ffab).
\label{eq:Supp_deltay_evap}
\end{align}
The lowest fixed-device term that can survive the $k_y$ integration is therefore linear in $\Ffab$. Its microscopic length and sign depend on the deposition angle, resist shadowing, sidewall coverage, edge fences, grain morphology, and corner potential. We denote the corresponding branch-resolved coefficient by $\ell_{\lambda'\lambda}^{({\rm ev})}$.

For a physical $C_{3v}$ domain at azimuth $\phi_c$, reflection maps the sample to a distinct domain at $-\phi_c$. The side-jump constraint is instead
\begin{align}
&\delta y_{\lambda'\lambda}(k_x',-k_y';k_x,-k_y;-\phi_c)
\nn\\
&\hspace{2cm}=-\delta y_{\lambda'\lambda}(k_x',k_y';k_x,k_y;\phi_c).
\label{eq:Supp_deltay_domain}
\end{align}
The lowest term that can survive the $k_y$ integration at fixed domain orientation is proportional to the mirror-odd harmonic $\mathcal M_3(\phi_c)=\sin3\phi_c$. Its microscopic length and sign depend on the atomistic corner potential, edge rounding, surface reconstruction, and inter-helicity mixing. We denote the corresponding branch-resolved coefficient by $\ell_{\lambda'\lambda}^{(3)}$.

\subsection{Coarse-grained nonlinear edge kernel}

The measured dc response is second order in the optical field. At the lowest angular order, the incoming longitudinal distortion is proportional to $\cos\varphi'$, while the transverse output channel is $\sin\varphi$. The leading paired-edge kernel with the required $x$-inversion, $xz$-mirror, and translation properties is
\begin{equation}
\begin{aligned}
\mathcal W^{\rm sj}_{\lambda\lambda'}(\varphi,\varphi';t)
&=2\Astr\left[
\gamma_{\lambda\lambda'}^{(\eta)}\lambda_\eta\eta+
\gamma_{\lambda\lambda'}^{({\rm ev})}\Ffab+
\gamma_{\lambda\lambda'}^{(3)}\Mbar
\right]\\
&\quad\times s_y^{(1)}(t)\sin\varphi\cos\varphi'.
\end{aligned}
\label{eq:Supp_Wkernel_correct}
\end{equation}
Here $\gamma_{\lambda\lambda'}^{(\eta)}$ is the projected SPP-dressed coefficient derived in Appendix~\ref{app:microscopic_vertex}, while $\gamma_{\lambda\lambda'}^{({\rm ev})}$ and $\gamma_{\lambda\lambda'}^{(3)}$ are the coarse-grained fabrication- and crystalline-anisotropic edge coefficients.  Each of $\lambda_\eta\eta\sin\varphi$, $\Ffab\sin\varphi$, and $\Mbar\sin\varphi$ is even under $\mathsf M_y$, so Eq.~\eqref{eq:Supp_Wkernel_correct} is a legitimate collision kernel.  The same expression without a mirror-odd factor is forbidden.

The first-order distribution is
\begin{equation}
\delta f_{\lambda'}^{(1)}(\varphi',t)
=-\partial_\epsilon f_{\lambda'}^{(0)}
a_{\lambda'}^{(1)}(t)\cos\varphi'.
\label{eq:Supp_deltaf_correct}
\end{equation}
Inserting Eqs.~\eqref{eq:Supp_Wkernel_correct} and~\eqref{eq:Supp_deltaf_correct} into
\begin{equation}
I_{{\rm edge},\lambda}^{(2,0)}
=\sum_{\lambda'}\mathcal N_{\lambda'}
\int_0^{2\pi}\frac{d\varphi'}{2\pi}
\mathcal W^{\rm sj}_{\lambda\lambda'}(\varphi,\varphi';t)
\delta f_{\lambda'}^{(1)}(\varphi',t)
\label{eq:Supp_Iedge_general_correct}
\end{equation}
and using $2\langle\cos^2\varphi'\rangle=1$ gives
\begin{align}
I_{{\rm edge},\lambda}^{(2,0)}
=&-\partial_\epsilon f_\lambda^{(0)}
\Astr s_y^{(1)}(t)\sin\varphi
\nn\\
&\times\sum_{\lambda'}
\left[
\gamma_{\lambda\lambda'}^{(\eta)}\lambda_\eta\eta+
\gamma_{\lambda\lambda'}^{({\rm ev})}\Ffab+
\gamma_{\lambda\lambda'}^{(3)}\Mbar
\right]\mathcal N_{\lambda'}
a_{\lambda'}^{(1)}(t).
\label{eq:Supp_Iedge_insert_correct}
\end{align}
The longitudinal sheet-current moment is
\begin{equation}
j_x^{(1)}(t)=-\frac e2\sum_{\lambda'}
\mathcal N_{\lambda'}v_{\lambda'}a_{\lambda'}^{(1)}(t).
\label{eq:Supp_jx_correct}
\end{equation}
Because the linear solution has $a_{\lambda'}^{(1)}\propto v_{\lambda'}$, we define
\begin{equation}
\resizebox{\columnwidth}{!}{$\displaystyle
\sum_{\lambda'}
\left[
\gamma_{\lambda\lambda'}^{(\eta)}\lambda_\eta\eta+
\gamma_{\lambda\lambda'}^{({\rm ev})}\Ffab+
\gamma_{\lambda\lambda'}^{(3)}\Mbar
\right]\mathcal N_{\lambda'}
a_{\lambda'}^{(1)}(t)
\equiv
\left[\eta_\lambda^{(\eta)}\lambda_\eta\eta+
\eta_\lambda^{({\rm ev})}\Ffab+
\eta_\lambda^{(3)}\Mbar\right]j_x^{(1)}(t).$}
\label{eq:Supp_eta3_def}
\end{equation}
The projected source is therefore
\begin{equation}
\begin{split}
I_{{\rm edge},\lambda}^{(2,0)}
=&-\partial_\epsilon f_\lambda^{(0)}\Astr
\left[\eta_\lambda^{(\eta)}\lambda_\eta\eta+
\eta_\lambda^{({\rm ev})}\Ffab+
\eta_\lambda^{(3)}\Mbar\right]
\\
&\times j_x^{(1)}s_y^{(1)}\sin\varphi,
\end{split}
\label{eq:Supp_Iedge_final_correct}
\end{equation}
which is the explicit dynamical-plus-static contribution to the generic response in Eq.~\eqref{eq:Supp_Jy_scope}.  The perturbative expression for the dynamical coefficient is Eq.~\eqref{eq:Supp_vertex_spectral}; its numerical value is not known.  The construction fixes the symmetry and angular dependence but does not claim that either static anisotropic coefficient is determined by the isotropic matrix element in Eq.~\eqref{eq:Supp_Mm_PM}. Cross-sectional morphology together with an atomistic edge calculation or an experimental calibration is required for their absolute values.

\section{Semiclassical and Kubo derivations of the Edelstein spin polarization}
\label{app:edelstein_derivation}

In this section we derive the current-induced spin polarization (Edelstein effect) for a
two-dimensional electron gas with Rashba spin-orbit coupling. We use the convention
\begin{equation}
H(\mathbf{k})
=
\frac{\hbar^2 k^2}{2m}\,\sigma_0
+
\alpha_R\left(k_x\sigma_y-k_y\sigma_x\right),
\label{eq:Supp_Rashba_H}
\end{equation}
where $m$ is the effective mass, $\alpha_R$ is the Rashba coefficient, and $e>0$ denotes
the magnitude of the electron charge, so that the electron charge is $-e$.

We write
\begin{equation}
\mathbf{k}=k(\cos\varphi,\sin\varphi),
\qquad
\hat{\mathbf{k}}=(\cos\varphi,\sin\varphi),
\label{eq:Supp_k_phi}
\end{equation}
so that the Rashba Hamiltonian becomes
\begin{equation}
H(\mathbf{k})
=
\frac{\hbar^2 k^2}{2m}\,\sigma_0
+
\alpha_R k
\left(
-\sin\varphi\,\sigma_x+\cos\varphi\,\sigma_y
\right).
\label{eq:Supp_Rashba_H_phi}
\end{equation}

Its eigenvalues are
\begin{equation}
\varepsilon_\lambda(k)
=
\frac{\hbar^2 k^2}{2m}
+
\lambda \alpha_R k,
\qquad
\lambda=\pm,
\label{eq:Supp_Rashba_dispersion}
\end{equation}
and a convenient choice of normalized eigenspinors is
\begin{equation}
u_{\lambda,\varphi}
=
\frac{1}{\sqrt{2}}
\begin{pmatrix}
1\\
i\lambda e^{i\varphi}
\end{pmatrix}.
\label{eq:Supp_Rashba_spinors}
\end{equation}

From Eq.~\eqref{eq:Supp_Rashba_spinors} one finds the spin expectation values
\begin{subequations}
\begin{align}
\innerproduct{u_{\lambda,\varphi}}{\sigma_x u_{\lambda,\varphi}}
&=
-\lambda\sin\varphi,
\\
\innerproduct{u_{\lambda,\varphi}}{\sigma_y u_{\lambda,\varphi}}
&=
\lambda\cos\varphi,
\\
\innerproduct{u_{\lambda,\varphi}}{\boldsymbol{\sigma} u_{\lambda,\varphi}}
&=
\lambda(-\sin\varphi,\cos\varphi,0)
=
\lambda\,(\hat{\mathbf z}\times \hat{\mathbf k}).
\end{align}
\label{eq:Supp_spin_expectation}
\end{subequations}
Therefore the physical spin carried by one state is
\begin{equation}
\mathbf{s}_\lambda(\varphi)
=
\frac{\hbar}{2}
\innerproduct{u_{\lambda,\varphi}}{\boldsymbol{\sigma} u_{\lambda,\varphi}}
=
\frac{\hbar}{2}\lambda(\hat{\mathbf z}\times \hat{\mathbf k}).
\label{eq:Supp_physical_spin_state}
\end{equation}

The group velocity on branch $\lambda$ is
\begin{equation}
\mathbf{v}_\lambda(\mathbf{k})
=
\frac{1}{\hbar}\nabla_{\mathbf{k}}\varepsilon_\lambda(k)
=
v_\lambda(k)\,\hat{\mathbf k},
\qquad
v_\lambda(k)
=
\frac{\hbar k}{m}
+
\frac{\lambda\alpha_R}{\hbar}.
\label{eq:Supp_branch_velocity}
\end{equation}

We now derive the Edelstein spin polarization in two ways.

\subsection{Semiclassical Boltzmann derivation}

In the relaxation-time approximation, a weak static electric field $\mathbf E$ produces
the nonequilibrium correction
\begin{equation}
\delta f_\lambda(\mathbf{k})
=
e\tau\,
\mathbf E\cdot \mathbf v_\lambda(\mathbf{k})\,
\frac{\partial f_0(\varepsilon_\lambda)}{\partial \varepsilon},
\label{eq:Supp_deltaf_Boltzmann}
\end{equation}
where $f_0$ is the equilibrium Fermi distribution and $\tau$ is the momentum relaxation time. Here $e>0$ is the magnitude of the electron charge and $\hbar\dot{\mathbf k}=-e\mathbf E$. Equivalently, $\delta f_\lambda=-e\tau\,\mathbf E\cdot\mathbf v_\lambda[-\partial f_0/\partial\varepsilon]$.

The induced spin density is
\begin{equation}
\mathbf S
=
\sum_{\lambda=\pm}
\int\frac{d^2k}{(2\pi)^2}\,
\mathbf s_\lambda(\varphi)\,
\delta f_\lambda(\mathbf{k}).
\label{eq:Supp_spin_density_def}
\end{equation}
Substituting Eqs.~\eqref{eq:Supp_physical_spin_state} and \eqref{eq:Supp_deltaf_Boltzmann},
\begingroup\small
\begin{align}
\mathbf S
&=
\sum_{\lambda=\pm}
\int\frac{d^2k}{(2\pi)^2}\,
\left[\frac{\hbar}{2}\lambda(\hat{\mathbf z}\times \hat{\mathbf k})\right]
\left[
e\tau\,(\mathbf E\cdot \hat{\mathbf k})\,v_\lambda(k)\,
\frac{\partial f_0(\varepsilon_\lambda)}{\partial \varepsilon}
\right]
\nonumber\\
&=
-\frac{\hbar e\tau}{2}
\sum_{\lambda=\pm}\lambda
\int\frac{d^2k}{(2\pi)^2}\,
(\hat{\mathbf z}\times \hat{\mathbf k})\,
(\mathbf E\cdot \hat{\mathbf k})\,
v_\lambda(k)\,
\left[-\frac{\partial f_0(\varepsilon_\lambda)}{\partial \varepsilon}\right].
\label{eq:Supp_spin_density_Boltzmann_step1}
\end{align}
\endgroup

At $T=0$,
\begin{equation}
-\frac{\partial f_0(\varepsilon_\lambda)}{\partial \varepsilon}
=
\delta(E_F-\varepsilon_\lambda(k)),
\label{eq:Supp_T0_derivative}
\end{equation}
so Eq.~\eqref{eq:Supp_spin_density_Boltzmann_step1} becomes
\begin{equation}
\resizebox{\columnwidth}{!}{$\displaystyle
\mathbf S
=
-\frac{\hbar e\tau}{2}
\sum_{\lambda=\pm}\lambda
\int_0^\infty \frac{k\,dk}{(2\pi)^2}
\int_0^{2\pi} d\varphi\,
(\hat{\mathbf z}\times \hat{\mathbf k})\,
(\mathbf E\cdot \hat{\mathbf k})\,
v_\lambda(k)\,
\delta(E_F-\varepsilon_\lambda(k)).$}
\label{eq:Supp_spin_density_Boltzmann_step2}
\end{equation}

Using
\begin{equation}
\hat{\mathbf z}\times \hat{\mathbf k}
=
(-\sin\varphi,\cos\varphi),
\qquad
\mathbf E\cdot \hat{\mathbf k}
=
E_x\cos\varphi+E_y\sin\varphi,
\label{eq:Supp_ang_vectors}
\end{equation}
the angular integral is
\begin{equation}
\resizebox{\columnwidth}{!}{$\displaystyle
\begin{aligned}
\int_0^{2\pi} d\varphi\,
(\hat{\mathbf z}\times \hat{\mathbf k})\,
(\mathbf E\cdot \hat{\mathbf k})
&=
\int_0^{2\pi} d\varphi\,
\begin{pmatrix}
-\sin\varphi\\
\cos\varphi
\end{pmatrix}
(E_x\cos\varphi+E_y\sin\varphi)
\\
&=
\begin{pmatrix}
-\displaystyle \int_0^{2\pi} d\varphi\,
\left(E_x\sin\varphi\cos\varphi+E_y\sin^2\varphi\right)
\\[1ex]
\displaystyle \int_0^{2\pi} d\varphi\,
\left(E_x\cos^2\varphi+E_y\sin\varphi\cos\varphi\right)
\end{pmatrix}
\\
&=
\begin{pmatrix}
-\pi E_y\\
\pi E_x
\end{pmatrix}
=
\pi(\hat{\mathbf z}\times \mathbf E).
\end{aligned}$}
\label{eq:Supp_angular_integral}
\end{equation}

Therefore,
\begin{equation}
\begin{aligned}
\mathbf S
&=
-\frac{\hbar e\tau}{2}\,
\frac{\pi}{(2\pi)^2}
\sum_{\lambda=\pm}\lambda\\
&\quad\times
\int_0^\infty k\,dk\,
v_\lambda(k)\,
\delta(E_F-\varepsilon_\lambda(k))
\,(\hat{\mathbf z}\times \mathbf E).
\end{aligned}
\label{eq:Supp_spin_density_Boltzmann_step3}
\end{equation}

To perform the radial integral, use
\begin{equation}
\delta(E_F-\varepsilon_\lambda(k))
=
\frac{\delta(k-k_{F\lambda})}
{\left|\partial_k\varepsilon_\lambda(k)\right|_{k_{F\lambda}}}
=
\frac{\delta(k-k_{F\lambda})}
{\hbar\,|v_\lambda(k_{F\lambda})|}.
\label{eq:Supp_delta_k}
\end{equation}
For the occupied Fermi circles considered here, $v_\lambda(k_{F\lambda})>0$, so
\begingroup\small
\begin{align}
\int_0^\infty k\,dk\,
v_\lambda(k)\,
\delta(E_F-\varepsilon_\lambda(k))
&=
\int_0^\infty k\,dk\,
v_\lambda(k)\,
\frac{\delta(k-k_{F\lambda})}{\hbar v_\lambda(k_{F\lambda})}
\nonumber\\
&=
\frac{k_{F\lambda}}{\hbar}.
\label{eq:Supp_radial_integral}
\end{align}
\endgroup

Substituting this into Eq.~\eqref{eq:Supp_spin_density_Boltzmann_step3},
\begin{equation}
\mathbf S
=
-\frac{e\tau}{8\pi}
\left(
\sum_{\lambda=\pm}\lambda k_{F\lambda}
\right)
(\hat{\mathbf z}\times \mathbf E).
\label{eq:Supp_spin_density_branchsum}
\end{equation}

This is already the general branch-sum expression. If both Rashba branches are occupied,
then the Fermi wavevectors are obtained from
\begin{equation}
E_F
=
\frac{\hbar^2 k_{F\lambda}^2}{2m}
+
\lambda\alpha_R k_{F\lambda},
\label{eq:Supp_Fermi_eqn}
\end{equation}
which gives
\begin{equation}
k_{F\lambda}
=
-\lambda\frac{m\alpha_R}{\hbar^2}
+
\sqrt{
\left(\frac{m\alpha_R}{\hbar^2}\right)^2
+
\frac{2mE_F}{\hbar^2}
}.
\label{eq:Supp_kF_lambda}
\end{equation}
Hence
\begin{equation}
k_{F+}-k_{F-}
=
-\frac{2m\alpha_R}{\hbar^2}.
\label{eq:Supp_kF_difference}
\end{equation}
Inserting Eq.~\eqref{eq:Supp_kF_difference} into Eq.~\eqref{eq:Supp_spin_density_branchsum},
we obtain
\begin{equation}
\mathbf S
=
\frac{e\tau m\alpha_R}{4\pi\hbar^2}\,
(\hat{\mathbf z}\times \mathbf E).
\label{eq:Supp_Edelstein_final_Boltzmann}
\end{equation}
where both Rashba branches are occupied.

For a field along $x$ only, $\mathbf E=E_x\hat{\mathbf x}$, this reduces to
\begin{equation}
S_y
=
\frac{e\tau m\alpha_R}{4\pi\hbar^2}\,E_x,
\qquad
S_x=0.
\label{eq:Supp_Edelstein_Sy_only}
\end{equation}

\subsection{Kubo-formula derivation}

We now derive the same result from linear response theory.

A spatially uniform electric field $E_x(\omega)$ can be represented by a vector potential
\begin{equation}
A_x(\omega)=\frac{E_x(\omega)}{i\omega},
\label{eq:Supp_A_from_E}
\end{equation}
so that the perturbation is
\begin{equation}
H'(\omega)
=
-j_x A_x(\omega),
\qquad
j_x=-e v_x.
\label{eq:Supp_Hprime}
\end{equation}
The induced spin density is
\begin{equation}
S_y(\omega)
=
\Pi^R_{s_y j_x}(\omega)\,A_x(\omega),
\label{eq:Supp_S_from_Pi}
\end{equation}
where
\begin{equation}
\Pi^R_{s_y j_x}(t)
=
-\frac{i}{\hbar}\,\Theta(t)\,
\left\langle
\left[
s_y(t),j_x(0)
\right]
\right\rangle .
\label{eq:Supp_retarded_correlator}
\end{equation}
Thus
\begin{equation}
S_y(\omega)
=
\frac{i}{\omega}\,
\Pi^R_{s_y j_x}(\omega)\,
E_x(\omega),
\label{eq:Supp_S_from_Pi_E}
\end{equation}
and the Edelstein coefficient is
\begin{equation}
\chi_{yx}
\equiv
\frac{S_y}{E_x}
=
\lim_{\omega\to 0}
\frac{i}{\omega}\,
\Pi^R_{s_y j_x}(\omega).
\label{eq:Supp_chi_def}
\end{equation}

Within the standard constant-$\tau$ approximation and neglecting additional disorder
vertex corrections, the dc Kubo-Greenwood expression is
\begin{equation}
\resizebox{\columnwidth}{!}{$\displaystyle
\chi_{yx}
=
-\frac{e}{2\pi}
\int\frac{d^2k}{(2\pi)^2}
\int d\epsilon\,
\left(-\frac{\partial f_0}{\partial\epsilon}\right)
\Tr\!\left[
s_y\,
G^R(\mathbf{k},\epsilon)\,
v_x\,
G^A(\mathbf{k},\epsilon)
\right].$}
\label{eq:Supp_Kubo_dc}
\end{equation}

For the Rashba Hamiltonian, the spin and velocity operators are
\begin{equation}
s_y
=
\frac{\hbar}{2}\sigma_y,
\qquad
v_x
=
\frac{1}{\hbar}\frac{\partial H}{\partial k_x}
=
\frac{\hbar k_x}{m}\sigma_0
+
\frac{\alpha_R}{\hbar}\sigma_y.
\label{eq:Supp_sy_vx_ops}
\end{equation}

It is convenient to write the retarded and advanced Green's functions in terms of the
helicity projectors
\begin{equation}
P_\lambda(\varphi)
=
u_{\lambda,\varphi}u_{\lambda,\varphi}^\dagger
=
\frac{1}{2}
\left[
\sigma_0
+
\lambda\left(
-\sin\varphi\,\sigma_x+\cos\varphi\,\sigma_y
\right)
\right].
\label{eq:Supp_projector}
\end{equation}
Then
\begin{equation}
G^{R/A}(\mathbf{k},\epsilon)
=
\sum_{\lambda=\pm}
\frac{P_\lambda(\varphi)}
{\epsilon-\varepsilon_\lambda(k)\pm i\Gamma},
\qquad
\Gamma=\frac{1}{2\tau}.
\label{eq:Supp_GRGA}
\end{equation}

Substituting Eq.~\eqref{eq:Supp_GRGA} into Eq.~\eqref{eq:Supp_Kubo_dc} yields
\begin{equation}
\resizebox{\columnwidth}{!}{$\displaystyle
\chi_{yx}
=
-\frac{e}{2\pi}
\sum_{\lambda,\lambda'=\pm}
\int\frac{d^2k}{(2\pi)^2}
\int d\epsilon\,
\left(-\frac{\partial f_0}{\partial\epsilon}\right)
\frac{
\Tr\!\left[s_y P_\lambda v_x P_{\lambda'}\right]
}{
(\epsilon-\varepsilon_\lambda+i\Gamma)
(\epsilon-\varepsilon_{\lambda'}-i\Gamma)
}.$}
\label{eq:Supp_Kubo_expand}
\end{equation}

In the weak-scattering dc limit, the intraband terms $\lambda=\lambda'$ scale as
$1/\Gamma\sim\tau$, whereas the interband terms $\lambda\neq\lambda'$ remain finite as
$\Gamma\to 0$. Therefore the leading Drude/Edelstein contribution is obtained by
retaining only $\lambda=\lambda'$:
\begin{equation}
\resizebox{\columnwidth}{!}{$\displaystyle
\chi_{yx}
\simeq
-\frac{e}{2\pi}
\sum_{\lambda=\pm}
\int\frac{d^2k}{(2\pi)^2}
\int d\epsilon\,
\left(-\frac{\partial f_0}{\partial\epsilon}\right)
\frac{
\Tr\!\left[s_y P_\lambda v_x P_\lambda\right]
}{
(\epsilon-\varepsilon_\lambda)^2+\Gamma^2
}.$}
\label{eq:Supp_Kubo_intraband}
\end{equation}

Because $P_\lambda O P_\lambda = \innerproduct{u_{\lambda,\varphi}}{O u_{\lambda,\varphi}}\,P_\lambda$
for any operator $O$, we have
\begin{equation}
\Tr\!\left[s_y P_\lambda v_x P_\lambda\right]
=
\innerproduct{u_{\lambda,\varphi}}{s_y u_{\lambda,\varphi}}\,
\innerproduct{u_{\lambda,\varphi}}{v_x u_{\lambda,\varphi}}.
\label{eq:Supp_trace_factorization}
\end{equation}
Using Eqs.~\eqref{eq:Supp_spin_expectation} and \eqref{eq:Supp_branch_velocity},
\begin{equation}
\begin{aligned}
\innerproduct{u_{\lambda,\varphi}}{s_y u_{\lambda,\varphi}}
&=\frac{\hbar}{2}\lambda\cos\varphi,\\
\innerproduct{u_{\lambda,\varphi}}{v_x u_{\lambda,\varphi}}
&=v_\lambda(k)\cos\varphi.
\end{aligned}
\label{eq:Supp_sy_vx_expectation}
\end{equation}
Therefore,
\begin{equation}
\Tr\!\left[s_y P_\lambda v_x P_\lambda\right]
=
\frac{\hbar}{2}\lambda\,v_\lambda(k)\cos^2\varphi.
\label{eq:Supp_trace_eval}
\end{equation}

In the limit $\Gamma\to 0$,
\begin{equation}
\frac{1}{(\epsilon-\varepsilon_\lambda)^2+\Gamma^2}
\longrightarrow
\frac{\pi}{\Gamma}\,\delta(\epsilon-\varepsilon_\lambda)
=
2\pi\tau\,\delta(\epsilon-\varepsilon_\lambda).
\label{eq:Supp_Lorentzian_to_delta}
\end{equation}
Thus Eq.~\eqref{eq:Supp_Kubo_intraband} becomes
\begin{align}
\chi_{yx}
&=
-e\tau
\sum_{\lambda=\pm}
\int\frac{d^2k}{(2\pi)^2}
\left(
\frac{\hbar}{2}\lambda\,v_\lambda(k)\cos^2\varphi
\right)
\left[
-\frac{\partial f_0(\varepsilon_\lambda)}{\partial\varepsilon}
\right]
\nonumber\\
&=
-\frac{e\tau\hbar}{2}
\sum_{\lambda=\pm}\lambda
\int\frac{d^2k}{(2\pi)^2}
v_\lambda(k)\cos^2\varphi
\left[
-\frac{\partial f_0(\varepsilon_\lambda)}{\partial\varepsilon}
\right].
\label{eq:Supp_Kubo_after_delta}
\end{align}

At $T=0$,
\begin{equation}
-\frac{\partial f_0(\varepsilon_\lambda)}{\partial\varepsilon}
=
\delta(E_F-\varepsilon_\lambda(k)),
\label{eq:Supp_Kubo_T0}
\end{equation}
so
\begin{equation}
\resizebox{\columnwidth}{!}{$\displaystyle
\chi_{yx}
=
-\frac{e\tau\hbar}{2}
\sum_{\lambda=\pm}\lambda
\int_0^\infty \frac{k\,dk}{(2\pi)^2}
\int_0^{2\pi} d\varphi\,
v_\lambda(k)\cos^2\varphi\,
\delta(E_F-\varepsilon_\lambda(k)).$}
\label{eq:Supp_Kubo_polar}
\end{equation}

The angular integral is
\begin{equation}
\int_0^{2\pi} d\varphi\,\cos^2\varphi = \pi,
\label{eq:Supp_cos2_integral}
\end{equation}
and the radial integral is the same as in Eq.~\eqref{eq:Supp_radial_integral},
\begin{equation}
\int_0^\infty k\,dk\,
v_\lambda(k)\,
\delta(E_F-\varepsilon_\lambda(k))
=
\frac{k_{F\lambda}}{\hbar}.
\label{eq:Supp_Kubo_radial}
\end{equation}
Hence
\begin{equation}
\chi_{yx}
=
-\frac{e\tau}{8\pi}
\sum_{\lambda=\pm}\lambda k_{F\lambda}.
\label{eq:Supp_Kubo_branchsum}
\end{equation}
Using Eq.~\eqref{eq:Supp_kF_difference},
\begin{equation}
\chi_{yx}
=
\frac{e\tau m\alpha_R}{4\pi\hbar^2},
\label{eq:Supp_Kubo_chi_final}
\end{equation}
so that
\begin{equation}
S_y
=
\frac{e\tau m\alpha_R}{4\pi\hbar^2}\,E_x
\label{eq:Supp_Kubo_Sy_final}
\end{equation}
and, by rotational symmetry,
\begin{equation}
\mathbf S
=
\frac{e\tau m\alpha_R}{4\pi\hbar^2}\,
(\hat{\mathbf z}\times \mathbf E).
\label{eq:Supp_Kubo_vector_final}
\end{equation}

\subsection{Final result and interpretation}

Both the semiclassical Boltzmann treatment and the Kubo-formula calculation lead to the same result:
\begin{equation}
\mathbf S
=
-\frac{e\tau}{8\pi}
\left(
\sum_{\lambda=\pm}\lambda k_{F\lambda}
\right)
(\hat{\mathbf z}\times \mathbf E).
\label{eq:Supp_Edelstein_general}
\end{equation}
When both Rashba branches are occupied,
\begin{equation}
\sum_{\lambda=\pm}\lambda k_{F\lambda}
=
k_{F+}-k_{F-}
=
-\frac{2m\alpha_R}{\hbar^2},
\label{eq:Supp_branch_difference_again}
\end{equation}
and therefore
\begin{equation}
\mathbf S
=
\frac{e\tau m\alpha_R}{4\pi\hbar^2}\,
(\hat{\mathbf z}\times \mathbf E).
\label{eq:Supp_Edelstein_main_result}
\end{equation}

Equation~\eqref{eq:Supp_Edelstein_main_result} makes the physical origin of the Edelstein
effect transparent. The two helicity branches have opposite spin textures, but they do not
cancel in the nonequilibrium state because the two Fermi radii are different:
$k_{F+}\neq k_{F-}$. The field shifts both Rashba circles in momentum space, and the
resulting imbalance of spin-momentum-locked states produces a net spin density.

\begin{figure*}[htb]
\centering
\includegraphics[width=0.95\linewidth]{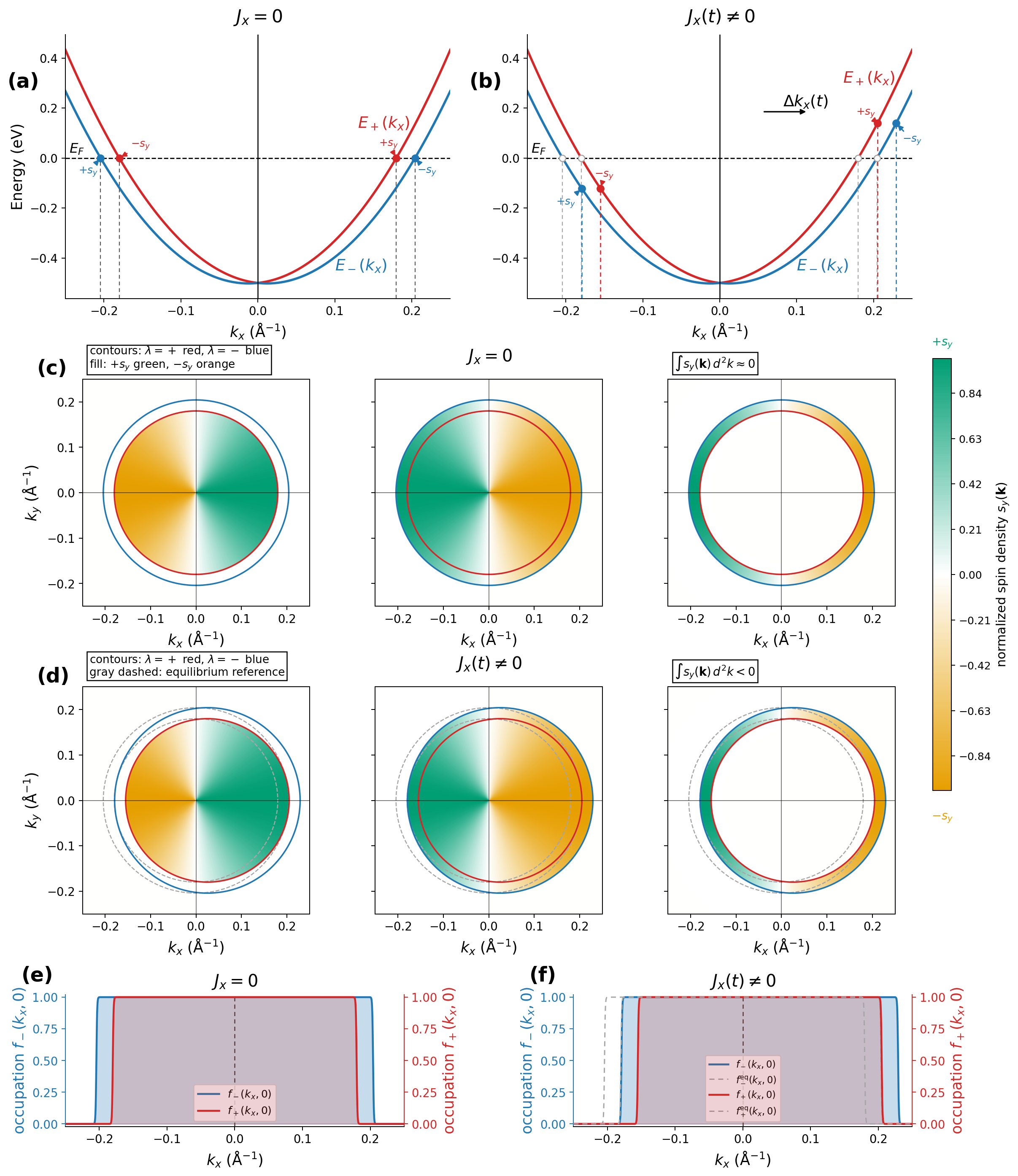}
\caption{
\textbf{Visualization of the Edelstein spin polarization in a Rashba-split surface state.}
\textbf{(a)} Rashba eigenenergies $E_\lambda(k_x)$ along the $k_y=0$ cut for the equilibrium state, $J_x=0$. The dashed horizontal line denotes the Fermi energy $E_F$. The two helicity branches carry opposite $y$-spin at a given $k_x$, and the equilibrium occupations of opposite momenta cancel, yielding zero net spin density.
\textbf{(b)} The same dispersion in the driven nonequilibrium state, $J_x(t)\neq 0$, modeled as a shift of the distribution by $\Delta k_x(t)$. For electrons, $\Delta k_x=-e\tau E_x/\hbar$; therefore the displayed rightward shift, $\Delta k_x>0$, corresponds to $E_x<0$ and conventional current $J_x<0$.
\textbf{(c)} Branch-resolved and total $k$-space spin-density maps at equilibrium, showing the $+$ branch contribution, the $-$ branch contribution, and their sum. The red and blue contours indicate the $\lambda=+$ and $\lambda=-$ Fermi contours, while the color scale indicates the sign of the spin density, with green for $+s_y$ and orange for $-s_y$. The total integrated spin density vanishes in equilibrium.
\textbf{(d)} Branch-resolved and total $k$-space spin-density maps in the driven nonequilibrium state. The gray dashed contours denote the equilibrium Fermi contours, while the solid contours denote the shifted nonequilibrium contours. The branch-resolved contributions no longer cancel; for the displayed rightward shift, the total spin density is negative.
\textbf{(e)} Equilibrium occupations along the $k_y=0$ cut, shown separately for the lower Rashba branch $f_-(k_x,0)$ and the upper Rashba branch $f_+(k_x,0)$.
\textbf{(f)} Nonequilibrium occupations along the $k_y=0$ cut, together with the equilibrium reference. 
}
\label{fig:Supp_Rashba_Edelstein_visualization}
\end{figure*}

\paragraph{Sign convention.}
Throughout these appendices we adopt the standard Rashba convention
\begin{equation}
H
=
\frac{\hbar^2 k^2}{2m}
+
\alpha_R(k_x\sigma_y-k_y\sigma_x),
\label{eq:Supp_standard_Rashba_convention}
\end{equation}
with $\hat{\mathbf z}$ pointing from the crystal toward vacuum. For Au(111), we take $\alpha_R>0$ in this convention. The corresponding Rashba helicities are such that, for $E_F$ above the Rashba crossing, the inner contour ($\lambda=+$) carries a counterclockwise tangential spin texture and the outer contour ($\lambda=-$) a clockwise one when viewed from the vacuum side.

To make the geometric origin of Eq.~\eqref{eq:Supp_Edelstein_main_result} more transparent, Fig.~\ref{fig:Supp_Rashba_Edelstein_visualization} shows the equilibrium contours and their driven displacement. The electric field does not change the Fermi energy, but shifts the distribution in momentum space. Because the two helicity branches have opposite spin textures and different Fermi radii, the shifted branch contributions no longer cancel, leading to a net Edelstein spin polarization. The plotted displacement is toward $+k_x$, so $\Delta k_x>0$ implies $E_x<0$ and, with the convention used here, $S_y<0$.
Fig.~\ref{fig:Supp_Rashba_Edelstein_visualization} makes explicit that the Edelstein effect originates from the nonequilibrium imbalance of the spin-momentum-locked Rashba branches, consistent with the analytical result
$\mathbf S = \frac{e\tau m\alpha_R}{4\pi\hbar^2}(\hat{\mathbf z}\times \mathbf E)$
for the Rashba convention used in this work.

\section{Derivation of the First-Harmonic Boltzmann and Moment Equations}
\label{sec:Supp_firstharmonic}
This section derives the $\Astr$-weighted $j_xs_y$ channel.  Accordingly, the symbol $j_y$ below denotes its transverse side-jump-like current moment.
\paragraph{Projection onto the first angular harmonics.}
The reduced branch-resolved kinetic equation is
\begin{equation}
\partial_t f_\lambda(\varphi,t)
-eE_x(t)v_\lambda\cos\varphi\,\partial_\epsilon f_\lambda^{(0)}
=I_{0,\lambda}[f]+I_{{\rm edge},\lambda}^{(2,0)}[f].
\label{eq:Supp_kinetic_reduced}
\end{equation}
In general, the nonequilibrium part of the branch-resolved distribution on the Fermi contour can be expanded in a complete angular Fourier series,
\begin{equation}
\begin{aligned}
\delta f_\lambda(k,\varphi,t)
&=-\partial_\epsilon f_\lambda^{(0)}\\
&\quad\times
\sum_{n=1}^{\infty}
\Big[
a_{\lambda n}(t)\cos(n\varphi)
+
b_{\lambda n}(t)\sin(n\varphi)
\Big].
\end{aligned}
\label{eq:Supp_general_harmonic_series}
\end{equation}
Equivalently,
\begin{equation}
\begin{aligned}
f_\lambda(k,\varphi,t)
&=f_\lambda^{(0)}-\partial_\epsilon f_\lambda^{(0)}\\
&\quad\times
\sum_{n=1}^{\infty}
\Big[
a_{\lambda n}(t)\cos(n\varphi)
+
b_{\lambda n}(t)\sin(n\varphi)
\Big].
\end{aligned}
\label{eq:Supp_full_harmonic_series}
\end{equation}
Here the coefficients $a_{\lambda n}(t)$ and $b_{\lambda n}(t)$ are the angular harmonics of the nonequilibrium distortion of the Fermi contour. In the minimal theory used in the main text and these appendices, we truncate this expansion to the first harmonic, $n=1$, because the longitudinal drift induced by $E_x$ appears first in the $\cos\varphi$ channel, while the leading transverse response appears in the $\sin\varphi$ channel. Higher harmonics can be generated microscopically, but they are neglected in the present lowest-order treatment.

Retaining only the first harmonic, we write
\begin{align}
f_\lambda(k,\varphi,t)&=f_\lambda^{(0)}(\epsilon_{\lambda k})
+\delta f_\lambda(k,\varphi,t),
\\
\delta f_\lambda
&=
-\partial_\epsilon f_\lambda^{(0)}
\Big[a_\lambda(t)\cos\varphi+b_\lambda(t)\sin\varphi\Big].
\nn
\label{eq:deltaf_expand}
\end{align}

Here $a_\lambda$ is the $\cos\varphi$ harmonic that carries the longitudinal drift and $b_\lambda$ is the $\sin\varphi$ harmonic that carries the transverse response. Since $f_\lambda^{(0)}$ depends only on $\epsilon_{\lambda k}$, one has
\begin{equation}
\nabla_{\bm k}f_\lambda^{(0)}
=
\partial_\epsilon f_\lambda^{(0)}\,\nabla_{\bm k}\epsilon_{\lambda k}
=
\hbar v_\lambda \hat{\bm k}\,\partial_\epsilon f_\lambda^{(0)},
\label{eq:gradf0}
\end{equation}
where $v_\lambda=\frac{1}{\hbar}\partial_k\epsilon_{\lambda k}$ and $\hat{\bm k}=(\cos\varphi,\sin\varphi)$.
Using the standard semiclassical acceleration $\dot{\bm k}=-(e/\hbar)\bm E$, the electric-field drive contributes
\begin{align}
-\frac{e}{\hbar}E_x(t)\,\partial_{k_x}f_\lambda^{(0)}
&=
-\frac{e}{\hbar}E_x(t)\cos\varphi\,\partial_k f_\lambda^{(0)}
\nonumber\\
&=
-eE_x(t)v_\lambda\cos\varphi\,\partial_\epsilon f_\lambda^{(0)}.
\label{eq:drive_term}
\end{align}
Likewise,
\begin{equation}
\partial_t \delta f_\lambda
=
-\partial_\epsilon f_\lambda^{(0)}
\Big[\dot a_\lambda(t)\cos\varphi+\dot b_\lambda(t)\sin\varphi\Big].
\label{eq:time_deriv_deltaf}
\end{equation}
In the reduced intraband treatment used here, the commutator term $(i/\hbar)[H_R,\hat f_{\bm k}]$ does not contribute to the diagonal helicity distribution $f_\lambda$ once inter-helicity coherences are neglected. The ordinary collision term and the edge conversion term are
\begin{equation}
I_{0,\lambda}[f]
=
\partial_\epsilon f_\lambda^{(0)}
\left[
\frac{a_\lambda}{\tau}\cos\varphi
+\frac{b_\lambda}{\tau_y}\sin\varphi
\right],
\label{eq:I0_repeat}
\end{equation}
and
\begin{equation}
\begin{aligned}
I_{{\rm edge},\lambda}^{(2,0)}[f]
&=-\partial_\epsilon f_\lambda^{(0)}\Astr
\left[\eta_\lambda^{({\rm ev})}\Ffab+\eta_\lambda^{(3)}\Mbar\right]\\
&\quad\times j_x(t)s_y(t)\sin\varphi.
\end{aligned}
\label{eq:Iext_repeat}
\end{equation}
Substituting Eqs.~(\ref{eq:drive_term})--(\ref{eq:Iext_repeat}) into Eq.~(\ref{eq:Supp_kinetic_reduced}), dividing by $-\partial_\epsilon f_\lambda^{(0)}$, and keeping only the $\cos\varphi$ and $\sin\varphi$ harmonics gives
\begingroup\small
\begin{align}
&\Bigg[
\dot a_\lambda+\frac{a_\lambda}{\tau}+ev_\lambda E_x(t)
\Bigg]\cos\varphi
\nonumber\\
&
+
\Bigg[
\dot b_\lambda+\frac{b_\lambda}{\tau_y}
-\Astr\left(\eta_\lambda^{({\rm ev})}\Ffab+
\eta_\lambda^{(3)}\Mbar\right)j_x(t)s_y(t)
\Bigg]\sin\varphi
=0.
\label{eq:harmonic_balance}
\end{align}
\endgroup
Now project onto the first angular harmonics using
\begin{equation}
\langle X\rangle_\varphi
\equiv
\int_0^{2\pi}\frac{d\varphi}{2\pi}\,X(\varphi),
\label{eq:angle_avg_def}
\end{equation}
together with the orthogonality relations
\begin{equation}
\langle \cos^2\varphi\rangle_\varphi
=
\langle \sin^2\varphi\rangle_\varphi
=
\frac12,
\qquad
\langle \sin\varphi\cos\varphi\rangle_\varphi=0.
\label{eq:trig_orthogonality}
\end{equation}
Multiplying Eq.~(\ref{eq:harmonic_balance}) by $2\cos\varphi$ and averaging over $\varphi$ gives
\begin{equation}
\dot a_\lambda+\frac{a_\lambda}{\tau}=-ev_\lambda E_x(t),
\label{eq:alambda}
\end{equation}
while multiplying by $2\sin\varphi$ and averaging gives
\begin{equation}
\dot b_\lambda+\frac{b_\lambda}{\tau_y}
=
\Astr\left(\eta_\lambda^{({\rm ev})}\Ffab+
\eta_\lambda^{(3)}\Mbar\right)j_x(t)s_y(t).
\label{eq:blambda}
\end{equation}
Thus Eq.~(\ref{eq:Supp_kinetic_reduced}) reduces to the pair of first-harmonic kinetic equations
\begin{subequations}
\begin{align}
\dot a_\lambda+\frac{a_\lambda}{\tau}&=-ev_\lambda E_x(t),
\label{eq:alambda_repeat}\\
\dot b_\lambda+\frac{b_\lambda}{\tau_y}&=\Astr
\left(\eta_\lambda^{({\rm ev})}\Ffab+
\eta_\lambda^{(3)}\Mbar\right)j_x(t)s_y(t),
\label{eq:blambda_repeat}
\end{align}
\label{eq:abproj}
\end{subequations}
with $v_\lambda=(1/\hbar)\partial_k\epsilon_{\lambda k}$.

\paragraph{Current and spin moments.}
At low temperature, the radial $k$ integration collapses onto the Fermi circles, so that for any smooth angular function $X_\lambda(\varphi)$,
\begin{equation}
\int \frac{d^2k}{(2\pi)^2}
\Bigl[-\partial_\epsilon f_\lambda^{(0)}\Bigr]
\,X_\lambda(\varphi)
=
\mathcal N_\lambda
\int_0^{2\pi}\frac{d\varphi}{2\pi}\,
X_\lambda(\varphi),
\label{eq:fs_projection_identity}
\end{equation}
where
\begin{equation}
\mathcal N_\lambda
=
\frac{k_{F\lambda}}{2\pi\hbar v_\lambda}
\label{eq:dos_branch}
\end{equation}
is the density of states on the helicity branch at the Fermi energy.

The longitudinal current is
\begin{align}
j_x
&=
-e\sum_{\lambda=\pm}
\int\frac{d^2k}{(2\pi)^2}
\,v_\lambda\cos\varphi\,\delta f_\lambda
\nonumber\\
&=
-e\sum_{\lambda=\pm}
\mathcal N_\lambda v_\lambda
\int_0^{2\pi}\frac{d\varphi}{2\pi}
\cos\varphi
\Big[a_\lambda\cos\varphi+b_\lambda\sin\varphi\Big]
\nonumber\\
&=
-e\sum_{\lambda=\pm}
\mathcal N_\lambda v_\lambda
\left[
a_\lambda\langle \cos^2\varphi\rangle_\varphi
+b_\lambda\langle \sin\varphi\cos\varphi\rangle_\varphi
\right]
\nonumber\\
&=
-\frac{e}{2}\sum_{\lambda=\pm}
\mathcal N_\lambda v_\lambda a_\lambda.
\label{eq:jxmoment_derivation}
\end{align}
Similarly, the transverse current is
\begin{align}
j_y
&=
-e\sum_{\lambda=\pm}
\int\frac{d^2k}{(2\pi)^2}
\,v_\lambda\sin\varphi\,\delta f_\lambda
\nonumber\\
&=
-e\sum_{\lambda=\pm}
\mathcal N_\lambda v_\lambda
\int_0^{2\pi}\frac{d\varphi}{2\pi}
\sin\varphi
\Big[a_\lambda\cos\varphi+b_\lambda\sin\varphi\Big]
\nonumber\\
&=
-e\sum_{\lambda=\pm}
\mathcal N_\lambda v_\lambda
\left[
a_\lambda\langle \sin\varphi\cos\varphi\rangle_\varphi
+b_\lambda\langle \sin^2\varphi\rangle_\varphi
\right]
\nonumber\\
&=
-\frac{e}{2}\sum_{\lambda=\pm}
\mathcal N_\lambda v_\lambda b_\lambda.
\label{eq:jymoment_derivation}
\end{align}

For the spin moment, the Rashba helicity eigenstate has
\begin{equation}
\ev{\sigma_y}_{\lambda,\varphi}=\lambda\cos\varphi.
\label{eq:sigma_y_expect}
\end{equation}
Defining the reduced scalar spin moment used in the main text as
\begin{equation}
s_y\equiv
\frac12
\sum_{\lambda=\pm}
\int\frac{d^2k}{(2\pi)^2}\,
\ev{\sigma_y}_{\lambda,\varphi}\,
\delta f_\lambda,
\label{eq:sy_def_reduced}
\end{equation}
one obtains
\begin{align}
s_y
&=
\frac12
\sum_{\lambda=\pm}
\lambda\,\mathcal N_\lambda
\int_0^{2\pi}\frac{d\varphi}{2\pi}
\cos\varphi
\Big[a_\lambda\cos\varphi+b_\lambda\sin\varphi\Big]
\nonumber\\
&=
\frac12
\sum_{\lambda=\pm}
\lambda\,\mathcal N_\lambda
\left[
a_\lambda\langle \cos^2\varphi\rangle_\varphi
+b_\lambda\langle \sin\varphi\cos\varphi\rangle_\varphi
\right]
\nonumber\\
&=
\frac14
\sum_{\lambda=\pm}
\lambda\,\mathcal N_\lambda a_\lambda.
\label{eq:symoment_derivation}
\end{align}
Collecting these results,
\begin{subequations}
\begin{align}
j_x&=-\frac{e}{2}\sum_{\lambda=\pm}\mathcal N_\lambda v_\lambda a_\lambda,
\label{eq:jxmoment}\\
j_y&=-\frac{e}{2}\sum_{\lambda=\pm}\mathcal N_\lambda v_\lambda b_\lambda,
\label{eq:jymoment}\\
s_y&=\frac{1}{4}\sum_{\lambda=\pm}\lambda\,\mathcal N_\lambda a_\lambda,
\label{eq:symoment}
\end{align}
\label{eq:kinetic_moments}
\end{subequations}
where $\mathcal N_\lambda=k_{F\lambda}/(2\pi\hbar v_\lambda)$.

\paragraph{Derivation of the Drude equation.}
Multiply Eq.~(\ref{eq:alambda}) by $-e\mathcal N_\lambda v_\lambda/2$ and sum over $\lambda$:
\begin{align}
-\frac{e}{2}\sum_\lambda \mathcal N_\lambda v_\lambda \dot a_\lambda
-
\frac{e}{2\tau}\sum_\lambda \mathcal N_\lambda v_\lambda a_\lambda
&=
\frac{e^2}{2}\sum_\lambda \mathcal N_\lambda v_\lambda^2 E_x(t).
\label{eq:drude_sum_step}
\end{align}
Using Eq.~(\ref{eq:jxmoment}), the left-hand side becomes
\begin{equation}
\dot j_x+\frac{j_x}{\tau}.
\label{eq:drude_lhs}
\end{equation}
Thus
\begin{equation}
\dot j_x+\frac{j_x}{\tau}
=
\left[
\frac{e^2}{2}\sum_\lambda \mathcal N_\lambda v_\lambda^2
\right]E_x(t).
\label{eq:drude_general}
\end{equation}
In the simplified Drude reduction used in the main text, the branch-summed prefactor is written as the usual Drude weight $ne^2/m$, so that
\begin{equation}
\dot j_x+\frac{j_x}{\tau}=\frac{ne^2}{m}E_x(t).
\label{eq:jx_eqn}
\end{equation}

\paragraph{Derivation of the transverse current equation.}
Likewise, multiply Eq.~(\ref{eq:blambda}) by $-e\mathcal N_\lambda v_\lambda/2$ and sum over $\lambda$:
\begin{equation}
\resizebox{\columnwidth}{!}{$\displaystyle
\begin{aligned}
-\frac{e}{2}\sum_\lambda \mathcal N_\lambda v_\lambda \dot b_\lambda
-
\frac{e}{2\tau_y}\sum_\lambda \mathcal N_\lambda v_\lambda b_\lambda
&=
-\frac{e}{2}\sum_\lambda
\mathcal N_\lambda v_\lambda
\Astr\left(\eta_\lambda^{({\rm ev})}\Ffab+
\eta_\lambda^{(3)}\Mbar\right)j_x(t)s_y(t).
\end{aligned}$}
\label{eq:jy_sum_step}
\end{equation}
Using Eq.~(\ref{eq:jymoment}), the left-hand side becomes
\begin{equation}
\dot j_y+\frac{j_y}{\tau_y},
\label{eq:jy_lhs}
\end{equation}
while the right-hand side can be written as
\begin{equation}
\Astr\left(\Gamma_{\rm ev}\Ffab+\Gamma_3\Mbar\right)j_x(t)s_y(t),
\label{eq:Gamma_source_step}
\end{equation}
where
\begin{equation}
\Gamma_{\rm ev}\equiv -\frac{e}{2}\sum_\lambda
\mathcal N_\lambda v_\lambda\eta_\lambda^{({\rm ev})},
\qquad
\Gamma_3\equiv -\frac{e}{2}\sum_\lambda
\mathcal N_\lambda v_\lambda\eta_\lambda^{(3)}.
\label{eq:Gamma_def_step}
\end{equation}
Hence
\begin{equation}
\dot j_y+\frac{j_y}{\tau_y}=\Astr
\left(\Gamma_{\rm ev}\Ffab+\Gamma_3\Mbar\right)j_xs_y.
\label{eq:jy_eqn}
\end{equation}

\paragraph{Derivation of the Edelstein equation.}
Finally, take the spin moment of the kinetic equation. Using Eq.~(\ref{eq:symoment}),
\begin{equation}
\dot s_y
=
\frac14\sum_{\lambda=\pm}\lambda\,\mathcal N_\lambda \dot a_\lambda.
\label{eq:sy_dot_start}
\end{equation}
If one multiplies Eq.~(\ref{eq:alambda}) by $\lambda\mathcal N_\lambda/4$ and sums over $\lambda$, one obtains
\begin{equation}
\dot s_y+\frac{s_y}{\tau}
=
-\frac{eE_x(t)}{4}\sum_{\lambda=\pm}\lambda\,\mathcal N_\lambda v_\lambda.
\label{eq:sy_field_driven}
\end{equation}
At the reduced-moment level used here, it is convenient to allow for a distinct spin-relaxation time $\tau_s$ and to rewrite the source term as a charge-to-spin conversion proportional to the longitudinal current. We therefore introduce the Edelstein conversion coefficient $\kappa$ through
\begin{equation}
\dot s_y+\frac{s_y}{\tau_s}=\kappa j_x,
\label{eq:sy_eqn}
\end{equation}
which is the form quoted in the main text. Equivalently, one may regard $\kappa$ as chosen so that the reduced equation reproduces the desired linear Edelstein susceptibility. In this way the detailed microscopic spin dynamics are absorbed into the phenomenological parameters $\tau_s$ and $\kappa$.

With Eqs.~(\ref{eq:jx_eqn}), (\ref{eq:jy_eqn}), and (\ref{eq:sy_eqn}), the coupled current-spin dynamics are fully reduced to the three scalar moments $j_x$, $j_y$, and $s_y$.

\section{Frequency dependence, magnitude, and optical magnetic field}
\label{sec:Supp_frequency_magnitude}

For $E_x^{\rm loc}(t)=\Re[E_x^{\rm loc}e^{-i\omega t}]$, the linear solutions are
\begin{equation}
j_x^{(1)}(\omega)=\sigma_D(\omega)E_x^{\rm loc},
\qquad
s_y^{(1)}(\omega)=\chi_{EE}(\omega)E_x^{\rm loc},
\label{eq:Supp_linear_frequency}
\end{equation}
with
\begin{equation}
\sigma_D(\omega)=\frac{\sigma_0}{1-i\omega\tau},
\qquad
\chi_{EE}(\omega)=\frac{\kappa\tau_s\sigma_D(\omega)}{1-i\omega\tau_s}.
\label{eq:Supp_sigma_chi_frequency}
\end{equation}
Time averaging Eq.~\eqref{eq:jy_eqn} gives
\begin{equation}
J_y^{(0)}=
\frac{\tau_y\Astr}{2}
\left(\Gamma_{\rm ev}\Ffab+\Gamma_3\Mbar\right)
\Re\!\left[\chi_{EE}(\omega)\sigma_D^*(\omega)\right]
|E_x^{\rm loc}|^2.
\label{eq:Supp_Jy_frequency_general}
\end{equation}
For real $\kappa$ this becomes
\begin{equation}
J_y^{(0)}(\omega)=
\frac{\tau_y\Astr(\Gamma_{\rm ev}\Ffab+\Gamma_3\Mbar)
\kappa\tau_s\sigma_0^2}
{2(1+\omega^2\tau^2)(1+\omega^2\tau_s^2)}
|F_x(\omega,\theta)E_x^{\rm inc}|^2,
\label{eq:Supp_Jy_frequency_explicit}
\end{equation}
where $E_x^{\rm loc}=F_xE_x^{\rm inc}$. The rational factors are the electronic Drude-Edelstein rolloff. The SPP resonances enter through the sample-specific local-field enhancement $|F_x|^2$. A universal frequency plot that omits $F_x$ would therefore not represent the grating experiment; Eq.~\eqref{eq:Supp_Jy_frequency_explicit} is the frequency-resolved prediction to be combined with RCWA, FDTD, or measured absorbance.

The experiment of Ref.~\onlinecite{Plumitallo2023} reported a peak current $I_y^{\rm pk}\simeq300~\mathrm{pA}$ for an illuminated width $w_x\simeq150~\mu\mathrm{m}$, incident power $P=85~\mathrm{mW}$, and continuous-film thickness $t=100~\mathrm{nm}$. These values correspond to
\begin{align}
J_y^{\rm pk}&=I_y^{\rm pk}/w_x\simeq2.0~\mu\mathrm{A\,m^{-1}},
\label{eq:Supp_sheet_current_estimate}\\
j_{y,3D}^{\rm pk}&=I_y^{\rm pk}/(w_xt)\simeq20~\mathrm{A\,m^{-2}},
\label{eq:Supp_volume_current_estimate}\\
I_y^{\rm pk}/P&\simeq3.5~\mathrm{nA\,W^{-1}}.
\label{eq:Supp_responsivity_estimate}
\end{align}
The theory does not determine $\Gamma_{\rm ev}\Ffab+\Gamma_3\Mbar$ without measured sidewall morphology, known deposition geometry, an atomistic edge model, and, for the crystalline term, an orientation distribution. The observed value is therefore used as a benchmark, not presented as a parameter-free prediction.

It is nevertheless useful to compare this benchmark with the longitudinal optical current available for rectification. Treating the reported $150\times300~\mu\mathrm{m}^2$ spot dimensions as the full axes of an ellipse gives
\begin{equation}
\begin{aligned}
I_{\rm inc}&\simeq
\frac{85~\mathrm{mW}}{\pi(75~\mu\mathrm{m})(150~\mu\mathrm{m})}
\simeq2.4\times10^6~\mathrm{W\,m^{-2}},\\
E_{\rm inc}&=\sqrt{\frac{2I_{\rm inc}}{c\epsilon_0}}
\simeq4.3\times10^4~\mathrm{V\,m^{-1}}.
\end{aligned}
\label{eq:Supp_incident_scale}
\end{equation}
Near $808~\mathrm{nm}$ the tabulated optical permittivity of Au is approximately $\epsilon_{\rm Au}\simeq-24+1.5i$~\cite{Johnson1972}. The associated conductivity magnitude is
\begin{equation}
|\sigma_{\rm Au}|=
\left|-i\omega\epsilon_0(\epsilon_{\rm Au}-1)\right|
\simeq5\times10^5~\mathrm{S\,m^{-1}}.
\label{eq:Supp_optical_conductivity}
\end{equation}
For a $100~\mathrm{nm}$ film, the incident-field longitudinal sheet-current scale is therefore
\begin{equation}
K_x^{\rm ac}\sim|\sigma_{\rm Au}|tE_{\rm inc}
\simeq2\times10^3~\mathrm{A\,m^{-1}},
\qquad
\frac{J_y^{\rm pk}}{K_x^{\rm ac}}\sim10^{-9}.
\label{eq:Supp_conversion_ratio}
\end{equation}
SPP enhancement increases the local longitudinal current and lowers the required ratio further. Thus only a weak second-order conversion is required to reach the observed current scale. This comparison is an order-of-magnitude consistency check rather than a microscopic calculation of $\Lamy$, $\lambda_\eta$, or any static-channel coefficient.

Here $K_x^{\rm ac}$ is an oscillatory sheet-current amplitude constructed from the complex optical conductivity at frequency $\omega$, not a dc longitudinal current and not the ordinary dc conductivity.  Consequently, $J_y^{\rm pk}/K_x^{\rm ac}$ is an available-current scale rather than a device efficiency or Hall angle.  It should not be confused with the ratio of the dimensionful voltage responsivities quoted in the two experiments.  Those reported values, termed efficiencies in Refs.~\onlinecite{Plumitallo2023,Moroshkin2021}, are
\begin{equation}
\begin{aligned}
\mathcal R_y&=1.25\times10^{-9}~\frac{\mathrm V}{\mathrm{W\,cm^{-2}}},\\
\mathcal R_x&=12\times10^{-9}~\frac{\mathrm V}{\mathrm{W\,cm^{-2}}},\\
\frac{\mathcal R_y}{\mathcal R_x}&\simeq0.104.
\end{aligned}
\label{eq:Supp_voltage_responsivity_ratio}
\end{equation}
Thus the reported transverse voltage responsivity is about $9.6$ times smaller than the longitudinal value, not nine orders of magnitude smaller.  Because the two responsivities were obtained in different device configurations and include sample resistance, contact geometry, illuminated area, and local-field enhancement, their ratio is a device-level comparison and does not define the microscopic conversion coefficient in Eq.~\eqref{eq:Jy_final}.

For the optical magnetic field, $B\simeq E/c$. A representative Au surface-state velocity $v_F\sim10^6~\mathrm{m\,s^{-1}}$ gives a Lorentz-to-electric force ratio $v_F/c\sim3\times10^{-3}$. Over a mean free path $\ell\sim10~\mathrm{nm}$, the Zeeman-to-electric work ratio is
\begin{equation}
\frac{\mu_BB}{eE\ell}\simeq\frac{\mu_B}{ec\ell}\sim2\times10^{-5}.
\label{eq:Supp_magnetic_ratio}
\end{equation}
For $p$ polarization the magnetic field points predominantly along $y$ and is even under the $xz$ mirror. It cannot supply the mirror-odd factor required for $J_y$. Magnetic corrections are thus parametrically smaller and symmetry-irrelevant to the leading term.

\section{Optional discrimination of static-bias channels}
\label{sec:Supp_tests}

The dynamical threshold and directional-field tests were stated in Appendix~\ref{sec:Supp_QNM}. If the optional static terms in Eq.~\eqref{eq:Supp_Lambda_static} are studied independently, the following sign and null tests can separate them without presuming that any fabrication step was erroneous:
\begin{enumerate}
\item Fabricate otherwise identical devices with the transverse evaporation azimuth reversed. This sends $\Ffab\to-\Ffab$ and reverses the fabrication term without changing the lithographic pattern.
\item Repeat the deposition at normal incidence or with continuous substrate rotation. The average $\Ffab$ and hence the proposed fabrication term should be strongly suppressed.
\item Image the sidewalls by tilted SEM or FIB cross-sectional SEM viewed from both $+y$ and $-y$ to determine whether a systematic transverse texture exists.
\item Mirror reverse the binary profile along $x$ while preserving the deposition geometry. This changes $\Astr\to-\Astr$ and reverses the coupled fabrication and crystalline terms.
\item On a single Au(111) domain, rotate the grating azimuth. The crystalline term follows $\sin3\phi_c$, vanishes at mirror alignments, and reverses under $\phi_c\to-\phi_c$. In a polycrystalline sample, electron-backscatter diffraction or orientation-resolved microscopy can test whether $\Mbar$ is nonzero.
\item Measure the transverse current versus groove-to-contact and groove-to-incidence-plane misalignment. A simple projection $J_y^{\rm meas}\simeq J_x^{\rm OR}\sin\delta$ follows the misalignment angle and must be separated from a side-jump contribution.
\item If a second structural asymmetry $\mathcal A_y$ is deliberately introduced along $y$, decompose
\begin{equation}
J_y^{(0)}=\Astr\left(\Gamma_{\rm ev}\Ffab+
\Gamma_3\Mbar+\Gamma_g\mathcal A_y\right)
|E_x^{\rm loc}|^2+\cdots .
\label{eq:Supp_y_asym_decomposition}
\end{equation}
Reversing the deposition azimuth, crystal azimuth, and deliberate $y$ geometry changes the three terms separately.
\end{enumerate}
Step rounding, surface adsorbates, or surface chemistry can change edge coefficients without producing the same shift of the SPP resonance. These optional tests distinguish static symmetry fields from a dynamical directional-SPP order parameter.

\section{Structural handedness from the first two \emph{spatial} harmonics of the grating profile}
\label{sec:structural_handedness_detailed}

In this section we make precise what is meant by the statement that ``the lowest inversion-odd translationally invariant quantity must involve at least two harmonics.'' The harmonics in question are \emph{not} optical harmonics and are \emph{not} harmonics of $\chi^{(2)}$. They are the first two \emph{spatial Fourier harmonics} of the periodic grating profile $h(x)$, or equivalently of its edge comb $\partial_x h(x)$. The optical rectification (OR) remains a second-order effect in the optical field, but the coefficient multiplying $|E_x|^2$ is controlled by an inversion-odd structural invariant built from the spatial Fourier components of the grating.

\subsection{Periodic profile and its Fourier harmonics}

Let $h(x)$ denote the periodic terrace profile with period $D$,
\begin{equation}
h(x+D)=h(x),
\end{equation}
and write its Fourier expansion as
\begin{equation}
h(x)=\sum_{m\in\mathbb Z} h_m e^{iG_m x},
\qquad
G_m=\frac{2\pi m}{D}.
\label{eq:hand_hx_fourier}
\end{equation}
For a real profile, the coefficients satisfy
\begin{equation}
h_{-m}=h_m^*.
\label{eq:hand_reality}
\end{equation}
Using the periodic Heaviside construction and edge-rounding convention of
Eqs.~\eqref{eq:Supp_binary_indicator_PM}--\eqref{eq:Supp_dhdx_step_PM},
\begin{equation}
\partial_x h(x)=\frac{H_1}{D}\sum_{m\in\mathbb Z}F_w(G_m)S_m e^{iG_m x},
\label{eq:hand_dhdx_fourier}
\end{equation}
so that for $m\neq 0$,
\begin{equation}
h_m=\frac{H_1}{iG_mD}F_w(G_m)S_m.
\label{eq:hand_hm_from_Sm}
\end{equation}
Thus the relevant ``first'' and ``second'' harmonics are simply
\begin{equation}
h_1 \quad\text{and}\quad h_2,
\end{equation}
the $m=1$ and $m=2$ spatial Fourier coefficients of the grating profile.

\subsection{Why a single harmonic cannot encode handedness}

A single Fourier harmonic does not define an origin-independent left-right handedness. To see this, consider the effect of translating the origin,
\begin{equation}
x\rightarrow x+x_0.
\end{equation}
Equation~\eqref{eq:hand_hx_fourier} becomes
\begin{equation}
h(x)=\sum_m h_m e^{iG_m(x+x_0)}
=\sum_m \underbrace{\bigl(e^{-iG_mx_0}h_m\bigr)}_{h_m'}e^{iG_mx},
\end{equation}
so the Fourier coefficients transform as
\begin{equation}
h_m\rightarrow h_m'=e^{-iG_mx_0}h_m.
\label{eq:hand_translation_hm}
\end{equation}
Therefore the phase of a single harmonic is not by itself physical, because it changes when the origin is shifted.

This is especially transparent for a one-harmonic profile:
\begin{equation}
h(x)=h_1e^{iG_1x}+h_1^*e^{-iG_1x}
=2|h_1|\cos(G_1x+\phi_1),
\end{equation}
where $h_1=|h_1|e^{i\phi_1}$. The phase $\phi_1$ can always be removed by shifting the origin. Hence a pure sinusoid is always mirror-symmetric about some point in the unit cell and cannot encode a genuine structural handedness.

\subsection{General criterion for a translationally invariant structural scalar}

Consider a general monomial built from the Fourier amplitudes,
\begin{equation}
M=\prod_m h_m^{p_m}(h_m^*)^{q_m},
\qquad
p_m,q_m\in \mathbb N_0.
\label{eq:hand_general_mon}
\end{equation}
Under translation, Eq.~\eqref{eq:hand_translation_hm} gives
\begin{equation}
M\rightarrow M'
=
\exp\!\left[-ix_0\sum_m G_m(p_m-q_m)\right]M.
\end{equation}
Therefore $M$ is translationally invariant if and only if
\begin{equation}
\sum_m m(p_m-q_m)=0.
\label{eq:hand_trans_inv_condition}
\end{equation}
This is the wave-vector balance condition.

Now consider inversion. For a real profile, mirror inversion acts as
\begin{equation}
h(x)\rightarrow h(-x),
\end{equation}
which implies
\begin{equation}
h_m\rightarrow h_m^*.
\label{eq:hand_inversion_hm}
\end{equation}
Hence any translationally invariant monomial $M$ transforms under inversion as
\begin{equation}
M\rightarrow M^*.
\end{equation}
Its inversion-even part is $\Re M$, while its inversion-odd part is $\Im M$.

We may now classify the lowest-degree possibilities.

At degree one there is no translationally invariant object, because $h_m\rightarrow e^{-iG_mx_0}h_m$.

At degree two, the only translationally invariant combinations are of the form
\begin{equation}
h_mh_m^*=|h_m|^2,
\end{equation}
because the balance condition requires equal total wave number on the holomorphic and antiholomorphic sides. Such quantities are manifestly real and therefore inversion-even. The candidate $h_1h_2^*$ is \emph{not} translationally invariant, because
\begin{equation}
h_1h_2^*\rightarrow e^{-i(G_1-G_2)x_0}h_1h_2^*
=e^{+iG_1x_0}h_1h_2^*.
\end{equation}

Thus the lowest-degree translationally invariant \emph{and} inversion-odd quantity must occur at degree three. The smallest solution of the balance condition is
\begin{equation}
2G_1-G_2=0,
\end{equation}
which gives the cubic monomial
\begin{equation}
M_{\min}=h_1^2h_2^*.
\end{equation}
Its inversion-odd part is therefore
\begin{equation}
\mathcal A_{\mathrm{str}}
\equiv
\Im\!\left[h_1^2h_2^*\right].
\label{eq:hand_Astr_def}
\end{equation}
This is what is meant by the statement that the ``lowest inversion-odd translationally invariant quantity must involve at least two harmonics.'' The relevant harmonics are the first two \emph{spatial} harmonics of the grating profile.

\subsection{Phase interpretation: why $\mathcal A_{\mathrm{str}}$ measures handedness}

Write
\begin{equation}
h_1=|h_1|e^{i\phi_1},
\qquad
h_2=|h_2|e^{i\phi_2}.
\label{eq:hand_h1h2_phase}
\end{equation}
Then
\begin{equation}
h_1^2h_2^*=|h_1|^2|h_2|e^{i(2\phi_1-\phi_2)},
\end{equation}
and therefore
\begin{equation}
\mathcal A_{\mathrm{str}}
=
|h_1|^2|h_2|\sin(2\phi_1-\phi_2).
\label{eq:hand_Astr_phase}
\end{equation}
The combination $2\phi_1-\phi_2$ is translationally invariant. Indeed, under
$\phi_m\rightarrow \phi_m-G_mx_0$ one has
\begin{equation}
2\phi_1-\phi_2
\rightarrow
2(\phi_1-G_1x_0)-(\phi_2-G_2x_0)
=
2\phi_1-\phi_2,
\end{equation}
because $G_2=2G_1$.

Equation~\eqref{eq:hand_Astr_phase} shows that $\mathcal A_{\mathrm{str}}$ measures the translation-invariant phase offset of the second spatial harmonic relative to the first-harmonic backbone. If the unit cell is inversion symmetric, one can choose the origin so that all $h_m$ are real, implying
\begin{equation}
2\phi_1-\phi_2=0\ \text{or}\ \pi
\qquad\Longrightarrow\qquad
\mathcal A_{\mathrm{str}}=0.
\end{equation}
By contrast, if the smaller-scale wall structure is shifted to one side or the other relative to the coarse first-harmonic profile, then $2\phi_1-\phi_2$ is nontrivial and $\mathcal A_{\mathrm{str}}$ acquires a sign. This sign is the one-dimensional analog of handedness.

\subsection{Binary cell with $A_1=D/2$: exact derivation in terms of $A_2$, $B_1$, and $B_2$}

We now specialize to the experimentally relevant binary unit cell
\begin{equation}
D=A_1+B_1+A_2+B_2,
\qquad
A_1=\frac{D}{2},
\label{eq:hand_binary_constraint}
\end{equation}
so that
\begin{equation}
A_2=\frac{D}{2}-B_1-B_2.
\label{eq:hand_A2_constraint}
\end{equation}
The four edge positions within one period are
\begin{equation}
x_1=0,
\qquad
x_2=\frac{D}{2},
\qquad
x_3=\frac{D}{2}+B_1,
\qquad
x_4=D-B_2,
\label{eq:hand_binary_edges}
\end{equation}
with edge signs
\begin{equation}
s_1=s_3=+1,
\qquad
s_2=s_4=-1.
\end{equation}
Hence the structure factor is
\begin{equation}
S_m
=
1-e^{-iG_mD/2}
+e^{-iG_m(D/2+B_1)}
-e^{-iG_m(D-B_2)}.
\label{eq:hand_Sm_raw}
\end{equation}

It is useful to rewrite the last two edges in terms of the center of the small terrace $A_2$. Define
\begin{equation}
x_c\equiv \frac{x_3+x_4}{2}
=
\frac{3D}{4}+\delta,
\label{eq:hand_xc_def}
\end{equation}
where
\begin{equation}
\delta \equiv x_c-\frac{3D}{4}
=\frac{B_1-B_2}{2}.
\label{eq:hand_delta_def}
\end{equation}
Thus $\delta$ measures the displacement of the small terrace away from the center of the second half-cell. The terrace edges are then
\begin{equation}
x_3=x_c-\frac{A_2}{2},
\qquad
x_4=x_c+\frac{A_2}{2},
\label{eq:hand_x3x4_center}
\end{equation}
because $x_4-x_3=A_2$.

Substituting Eq.~\eqref{eq:hand_x3x4_center} into Eq.~\eqref{eq:hand_Sm_raw} gives
\begin{align}
S_m
&=
1-(-1)^m
+e^{-iG_m(x_c-A_2/2)}
-e^{-iG_m(x_c+A_2/2)}
\nonumber\\
&=
1-(-1)^m
+e^{-iG_mx_c}
\left(e^{+iG_mA_2/2}-e^{-iG_mA_2/2}\right)
\nonumber\\
&=
1-(-1)^m
+2i\,e^{-iG_mx_c}\sin\!\left(\frac{G_mA_2}{2}\right).
\label{eq:hand_Sm_center}
\end{align}

Using $x_c=3D/4+\delta$, we obtain
\begin{equation}
e^{-iG_mx_c}
=
e^{-i3\pi m/2}e^{-iG_m\delta}.
\label{eq:hand_phase_center}
\end{equation}
Now evaluate the first two harmonics explicitly.

For $m=1$, since $1-(-1)^1=2$ and $e^{-i3\pi/2}=i$,
\begin{align}
S_1
&=
2+2i\left(i\right)e^{-iG_1\delta}
\sin\!\left(\frac{G_1A_2}{2}\right)
\nonumber\\
&=
2-2e^{-iG_1\delta}\sin\!\left(\frac{\pi A_2}{D}\right).
\label{eq:hand_S1_exact}
\end{align}

For $m=2$, since $1-(-1)^2=0$ and $e^{-i3\pi}=-1$,
\begin{align}
S_2
&=
2i(-1)e^{-iG_2\delta}
\sin\!\left(\frac{G_2A_2}{2}\right)
\nonumber\\
&=
-2i\,e^{-iG_2\delta}\sin\!\left(\frac{2\pi A_2}{D}\right).
\label{eq:hand_S2_exact}
\end{align}

Define
\begin{equation}
\beta \equiv \frac{\pi A_2}{D},
\qquad
\phi \equiv G_1\delta=\frac{2\pi\delta}{D}
=\frac{\pi(B_1-B_2)}{D}.
\label{eq:hand_alpha_phi}
\end{equation}
Then Eqs.~\eqref{eq:hand_S1_exact}--\eqref{eq:hand_S2_exact} become
\begin{equation}
S_1 = 2\left(1-\sin\beta\,e^{-i\phi}\right),
\label{eq:hand_S1_compact}
\end{equation}
\begin{equation}
S_2 = -2i\,\sin(2\beta)\,e^{-i2\phi}.
\label{eq:hand_S2_compact}
\end{equation}

\subsection{Exact formula for $\mathcal A_{\mathrm{str}}$}

For clarity, first take the sharp-edge limit $F_w(G_m)\to 1$. Using Eq.~\eqref{eq:hand_hm_from_Sm},
\begin{equation}
h_1
=
\frac{H_1}{iG_1D}S_1
=
\frac{H_1}{i2\pi}S_1
=
-\frac{iH_1}{\pi}\left(1-\sin\beta\,e^{-i\phi}\right),
\label{eq:hand_h1_exact}
\end{equation}
and
\begin{equation}
h_2
=
\frac{H_1}{iG_2D}S_2
=
\frac{H_1}{i4\pi}S_2
=
-\frac{H_1}{2\pi}\sin(2\beta)\,e^{-i2\phi}.
\label{eq:hand_h2_exact}
\end{equation}
Therefore
\begin{align}
h_1^2h_2^*
&=
\left[-\frac{iH_1}{\pi}\left(1-\sin\beta\,e^{-i\phi}\right)\right]^2
\left[-\frac{H_1}{2\pi}\sin(2\beta)\,e^{+i2\phi}\right]
\nonumber\\
&=
\frac{H_1^3}{2\pi^3}\sin(2\beta)\,
e^{i2\phi}
\left(1-2\sin\beta\,e^{-i\phi}+\sin^2\beta\,e^{-i2\phi}\right)
\nonumber\\
&=
\frac{H_1^3}{2\pi^3}\sin(2\beta)
\left(e^{i2\phi}-2\sin\beta\,e^{i\phi}+\sin^2\beta\right).
\label{eq:hand_h1sqh2star}
\end{align}
Taking the imaginary part yields
\begin{equation}
\mathcal A_{\mathrm{str}}
=
\frac{H_1^3}{2\pi^3}\sin(2\beta)
\left[\sin(2\phi)-2\sin\beta\,\sin\phi\right].
\label{eq:hand_Astr_exact_1}
\end{equation}
Using $\sin(2\phi)=2\sin\phi\cos\phi$, this can also be written as
\begin{equation}
\mathcal A_{\mathrm{str}}
=
\frac{H_1^3}{\pi^3}\sin(2\beta)\sin\phi
\left[\cos\phi-\sin\beta\right].
\label{eq:hand_Astr_exact_2}
\end{equation}

If the finite edge width is retained, one simply multiplies by the edge form factors:
\begin{equation}
\mathcal A_{\mathrm{str}}
=
\frac{H_1^3F_w(G_1)^2F_w(G_2)}{\pi^3}
\sin(2\beta)\sin\phi
\left[\cos\phi-\sin\beta\right].
\label{eq:hand_Astr_exact_fw}
\end{equation}

Equation~\eqref{eq:hand_Astr_exact_fw} is the desired explicit expression in terms of the binary-cell geometry. It shows that the sign of the structural invariant is controlled by the sign of $\phi$, hence by the sign of $B_1-B_2$.

\subsection{Mirror operation and the sign of the Hall current}

The mirror-related binary cell is obtained by exchanging the left and right groove widths,
\begin{equation}
B_1 \longleftrightarrow B_2.
\end{equation}
From Eq.~\eqref{eq:hand_delta_def},
\begin{equation}
\delta \rightarrow -\delta,
\qquad
\phi\rightarrow -\phi.
\label{eq:hand_mirror_phi}
\end{equation}
Since $\sin\phi$ is odd and $\cos\phi$ is even, Eq.~\eqref{eq:hand_Astr_exact_fw} immediately gives
\begin{equation}
\mathcal A_{\mathrm{str}}(\delta)
=
-\mathcal A_{\mathrm{str}}(-\delta).
\label{eq:hand_Astr_odd_delta}
\end{equation}
Therefore mirror-related gratings have equal-magnitude and opposite-sign structural handedness.

For the explicitly projected static-bias channel, the contribution has the form
\begin{equation}
J_y^{(0)}
=
\frac{\tau_y\,\mathcal A_{\mathrm{str}}}{2}
\left(\Gamma_{\rm ev}\Ffab+\Gamma_3\Mbar\right)\,
\Re\!\left[\chi_{EE}(\omega)\sigma_D^*(\omega)\right]|E_x|^2,
\label{eq:hand_Jy_main}
\end{equation}
the sign of the transverse current is determined by
\begin{equation}
\operatorname{sgn}J_y^{(0)}
=
\operatorname{sgn}\!\left[\mathcal A_{\mathrm{str}}
\left(\Gamma_{\rm ev}\Ffab+\Gamma_3\Mbar\right)\right].
\label{eq:hand_sign_Jy}
\end{equation}
Within that optional channel, fixing the axes and all static-bias coefficients makes the mirror-related choice $B_1>B_2$ versus $B_1<B_2$ reverse the sign of its Hall-current contribution.  For example, if $\Mbar=0$ because the grains are randomly oriented, a distinct deposition-correlated term may remain finite when $\Ffab\neq0$.  This sign rule neither identifies a dominant microscopic origin nor implies a fabrication error: the full phenomenology is written in terms of the generic coefficient $\Lambda_y=\Lambda_y^{\rm stat}+\Lambda_y^{\rm dyn}$, and either part may vanish.  If both mirror-odd static factors vanish, both mirror-related gratings give zero for this particular static mechanism.

\subsection{Small-displacement limit and a simple design rule}

The centered configuration corresponds to
\begin{equation}
B_1=B_2
\qquad\Longleftrightarrow\qquad
\delta=0
\qquad\Longleftrightarrow\qquad
\phi=0.
\end{equation}
Then Eq.~\eqref{eq:hand_Astr_exact_fw} gives
\begin{equation}
\mathcal A_{\mathrm{str}}=0.
\label{eq:hand_Astr_centered_zero}
\end{equation}
Hence merely having a nonzero small terrace $A_2$ is not enough; the small terrace must be \emph{displaced} from the center of the second half-cell.

For $|\delta|\ll D$, equivalently $|\phi|\ll 1$, expand
\begin{equation}
\sin\phi=\phi+\mathcal O(\phi^3),
\qquad
\cos\phi=1-\frac{\phi^2}{2}+\mathcal O(\phi^4).
\end{equation}
Keeping the leading term gives
\begin{align}
\mathcal A_{\mathrm{str}}
&\approx
\frac{H_1^3F_w(G_1)^2F_w(G_2)}{\pi^3}
\sin(2\beta)\left[1-\sin\beta\right]\phi
\nonumber\\
&=
\frac{2H_1^3F_w(G_1)^2F_w(G_2)}{\pi^2D}
\sin\!\left(\frac{2\pi A_2}{D}\right)
\nonumber\\
&\quad\times
\left[1-\sin\!\left(\frac{\pi A_2}{D}\right)\right]\delta.
\label{eq:hand_Astr_small_delta}
\end{align}
Since
\begin{equation}
\delta=\frac{B_1-B_2}{2},
\end{equation}
we obtain the explicit leading-order design rule
\begin{equation}
\mathcal A_{\mathrm{str}}
\propto
(B_1-B_2).
\label{eq:hand_design_rule}
\end{equation}
For the physically relevant range
\begin{equation}
0<A_2<\frac{D}{2},
\end{equation}
one has
\begin{equation}
0<\beta<\frac{\pi}{2},
\qquad
\sin(2\beta)>0,
\qquad
1-\sin\beta>0,
\end{equation}
so the prefactor in Eq.~\eqref{eq:hand_Astr_small_delta} is positive. Therefore, in this regime,
\begin{equation}
\operatorname{sgn}\mathcal A_{\mathrm{str}}
=
\operatorname{sgn}(B_1-B_2).
\label{eq:hand_sign_rule}
\end{equation}

This gives a simple geometrical interpretation:
if $B_1>B_2$, the small terrace is displaced to one side of the second half-cell and the unit cell has one handedness;
if $B_1<B_2$, the displacement is reversed and the handedness flips.
The parameter $A_2$ mainly controls the \emph{magnitude} of the effect through the factor
\begin{equation}
\sin\!\left(\frac{2\pi A_2}{D}\right)
\left[1-\sin\!\left(\frac{\pi A_2}{D}\right)\right],
\end{equation}
while $B_1-B_2$ selects the sign.

\subsection{Connection to optical rectification and effective $\chi^{(2)}$ language}

It is important to distinguish the role of the spatial harmonics of the grating from the optical nonlinearity itself. The OR current is still second order in the optical field because
\begin{equation}
j_x^{(1)}\propto E_x,
\qquad
s_y^{(1)}\propto E_x,
\qquad
J_y^{(0)}\propto j_x^{(1)}s_y^{(1)}\propto |E_x|^2.
\end{equation}
The structural invariant $\mathcal A_{\mathrm{str}}$ is \emph{not} a second-order susceptibility by itself. It supplies the $x$-inversion-odd geometric factor, while $\Lamy$ supplies the independently required $xz$-mirror-odd factor. In this sense one may write schematically
\begin{equation}
\begin{aligned}
J_y^{(0)}
&=\beta_{yxx}^{\mathrm{eff}}|E_x|^2,\\
\beta_{yxx}^{\mathrm{eff}}
&=
\frac{\tau_y\,\mathcal A_{\mathrm{str}}}{2}
\Lamy
\Re\!\left[\chi_{EE}(\omega)\sigma_D^*(\omega)\right].
\end{aligned}
\label{eq:hand_beta_eff}
\end{equation}
Thus the first two \emph{spatial} harmonics determine the structural handedness through $\mathcal A_{\mathrm{str}}$, $\Lamy$ supplies the transverse mirror breaking, and the OR character of the response comes from the quadratic dependence on the optical field.

In summary, the first and second spatial harmonics of the grating profile are the minimal ingredients required to form a translationally invariant, inversion-odd scalar. For the binary cell with $A_1=D/2$, the sign of this scalar is controlled by the displacement
\begin{equation}
\delta=\frac{B_1-B_2}{2}
\end{equation}
of the small terrace from the center of the second half-cell. Hence the choice $B_1>B_2$ versus $B_1<B_2$ directly selects the sign of $\mathcal A_{\mathrm{str}}$. A transverse OR Hall current additionally requires $\Lamy\neq0$.

\section{Summary of what is rigorous and what is model dependent}
\label{app:rigorous_summary}

The symmetry and projected kinetic derivations above fix the following points:
\begin{enumerate}
\item The ideal one-dimensional profile combined with the isotropic Rashba model has the $xz$ mirror and gives zero transverse current in a mirror-even optical state. The $\sigma_y$ edge term alone cannot produce a $y$-directed side jump.
\item The leading physical measure of binary-cell handedness is the translation-invariant inversion-odd invariant
\(
\mathcal A_{\rm str}=\Im[h_1^2h_2^*].
\)
\item Any nonzero response requires an additional mirror-odd quantity. The directional-QNM imbalance $\eta$ is mirror odd and can supply $\Ldyn=\lambda_\eta\eta$ if the Kerr instability occurs.  The coefficient $\lambda_\eta$, not $\eta$ alone, contains the total directional-mode population and electron--SPP coupling scale.
\item A mirror-related two-QNM model permits an antisymmetric instability when Eq.~\eqref{eq:Supp_QNM_threshold} is satisfied. This is an effective-model result; its occurrence in the Au grating has not yet been computed.
\item In the perturbative model of Appendix~\ref{app:microscopic_vertex}, the $\Astr\eta\sigma_y$ vertex contains two electron--SPP vertices and three grating vertices, one of which is a projected Pauli vertex; in the shallow-height limit its order is $g_{\rm SPP}^2\lambda_Ph^3$.  The factor $\Astr$ comes from interference of the $G+G$ and $2G$ paths, $\eta$ from the mirror-odd optical self-energy, and $\sigma_y$ from the complete Hermitian terrace--sidewall Pauli insertion.  The factor $\sin\varphi\cos\varphi'$ is the $y\leftarrow x$ first-harmonic projection, not a bare local potential.
\item A direct $\Astr$-independent current proportional to $\eta$ is symmetry allowed; in particular, the $yz$ mirror does not forbid either $S_x$ or $J_y$.  The Keldysh derivation in Appendix~\ref{app:direct_keldysh} nevertheless gives $\mathcal I_\eta=0$ and hence $S_{x,{\rm dir}}=J_{y,{\rm dir}}^{(0)}=0$ for the $\eta$-linear one-SPP channel at order $g_{\rm SPP}^2$ in the clean, low-harmonic, single-Au(111)-surface-band model at $808~\mathrm{nm}$.  This zero follows from the absence of on-shell energy--momentum support, not from mirror symmetry.  Mode-off-diagonal terms proportional to $a_+a_-^*$ evade the one-photon mismatch but create a finite-$2q_y$ electron coherence, not a uniform order-$g_{\rm SPP}^2$ population source; a spatially averaged contribution requires boundaries, disorder, or another momentum-mixing insertion.
\item For this SPP-dressed route $\lambda_\eta\eta\propto N_+-N_-$.  Its contribution is therefore generally higher than second order in the incident optical field and can acquire nonanalytic power dependence at the bifurcation.  It is a dc rectification channel, but not a power-independent material $\chi^{(2)}$.
\item Sidewall texture, deposition direction, Au(111) texture, and deliberate transverse geometry are nonexclusive possible contributions to $\Lstat$ or to the branch-selection field $h_y$. None is established as dominant. For the explicitly projected dynamical and static edge channels,
\(
I_{{\rm edge},\lambda}^{(2,0)}
\propto
-\partial_\epsilon f^{(0)}_\lambda
\,\mathcal A_{\rm str}
[\eta_\lambda^{(\eta)}\lambda_\eta\eta
+\eta_\lambda^{({\rm ev})}\Ffab+\eta_\lambda^{(3)}\Mbar]
j_x s_y \sin\varphi.
\)
\item A spin-dependent change of $|T|^2$ is skew-scattering-like.  Calling the resulting current a genuine side-jump contribution additionally requires the phase-gradient coordinate shift and its companion anomalous-distribution term.  The manuscript therefore uses ``side-jump-like'' until those finite-device quantities are computed.
\item The frequency dependence of the longitudinal $j_xs_y$ channel follows Eq.~\eqref{eq:Supp_Jy_frequency_explicit}; the SPP spectrum enters through $|F_x|^2$.
\end{enumerate}

By contrast, the QNM frequencies, Kerr coefficients, instability threshold, $\lambda_\eta$, $C_{\lambda\lambda'}^{yx}$, and all static edge coefficients are not fixed by symmetry. They require a nonlinear three-dimensional Maxwell--material calculation, microscopic electronic theory, or controlled experiments.  In particular, the perturbative path establishes the form and order of the projected vertex but does not prove that helicity and path sums are nonzero for the fabricated sample.  Likewise, finite broadening and surface-to-bulk, interband, disorder-, phonon-, edge-, or contact-assisted absorption can restore a nonzero direct coefficient beyond the controlled clean-limit result of Appendix~\ref{app:direct_keldysh}.  The manuscript therefore presents the Kerr--QNM bifurcation and its structure-dependent electronic readout as falsifiable effective mechanisms, not as completed ab-initio results.

\section{Remarks on citations and related literature}
\label{app:literature_remarks}

The side-jump displacement formula, its gauge-invariant meaning, and its role in Hall transport are standard and discussed extensively in the anomalous and spin Hall literature \cite{Berger1970,Sinitsyn2008,Nagaosa2010,Sinova2015,Atencia2022}. For broader second-order Hall-type transport, modern semiclassical perspectives, including disorder, side-jump, and skew scattering, are developed in Refs.~\onlinecite{Xiao2019,Du2019,Du2021,Koenig2021}. QNM expansions and nonlinear plasmonics are reviewed in Refs.~\onlinecite{Sauvan2013,Lalanne2018,Kauranen2012}; counterpropagating Kerr symmetry breaking is demonstrated and analyzed in Refs.~\onlinecite{DelBino2017,Hill2020}. Directional guided-mode excitation is discussed in Ref.~\onlinecite{RodriguezFortuno2013}. The experimental motivation is the observation of longitudinal and transverse optical rectification in structurally asymmetric gold gratings~\cite{Moroshkin2021,Plumitallo2023}, while the Au(111) Rashba surface-state context is summarized in Refs.~\onlinecite{Yan2015,Henk2003}.

\ifrevtex\else\small\fi
\ifrevtex\bibliographystyle{apsrev4-2}\else\bibliographystyle{unsrt}\fi
\bibliography{ORHall_Edelstein_PRB_refs_v4}

\end{document}